\documentclass{article}
\usepackage[utf8]{inputenc}
\usepackage{graphicx}
\usepackage{multicol}
\usepackage{authblk}
\usepackage[margin=0.8in]{geometry}
\usepackage{array}
\usepackage{hyperref}
\usepackage{tcolorbox}
\usepackage{markdown}
\usepackage{xcolor}
\usepackage{parcolumns}
\usepackage{listings}
\usepackage{adjustbox}
\usepackage{float}
\usepackage{enumitem}
\lstdefinestyle{base}{
  breaklines=true,
  basicstyle=\tiny\ttfamily\color{black},
  moredelim=**[is][\color{red}]{@}{@},
}

\usepackage[sorting = none, backend = bibtex, style=numeric-comp]{biblatex}
\title{Automated Creation of Source Code Variants of a Cryptographic Hash Function Implementation Using Generative Pre-Trained Transformer Models}

\author[1]{Elijah Pelofske\thanks{E-mail: elijah.pelofske@protonmail.com}}
\author[1]{Vincent Urias}
\author[2,1]{Lorie M. Liebrock}
\affil[1]{Sandia National Laboratories}
\affil[2]{New Mexico Cybersecurity Center of Excellence, New Mexico Tech}

\date{\vspace{-6ex}}

\begin{document}

\maketitle

\begin{abstract}
Generative pre-trained transformers (GPT's) are a type of large language machine learning model that are unusually adept at producing novel, and coherent, natural language. Notably, these technologies have also been extended to computer programming languages with great success. However, GPT model outputs in general are stochastic and not always correct. For programming languages, the exact specification of the computer code, syntactically and algorithmically, is strictly required in order to ensure the security of computing systems and applications. Therefore, using GPT models to generate computer code poses an important security risk -- while at the same time allowing for potential innovation in how computer code is generated. In this study the ability of GPT models to generate novel and correct versions, and notably very insecure versions, of implementations of the cryptographic hash function SHA-1 is examined. The GPT models \texttt{Llama-2-70b-chat-hf}, \texttt{Mistral-7B-Instruct-v0.1}, and \texttt{zephyr-7b-alpha} are used. The GPT models are prompted to re-write each function using a modified version of the localGPT framework and langchain to provide word embedding context of the full source code and header files to the model, resulting in over $150,000$ function re-write GPT output text blocks (that are potentially correct source code), approximately $50,000$ of which were able to be parsed as C code and subsequently compiled. The generated code is analyzed for being compilable, correctness of the algorithm, memory leaks, compiler optimization stability, and character distance to the reference implementation. Remarkably, several generated function variants have a high implementation security risk of being correct for some test vectors, but incorrect for other test vectors. Additionally, many function implementations were not correct to the reference algorithm of SHA-1, but produced hashes that have some of the basic characteristics of hash functions. Many of the function re-writes contained serious flaws such as memory leaks, integer overflows, out of bounds accesses, use of uninitialised values, and compiler optimization instability. 
Compiler optimization settings and SHA-256 hash checksums of the compiled binaries are used to cluster implementations that are equivalent but may not have identical syntax - using this clustering over $100,000$ distinct, novel, and correct versions of the SHA-1 codebase were generated where each component C function of the reference implementation is different from the original code. 
\end{abstract}


\section{Introduction}
\label{section:introduction}

Generative Pre-Trained Transformer (GPT) models are a type of Large Language Model that has shown to be highly capable at a large number of natural language processing tasks, including computer code \cite{chen2021evaluating, vaswani2023attention, touvron2023llama, brown2020language, openai2023gpt4, dao2022flashattention, su2023roformer, shazeer2019fast}.

In this study we explore whether current GPT models can be used to generate correct algorithmic invariant implementations of a cryptographic hash function in C code. Specifically, we examine the task of rewriting an implementation of the cryptographic hash function known as SHA-1 \cite{SHA1_NIST_specifications}. An interesting byproduct of code-rewriting with the GPT models is that a large number of the implementations are wrong in a variety of surprising ways, including containing software risks. These include compiler optimization instability (meaning that the output changes based on what compiler optimization settings were used), memory leaks, integer overflows, out of bounds writes, and implementations that are correct for some test vectors but not correct for other test vectors. Implementation risks in cryptographic algorithms is a critically important type of bug that exists in cryptography library implementations \cite{cryptoeprint:2023/331, 9426079, 6494288, 8405614}. This study serves to caution that arbitrarily using GPT models for creating, or rewriting, source code can introduce serious flaws. GPT models, with their current capabilities, work well as research tools to study interesting source code variants, but using them for practical code generation poses a software security risk. 

Importantly, SHA-1 \cite{SHA1_NIST_specifications} is considered \emph{broken} since at least one method of generating a SHA-1 hash collision has been demonstrated \cite{stevens2017first}, and is generally considered not secure for validating the integrity of data based on a variety of attacks \cite{wang2005finding, biham2005collisions, karpman2015practical, stevens2013new, 255248}. However, SHA-1 is still widely deployed in information technology systems, and moreover its implementation is relatively short compared to other cryptographic algorithms, making it a good candidate for a proof of concept study on the capabilities of using GPT models for rewriting a cryptographic algorithm implementation. 

In the context of malware detection and analysis, polymorphic versions of malware binaries can be very easily constructed \cite{10.1145/1315245.1315312}. However, changing the underlying syntax while maintaining the same functionality is harder to do, especially in an automated manner. Here, we show that GPT models can be used as tools for creating unique versions of a codebase. This is important because this capability of GPT models can, and likely will, be used by malware developers to create software implementations that have distinct behaviors and syntax, thus making their detection more difficult. 

Automated code re-writing, in particular of cryptographic algorithm implementations, is an excellent test case for evaluating GPT models capabilities involving computer code because the implementation must be exactly correct or software implementation risks will be introduced. The validity of GPT produced cryptographic function re-writes is highly testable for correctness and implementation flaws such as being incorrect for some proportion of inputs or memory leaks. A cryptographic algorithm serves as a good reference benchmark for the capabilities of generative machine learning to correctly rewrite computer code because they need to implemented exactly correctly or they do not work -- and in particular the code is therefore highly testable for correctness. Although there exist many instances of implementations of cryptographic algorithms in the codebases that very likely were used to train many of the GPT models in existence today, they are not as common (and in as many different varieties) as other foundational functions in computing (e.g., especially for teaching programming, such as sorting algorithms), and moreover their secure and correct implementation is extremely consequential. 

Importantly, code re-writing and synthesis tasks are severely constrained by the \emph{token context window size} of the given GPT model. In this case, the cryptographic source code functions and the input prompt do fit within the context window, but depending on the given run the GPT model output may overrun that context window and begin to generate incoherent text. Future evaluations of code re-writing will also be limited by the context window of the GPT models, and therefore GPT models with larger context windows will be needed for synthesizing larger pieces of computer code. 

When parsing the generated code, we do not impose any further post-processing beyond attempting to extract the code from assumed markdown-style code formatting. In particular, the extracted strings are directly substituted in the source code for the original function implementation. This means that if the format was correct, we allow the GPT model output to include additional function definitions or even additional standard libraries -- the test of whether any of such code modifications succeed is determined by the ensemble of compile attempts and algorithmic correctness tests.

The re-written C code functions are evaluated in a number of ways, most importantly by being compilable by both \texttt{gcc} and \texttt{clang} \cite{Clang, 1281665, lattner2008llvm} with a variety of optimization levels, and by algorithmic correctness. Memory leaks, memory allocation flaws, and out of bounds writing are checked using the address sanitizer in clang and gcc \cite{37752}, as well as Valgrind with memcheck \cite{nethercote2003valgrind, nethercote2004dynamic, 10.1145/1273442.1250746, 10.1145/1254810.1254820, seward2005using}.

\subsection{Brief Literature Overview of GPT Model Code Generation}
\label{section:introduction_literature_overview}

Ref. \cite{cryptoeprint:2023/606} used chatGPT to implement cryptographic algorithm source code, however it was in a semi-supervised chat interaction, not in an automated or systematic study. 

There exist several previous studies on measuring the capability and accuracy for code generated using GPT models \cite{valerolara2023comparing, 10.1145/3368089.3417058, narasimhan2021cgems, perez2021automatic, 10.1145/3643681, 10.14778/3551793.3551841, li2023large}, however testing for accuracy and completeness can be difficult especially for extremely complicated computer code. Previous studies have also used LLMs for helping with fixing compilation errors \cite{deligiannis2023fixing} and generally as an assistant for writing code \cite{agarwal2024copilot}. 

Several studies have investigated the ability of GPT models to repair source code that contains flaws \cite{10.1145/3524459.3527350, 10.1145/3623476.3623522, 10179324, lajko2022fine, 10.1145/3626252.3630875, 10.1145/3611643.3613892}. The more general task of using GPT models to generate source code has also been studied in several contexts \cite{narasimhan2021cgems, 10.14778/3551793.3551841, huang2024generative, siddiq2023lightweight, spiess2024calibration}; overall such generated code has the same problem as human developer written code which is properly unit-testing for correctness. The task of automatically producing unit tests for source code and software using GPT models has also been studied \cite{tufano2021unit, 10.1145/3611643.3617850, 10.1145/3624032.3624035, 10485640, li2023prompting}.

\section{Methods}
\label{section:methods}

\subsection{GPT Model Implementation}
\label{section:methods_GPT_implementation}

For this study, langchain \cite{Chase_LangChain_2022}, and in particular the software codebase localGPT \cite{localGPT}, is used with the goal of \emph{anchoring} the GPT output within the reference hash function code -- the entirety of the code, not just the single function being re-written within each inference call. This technique is generally known as retrieval augmented generation \cite{lewis2021retrievalaugmented}, and the goal is to provide sufficient context via word embeddings of a corpus of text we wish to extract information from such that the GPT models will generate text that is grounded in the content of those documents.

localGPT is used to first create word embeddings, tailored for text generation using the \texttt{instructor-xl} model \cite{su2023embedder}, of the entirety of the original source code, which is comprised specifically of the SHA-1 reference implementation and the corresponding header file (shown in Appendix~\ref{section:appendix_reference_source_code_SHA-1} as Code Listing~\ref{source_code:SHA-1_reference_functions} and Code Listing~\ref{source_code:SHA-1_reference_header_file}). Note that the original macro and definition comments are left in the header file, meaning that the text will be in the generated word embeddings, so as to provide better context for the functionality of the code. These files are parsed as raw text files. 

The localGPT inference calls are performed on a small local cluster on four Nvidia A100 GPU's~\cite{9361255} with 82 Gigabytes of memory, with CUDA Version 12.4, and the GPT models are all obtained from the huggingface GPT repository \cite{wolf2020huggingfaces}. The GPT models were trained and run using the Python 3 library PyTorch \cite{Paszke_PyTorch_An_Imperative_2019}. 

The localGPT prompt template used is a combination of \emph{Context} and the user-facing \emph{question}. No chat history was used for the prompting of the re-written code. Regardless of the type of the underlying model, the high-level organization of the prompt is the system prompt \cite{localGPT}, followed by the context of the generated word embeddings, followed by the user prompt. The system prompt used in this study, which is from a version of the localGPT codebase \cite{localGPT}, is the following:

\begin{tcolorbox}
You are a helpful assistant, you will use the provided context to answer user questions.
Read the given context before answering questions and think step by step. If you can not answer a user question based on the provided context, inform the user. Do not use any other information for answering user. Provide a detailed answer to the question.
\end{tcolorbox}

A total of three pre-trained language models are used in this study;~\texttt{Llama-2-70b-chat-hf}~\cite{touvron2023llama}, \texttt{Mistral-7B-Instruct-v0.1}~\cite{jiang2023mistral}, and \texttt{zephyr-7b-alpha}~\cite{tunstall2023zephyr}. \texttt{Llama-2-70b-chat-hf}~has a maximum token context window of 4,096, \texttt{Mistral-7B-Instruct-v0.1} and \texttt{zephyr-7b-alpha} both have a maximum token context window of 32,768. These GPT models are intended to be prompted in a chat-type manner of interaction. These three GPT models were chosen as a representative group for their relatively large context window, and their overall good performance for handling computer code. However, there are a very large number of GPT models in general and there are likely many other models that may perform very well, or even better, than these three.

For each inference call to the GPT model, the source code of the function is appended (along with a newline character) after the prompt text. Each prompt needs to ensure a few things. The first is that the code needs to be enclosed in triple backticks (also known as triple backquotes) so that the code can be automatically parsed from the output. The second is that the code needs to be compatible with the rest of the SHA-1 codebase, including usage of macros, functions from library imports when required, and using consistent function naming schemes so that the algorithm can be automatically tested. In cases where these requirements are not adhered to, the result is either a failure to compile or the compiled binary having a critical error when executed. The prompts are also intended to be code-agnostic; for example the prompts are not requesting a specific algorithm or type of syntax be used. Lastly, the primary intention is for the generated code to be correct, but to have different syntax than the original. A total of $10$ prompts are tested which aim to produce generative text output that has these desired properties. The exact text of these $10$ prompts are given below:

\begin{tcolorbox}[title=Prompt 1]
Re-write this C code function into an entirely different function that maintains the same functionality as the original code and uses the same function name. Enclose the code in triple backquotes. 
\end{tcolorbox}

\begin{tcolorbox}[title=Prompt 2]
Re-write this C code function into an entirely different function that maintains the same functionality as the original code and uses the same function name. Use different syntax choices when re-writing the source code including but not limited to different control flow, equivalent but different array indexing, different logic operations, different variable types, different algorithm choices, and different variable names. Be creative! Try to obfuscate the intended functionality of the code as much as possible while retaining the same functionality. Enclose the code in triple backquotes. 
\end{tcolorbox}

\begin{tcolorbox}[title=Prompt 3]
Re-write this C code function using different variable names, control flow, and array indexing, so that the functionality of the code is obfuscated, but the functionality is the same as the original code. Use the same function name as the original code and enclose the rewritten code in triple backquotes. 
\end{tcolorbox}

\begin{tcolorbox}[title=Prompt 4]
Obfuscate this C code function by rewriting the syntax and making the code more complicated than it needs to be while performing the same functionality as the original code. Use the same function name as the original code and enclose the rewritten code in triple backquotes.
\end{tcolorbox}

\begin{tcolorbox}[title=Prompt 5]
Obfuscate this C code. Enclose the code in triple backquotes. 
\end{tcolorbox}

\begin{tcolorbox}[title=Prompt 6]
Re-implement this C function using a different implementation with changed logic and variable names. Use the same function names as the original, and enclose the code in triple backquotes. 
\end{tcolorbox}

\begin{tcolorbox}[title=Prompt 7]
Rewrite this C code using different variable names and different control flow logic, but keep the function name the same. Enclose the code in triple backquotes. 
\end{tcolorbox}

\begin{tcolorbox}[title=Prompt 8]
Re-implement this C function using different logic and variable names. Use the same function names as the original code, and enclose the new code in triple backquotes. 
\end{tcolorbox}

\begin{tcolorbox}[title=Prompt 9]
Act as a professional C code developer. Re-implement this C function using different logic and variable names. Use the same function names as the original code, and enclose the new code in triple backquotes. 
\end{tcolorbox}

\begin{tcolorbox}[title=Prompt 10]
Please alter this C function so that it uses completely different, but still valid, C syntax such that it performs the same computations as the original code. Surround the new C function in triple backticks and use the same function names as the original code. Do not write explanations or justifications in your reply; write only the new C function and nothing else. 
\end{tcolorbox}

For each of the $10$ prompts, a total of $100$ text-generation calls are performed, and the inference temperature is varied across $11$ temperature settings between the maximum of $1$ and $0.01$; $\{1, 0.9, 0.8, 0.7, 0.6, 0.5, 0.4, 0.3, 0.2, 0.1, 0.01\}$. The inference temperature of $1$ gives the generative text calls higher stochasticity, meaning that each of the $100$ inference calls will very likely output very different source code. The inference temperature of $0.01$ gives nearly deterministic code generation. Temperature here refers to the type of distribution that is sampled from whenever each next token is chosen. Here, sampling is always on meaning that the temperature can not be exactly $0$, but $0.01$ means that the next token that is chosen is almost always the most likely choice (based on the pre-training of the model), and higher temperatures mean that the sampling has some higher likelihood of being a not very likely token. An inference temperature of $100$ would give closer to uniform sampling across the tokens when making the choice of what token to produce next. Because of the nature of computer code requiring very exact syntax, we opted for temperature settings starting at $1$ and lower so as to (hopefully) generate coherent code. The wide range of inference temperatures gives provides output code that is highly diverse. This procedure is then repeated for each of the $3$ GPT models, and is also repeated for the four component functions of the chosen SHA-1 algorithm implementation \cite{cryptographic_algorithms_Github}. Therefore, in total there are $10 \cdot 100 \cdot 11 \cdot 4 \cdot 3 = 132,000$ function re-write attempts that are generated and then parsed. Although, the experiments shown in this study do not fully cover this parameter space, only about 90\% of the parameters are covered -- however due to the \emph{extra} function re-writes that are parsed, over $132,000$ function re-writes are produced in total. This large number of function re-writes is motivated by the inherent randomness of the GPT model output, and therefore it is important to quantify a large \emph{distribution} of code samples so as to understand the variability in GPT produced code.

\subsection{GPT Output Parsing and Code Testing}
\label{section:methods_GPT_output_parsing_and_testing}

The SHA-1 implementation re-writing is performed entirely on the four component functions of the SHA-1 C code given by this Github repository \cite{cryptographic_algorithms_Github}. This particular SHA-1 implementation was chosen because it is relatively self contained in terms of required external libraries, and is relatively short. However, the same methodology described in this study could be easily applied (with a reasonable amount of GPU compute time) to other SHA-1 implementations, or any other source code implementation. 

With the goal of fully evaluating each of the GPT re-written functions, we execute the following series of parsing attempts and tests for each re-written function, in sequence. 

\begin{enumerate}[noitemsep]
    \item The first step is to parse the GPT output to extract the source code. GPT output, at least for the models we tested, can be quite unstructured and not adhere to prompted formats. Here we apply a reasonably black-box approach, in that we do not apply extensive natural language processing to separate out natural language descriptions from source code. The requested format for all of the prompts is to output the re-written C code enclosed in triple backquotes (also known as triple back-ticks). The triple backquotes formatting is chosen because it is quite distinguishable and parseable for the automated execution of the source code. GPT models typically output natural language text in addition to generated code, and therefore we need to use a mechanism to (hopefully) differentiate the source code. It is also markdown formatting syntax, commonly used for code formatting in software development documentation, so it should be a reasonably consistent formatting choice with the source code the GPT models were trained on. Therefore, the following sequence of initial parsing is attempted:
    \begin{enumerate}[noitemsep]
        \item Apply a regular expression to split the output string into an array at all instances of triple backquotes, and return the string at index 1 (indexing beginning at 0). Note that if the beginning of the string is the first triple backquotes, the string at index 1 is the intended source code function text. If there are not triple backquotes, proceed to step 2. If the output text contained two triple backquote lines enclosing some source code (with anything else before or after that block), then this parsing will succeed correctly, as in the generated code will be extracted. This parsing method specifically extracts the strings in between the delimiter of a pair of triple backticks -- this allows the GPT output to have any amount of further text or backticks in the output as this will not be parsed. 
        \item Apply a regular expression to split the output string into an array at all instances of a \emph{single} backquote, and return the string at index 1, as before. If there are not backquotes in the output string, proceed to step 3. In some instances, the GPT model output would enclose the function code in single backquotes instead of the requested triple backquotes. This is not entirely unreasonable since this is also a used in-line markdown formatting, usually for code syntax. Therefore, if step 1 fails we do apply this step in order to maximize the parse-able test cases. This case does not occur very frequently, but it does happen in various GPT model outputs. 
        \item Return the raw source code with no parsing modifications. If we reach this step, most likely the generated text output does include some non C-syntax characters. But, in some cases the GPT output is actually just re-written C code with no ancillary text, and in those cases this parsing works. 
    \end{enumerate}
    \item If in the previous parsing step, the first or second step applied, then we will apply a further post-processing step here. This step was motivated by some of the GPT model output adhering to markdown formatting in a somewhat strange way, which is by including the computer code language identifier immediately following the markdown triple, or single, back-quotes, such as \texttt{\textasciigrave \textasciigrave \textasciigrave Python}. Therefore, this parsing step is to split remaining text by newlines into an array and check if the first index is the same as any of the standard computer language identifiers. If it is, then we remove that text, and proceed. We also check minor edge cases such as computer language code identifiers followed by a single whitespace, and remove those cases as well. The list of computer code language identifiers that are checked for are given in Appendix \ref{section:appendix_markdown_code_identifiers}. This parsing step is safe, in the sense that it does not remove any potentially valid C code syntax, since all of the language identifier extensions are not valid C code for the beginning of a function. 
    \item Since we are considering each function individually, we want to evaluate that function by itself, with no other changes to the source code. Therefore, we directly substitute this re-written function (which, at this stage is simply the post-processed string from the GPT model) for the correct function from the reference implementation, concatenate the other reference implementation C code with the re-written C code, and write that combined string to a new C file. Since the function re-writing does not include the macros, and library imports, those strings are also written to the new C file. 
    \item The new C file is compiled, along with the reference header file (none of which is ever re-written by the GPT models), along with the reference test C code file (which contains the \texttt{main} function as the entrypoint for the program) that will call the functions on test vectors and check if the output is correct. This codebase is compiled using both \texttt{clang} and \texttt{gcc}. For \texttt{gcc}, the optimization levels of 0/1/2/3/fast/s are used (and all default settings otherwise). For \texttt{clang}, the optimization levels of 0/1/2/3/fast/s/z are used (and all default settings otherwise). s denotes space optimization compilation, z denotes heavier space reduced optimization. This codebase is compiled on the same computer, and using all identical settings to ensure maximum similarity for the eventual checking of identical generated binaries. The large number of compiler options at this step can produce a variety of outputs and behaviors, including a binary being generated with no errors, a binary being generated but with warnings, a binary not being generated at all, and various compiler optimization instabilities. Importantly, note that the original source code is compiler optimization stable. 
    \item If the compilation step (this step is tested for each compiler setting) produces a compiled binary, then the binary is executed. This binary is first a check of the correctness of the code, but it also outputs what the computed cryptographic function output is on the input of the test vectors (for example, even if it is incorrect) which can be analyzed at a later point as well. Note that the GPT re-write can introduce ancillary text as well besides simply computing the cryptographic function (which is almost certainly measured as a fail case because the code can not be compiled). At this point of executing the binary, there is no guarantee that it is correct -- for example, it can (and we found examples that have) return an error code, it can enter into an apparent infinite loop (which is checked by a timeout), or it can have memory leaks. 
    \item The last step is is to perform a series of automated memory leak, memory allocation error, and out of bound write checks. These steps are attempted on every function re-write; in the cases where the code does not compile, or the binary throws an error when executed (or times out due to an apparent infinite loop), the test aborts and no memory leak or out of bounds information is obtained. This involves four separate compilation and binary execution steps. All of these four compilation attempts use an optimization level of $0$ so as to minimize potential errors, such as false positives, for the automated tools evaluation. The binary output from these tools executing and or compiling the code is recorded, but is not analyzed for correctness or interesting outputs; the output of the tools however is parsed in an automated system to find specific phrases that indicate the lack or presence of certain errors. 
    \begin{itemize}[noitemsep]
        \item Compile the code using \texttt{gcc} with flags \texttt{-g} and \texttt{-fno-omit-frame-pointer}, with the full address, undefined, and bounds sanitizer check flags enabled \cite{37752}: \texttt{-fsanitize=address}, \texttt{-fsanitize=bounds}, \texttt{-fsanitize=undefined} (and statically linked with \texttt{-static-libasan}). Then execute the compiled binary, if it was compiled, and record the output. The automated sanitation checks will print metadata about flaws in the code such as memory leaks or out of bounds accesses. 
        \item Compile the code using \texttt{clang}, and supply the same address, bounds, and undefined sanitize checks \cite{37752} as in the above case. 
        \item Compile the code using \texttt{gcc} and all default flags, then if the binary was produced run it through valgrind and record the output (catch exceptions, and set a timeout check usual to handle the binaries with fatal errors or apparent infinite loops). The valgrind flags that are specified are: \texttt{--tool=memcheck, -s, --leak-check=full, --leak-check=yes, --show-leak-kinds=all, --error-limit=no}
        \item Perform the same steps as the previous test, but using \texttt{clang}
    \end{itemize}
    \item Finally, this last step is not actually parsing the primary function, as described in all of the previous steps. Instead, this step allows us to obtain \emph{extra} function re-writes that were generated by the GPT output, but were not parsed in any of the previous steps. This step was implemented because it was found that in many of the GPT outputs, there were additional markdown formatted fields of code in the output strings. This step splits the output strings based on all occurrences of triple back-ticks, then reading in all resulting split substrings (skipping the first one at index 1, since this was the function parsed in Step 1). Then, for each of these substrings, they are considered potential function re-writes if they contain at least one of each of the following characters; \{, \}, (curly braces) (, ), (parentheses) \_ (underscore). These checks are used because any valid C functions needs the curly braces and the parentheses, and then all of the function names used in this codebase contain an underscore. If these checks pass, then the string is parsed and is considered as a potential function rewrite - this means specifically that it then is processed by steps 2-6. The number of extra functions that this step can produce is overall somewhat small (this usually adds about 10 percent additional functions), but some of the GPT outputs can have upwards of 100 function re-writes that get added at this step. 
\end{enumerate}

Note that all of these steps (not including the actual GPT inference calls) are intended to be deterministic -- these steps are intended to be fully reproducible in order to analyze the GPT source code outputs. 

Using the metadata generated from these tests and parsing steps, we compute the following aggregated metrics on the generated source code function variants. Note that these metrics are computed specifically on the individual re-written functions, while leaving all other aspects of the codebase unchanged. The GPT generated code can, and often does, include function names which are not compatible with the assumed function names of the component functions, adds extra functions, among a variety of other interesting outputs. In such cases, no extra parsing is performed -- the above tests are executed in a fully automated manner, and the output is then quantified by the following metrics so as to obtain a high-level summary of the types of code produced.

\begin{itemize}[noitemsep, wide, labelwidth=!, labelindent=0pt]
    \item[\textbf{Metric 1.}] Count of how many of the function variants were able to be compiled, in the sense that a binary was produced from the compilation even if there were warnings, for all compiler settings. 
    \item[\textbf{Metric 2.}] Count of how many of the function variants were \emph{compiler optimization unstable}, meaning that the compilation was successful for at least one, but not all, of the compiler settings. 
    \item[\textbf{Metric 3.}] Count of how many of the function variants were output-verified (e.g., the implementation was correct for the test vectors) for all compiler optimization settings. 
    \item[\textbf{Metric 4.}] Count of how many of the function variants were algorithmically incorrect in some way for all compiler settings. In particular, this means that for all compiler variants output was produced (meaning that the binary did not crash, for example, for any of the compiler settings), but the output was incorrect. The way that the output is incorrect can vary - from being off by a single character, to adding large amounts of output that is ancillary. This count is strictly for the cases where the compiled binaries were able to be produced for all compiler optimization settings. 
    \item[\textbf{Metric 5.}] Count of how many of the function variants, for which binaries could be compiled, were compiler optimization unstable for their algorithmic correctness -- meaning that for some compiler optimization settings the code passed all cryptographic algorithm test vectors, but for others it failed. This test allows other optimization settings to cause the binary to not be compiled or executed with a critical or timeout error; the relevant count here measures purely if there were two compiler optimization settings where one resulted in a successful algorithmic test check, and the other resulted in a failed algorithmic test check. 
    \item[\textbf{Metric 6.}] Count of how many of the function variants that were correct for all compiler settings and had a Levenshtein character distance of 0 to the original source code (meaning, that the source code is strictly identical to this function variant), after repeated whitespace was removed and all comments were removed. 
    \item[\textbf{Metric 7.}] Count of how many of the function variants that were correct for compiler settings had a Levenshtein distance greater than 0 with respect to the original source code after repeated whitespace was removed and all comments were removed. This metric shows how many correct function re-writes there were, in the sense that the underlying code was changed in some way, and the code was correct (and not compiler optimization unstable). 
    \item[\textbf{Metric 8.}] Count of how many of function variants produced a compiled binary that crashed due to a timeout error (e.g., a presumed infinite loop) for all compiler settings. The timeout threshold was set at 10 seconds, and for reference, the original implementation completed all tests in less than 1 second of CPU time.
    \item[\textbf{Metric 9.}] Count of how many of the function variants produced a compiler binary that crashed due to critical error, for all compiler settings. 
    \item[\textbf{Metric 10.}] Count of how many of function variants produced a compiled binary that crashed due to a timeout error (e.g., a presumed infinite loop) for at least one, but not all, compiler setting. Meaning that the occurrence of this error is unstable and dependent on the compiler optimizations. 
    \item[\textbf{Metric 11.}] Count of how many of the function variants produced a compiler binary that crashed due to a critical error (for example, a segmentation fault), for at least one, but not all compiler setting. Meaning that the occurrence of this error is unstable and dependent on the compiler optimizations.
    \item[\textbf{Metric 12.}] For each compiler setting, a count of how many unique (identical) SHA-256 hash clusters exist for the binaries that were output-verified for all compiler settings and whose source code was distance 1 or greater away from the original (with comments and repeated whitespace removed). One integer is reported for each of the 13 compiler settings (separated by dashes). The hash clusters were computed by taking the hash of each compiled binary; if that hash was the same as other compiled binaries, then they were assigned the same cluster. This hash based clustering abstracted away the problem of similar source code (away from directly computing text distance) to compiled binary similarity. This works especially well thanks to compiler optimization, which can elucidate cases where two pieces of source code are really doing the same computations but just with slightly different syntax (and not entirely different datastructures or control flow). The order of the reported clusters is; gcc level 0, gcc level 1, gcc level 2, gcc level 3, gcc level s, gcc level fast, clang level 0, clang level 1, clang level 2, clang level 3, clang level s, clang level fast, clang level z. 
    \item[\textbf{Metric 13.}] This step aggregates the compiler setting hash clustering into a unified graph (e.g., network) that reveals even more underlying clustering of the source code variants. The graph is defined by each binary being represented by a node, and edges are formed between nodes if the SHA-256 checksum of the two binaries is equal. Next, we check within all of these existing clusters if there exist any with hashes that are equal to the implementations with text distance of 0 to the original source code. This cluster is removed, and not included in the returned counts. In practice, we found there was always exactly 1 such cluster, and it was the largest cluster. The final metric that is reported is an aggregated number of how many disconnected components of the meta-graph exist (this combines binaries that were compiled with the same source code, in addition to the SHA-256 checksum formed graph), which corresponds to showing how many \emph{actually unique} variants of this C code function were generated by the GPT models. Here we also report how many source code versions exist in each of the clusters of the meta-grouping. Note that function variants within each cluster may share identical syntax to each other -- this clustering is specifically intended to delineate unique algorithmic invariant implementations of the original code. The integer counts of clustered group sizes are given in an unsorted sequence, separated by dashes. 
    \item[\textbf{Metric 14.}] The number of source code variants that were found to be duplicates of the original source code, but only found via hashing of the compiled binary in Metric 13. 
    \item[\textbf{Metric 15.}] Count of how many of the function variant and compiler setting tuples that did not adhere to the basic format of the algorithm output, and produced some ancillary output strings or non-unicode characters (an example of this could be appending print messages on the internal state of the algorithm). All of these variants are decidedly incorrect, but they are incorrect for potentially additional reasons besides implementing the algorithm incorrectly. 
    \item[\textbf{Metric 16.}] Count of how many of the function variants that under some compiler settings will not compile, whereas for any other compiler setting the code can be compiled and is output-verified. Note that in practice, we never observed an example of this case ever occurring. 
    \item[\textbf{Metric 17.}] Count of how many of the function variants produce binaries that are correct for at least one, but not all, of the test vectors, and this behavior (of some outputs being correct, but others are incorrect ) is the same for all compiler optimization settings. In particular, all compiler optimization settings produced a compiled binary, all of the binaries executed without an error, and the compiled binaries failed to pass all of the SHA-1 tests. \textbf{These instances are examples of implementation risks, since if these are under-tested they could pass for being correct and then subsequently fail to authenticate the integrity of data. }
    \item[\textbf{Metric 18.}] Count of how many of the function variants produce binaries that are correct for at least one, but not all, of the test vectors, and this behavior (of some outputs being correct, but others are incorrect ) is compiler optimization unstable (meaning that this occurs for at least one, but not all, of the compiler optimization settings). The cases which are not correct could be because that optimization setting caused the binary to not be compiled or to result in an error status. \textbf{These instances are examples of implementation risks, since if these are under-tested they could pass for being correct and then subsequently fail to authenticate the integrity of data. }
    \item[\textbf{Metric 19.}] Count of how many of the function variants produce binaries that are incorrect for all test vectors, but for at least one test vector the output hash is 5 characters or less away from the correct hash, and the output hashes are deterministic regardless of the compiler optimization used. This character distance measure is strictly from the generated hash - if the generated hash contains fewer characters than the correct hash, the missing characters are not counted towards the character distance. In particular, all compiler optimization settings produced a compiled binary, all of the binaries executed without an error, and the compiled binaries failed to pass the SHA-1 tests. The choice of distance of $5$ characters is arbitrary - it was selected to identify clear cases where there was very minimal change to the output hash compared to the correct SHA-1 implementation. Notably, instances found by this metric are interesting because they could fail human visual authenticity checks. 
    \item[\textbf{Metric 20.}] Count of how many of the function variants produce binaries that are incorrect for all test vectors, but for at least one test vector the output hash is 5 characters or less away from the correct hash, and the output hashes are inconsistent across different compiler optimization used. The cases which do are not correct could be because that optimization setting caused the binary to not be compiled, or to result in an error status. Like Metric 19, instances found by this metric are notable because they could fail to be found to be incorrect from human visual authenticity checks. 
    \item[\textbf{Metric 21.}] Count of function variants that were incorrect, but the output (e.g., the raw output of the test functions) changed, in at least one way, depending on the compiler optimization level. This count is specifically for the function variants where an executable was able to be compiled and executed without critical or timeout errors for all compiler optimization settings. 
    \item[\textbf{Metric 22.}] Count of how many of the function variants that are incorrect for all test vectors are compiler optimization stable, meaning that the output is the same for all of the tested compiler settings. This count is specifically for the function variants where an executable was able to be compiled and executed without critical or timeout errors for all compiler optimization settings. 
    \item[\textbf{Metric 23.}] Count of how many functions variants were optimization unstable in the sense that for some settings there was a critical error, but for others the resulting binary was correctly output-verified. 
    \item[\textbf{Metric 24.}] Count of how many functions variants were optimization unstable in the sense that for some settings there was a timeout error (likely infinite loop), but for others the resulting binary was output-verified. 
    \item[\textbf{Metric 25.}] Count of function variants that resulted in any detected memory leak using Valgrind (detected using either gcc or clang compiled binaries or both). Note that necessarily these counts are only for the cases where the binaries could be compiled. 
    \item[\textbf{Metric 26.}] Count of the function variants, out of the function variants that were true for Metric 25, that for any compiler optimization level (without the memory checks or memory address sanitizer) was output-verified. Note that in practice, we never observed an example of this case ever occurring. 

    \item[\textbf{Metric 27.}] Count of function variants that had any Valgrind detected \texttt{Invalid free() / delete / delete[] / realloc()} error (detected using either gcc or clang compiled binaries or both). Note that necessarily these counts are only for the cases where the binaries could be compiled. 
    \item[\textbf{Metric 28.}] Count of the function variants, out of the function variants that were true for Metric 27, that for any compiler optimization level (without the memory checks or memory address sanitizer utility used in the compilation) was output-verified. Note that in practice, we never observed an example of this case ever occurring. 
    
    \item[\textbf{Metric 29.}] Count of function variants that had any Valgrind detected \texttt{Invalid read} error (detected using either gcc or clang compiled binaries or both). Note that necessarily these counts are only for the cases where the binaries could be compiled. 
    \item[\textbf{Metric 30.}] Count of the function variants, out of the function variants that were true for Metric 29, that for any compiler optimization level (without the memory checks or memory address sanitizer) was output-verified. Note that in practice, we never observed an example of this case ever occurring. 

    \item[\textbf{Metric 31.}] Count of function variants that had any Valgrind detected \texttt{Use of uninitialised value} error (detected using either gcc or clang compiled binaries or both). Note that necessarily these counts are only for the cases where the binaries could be compiled. 
    \item[\textbf{Metric 32.}] Count of the function variants, out of the function variants that were true for Metric 31, that for any compiler optimization level (without the memory checks or memory address sanitizer) was output-verified. 

    \item[\textbf{Metric 33.}] Count of function variants that had any Valgrind detected \texttt{Conditional jump or move depends on uninitialised value} error (detected using either gcc or clang compiled binaries or both). Note that necessarily these counts are only for the cases where the binaries could be compiled. 
    \item[\textbf{Metric 34.}] Count of the function variants, out of the function variants that were true for Metric 33, that for any compiler optimization level (without the memory checks or memory address sanitizer) was output-verified. 

    \item[\textbf{Metric 35.}] Count of function variants that had any clang or gcc memory sanitizer detected integer overflow error (detected using either gcc or clang compiled binaries or both). Note that necessarily these counts are only for the cases where the binaries could be compiled. 
    \item[\textbf{Metric 36.}] Count of the function variants, out of the function variants that were true for Metric 35, that for any compiler optimization level (without the memory checks or memory address sanitizer) was output-verified. Note that in practice, we never observed an example of this case ever occurring. 

    \item[\textbf{Metric 37.}] Count of function variants that had any clang or gcc memory sanitizer detected out of bounds error (detected using either gcc or clang compiled binaries or both). Note that necessarily these counts are only for the cases where the binaries could be compiled, and the memory sanitizer check is performed on the binaries compiled using an optimization level of $0$. 
    \item[\textbf{Metric 38.}] Count of the function variants, out of the function variants that were true for Metric 37, that for any compiler optimization level (without the memory checks or memory address sanitizer) was output-verified. 
    
    \item[\textbf{Metric 39.}] Count of how many of the function variants produce binaries that output hashes that are not correct to the SHA-1 implementation (in particular none of the hashes for any of the four test vectors are correct, and all of the hashes have an absolute character distance greater than 5 away from the correct SHA-1 hash), the output does not change depending on the compiler optimization settings that were used (e.g., it is compiler optimization stable), and the output conforms to the basic requirements of a hash function - in this case meaning that there are $40$ hexadecimal characters produced for each test vector (which is the same length as the correct SHA-1 hash digests), and the output hashes change (by at least one character) for each of the four test vectors. This case is designed to be disjoint to Metrics 17, 18, 19 and 20, meaning there is no overlap between this Metric and those. These cases are interesting because some of these produce very bad checksums (e.g., clear repeating patterns), but others produce ``hashes'' that appear to be reasonably high entropy. These test cases are not further analyzed in detail for how secure they are (for example, if there are clear correlations between the input and the output), but are notable because these could in theory be used as (likely bad and non-secure) hash functions -- which were produced as a byproduct of the high variability GPT code re-writing output. These counts include only the cases that are not optimization unstable so as to simplify the example test cases. 
    \item[\textbf{Metric 40.}] Count of how many of the function re-writes, for any compiler optimization setting, where there was any hash output that produced a number of characters not equal to $40$. 
\end{itemize}

Note that many of the function re-writes may fall into more than one of these categories (Metrics). Also note that the counts of the various function variants that are incorrect can contain duplicate source code, similarly to the correct function rewrites. Duplicates of incorrect function versions is not checked for, but it does occur in at least a few instances.

The use of varying compiler optimization levels is motivated by the following points:

\begin{itemize}[noitemsep]
    \item Higher optimization could uncover source code variants which are quite similar with some minimal changes thus making them not direct copies, but have sufficient similarity that the optimization can produce identical binaries. 
    \item The low optimization level shows a baseline equivalence of the original source code to very minimally modified code (such as minimally changing variable names), thus not producing a meaningful substantive syntax change. 
    \item The fast code optimization option is tested because it can yield even more heavily modified binaries undefined behavior can be revealed 
    \item Interestingly, in some instance the higher optimization levels allow the compiler to generate binaries whereas for the no-optimization level the compiler was not able to produce a binary. 
\end{itemize}

In summary, optimization in the compiler can uncover cases where although the source syntax is different, the underlying logic and algorithmic choices are the same. This allows us to use compiler optimization as a tool to differentiate genuine source code alterations that are meaningful. With the intention of thoroughly checking for variant equivalences that may be difficult to arrive at, both the compiler tools \texttt{gcc} and \texttt{clang} are executed with all available optimization levels.

The motivation for the test cases that detect binaries whose output is unstable based on different compiler optimization settings is that a reasonably large number of these cases were found in the GPT function re-writes. In particular, many of the function re-writes have undefined behaviors. This then causes the compiler to have some freedom in how to interpret the undefined source code, and this can result in compiler optimization instability (also known as undefined behavior), which is a well-studied aspect of the C language \cite{baev2022preventing, 8802866, 10.1145/2737924.2737979, 10.1145/2699678, 10.1145/3501774.3501781, 8905996, 10.1145/3582016.3582053, 10.1145/3062341.3062343, dahiya2017modeling, 10.1145/2517349.2522728}. Additionally, these tests are performed to categorize in what ways the output changes based on the compiler settings because there have been examples of vulnerabilities introduced by compilers \cite{7163211, doi:10.2514/1.I010699, nurmukhametov2015application, baev2022preventing}, and therefore it is of considerable interest to determine what is being affected by these compiler optimization unstable GPT function re-writes when different optimization levels are applied. It is difficult to systematically categorize undefined behavior and undefined syntax in a piece of C code, but what we found is that the binaries with compiler optimization instability often threw various compiler warnings including incorrect C syntax, such as incorrect type conversions. We leave more extensive analysis of undefined C syntax produced from GPT models to future research.

The hash output correctness is measured by the testing source code (given in Appendix \ref{section:appendix_reference_source_code_SHA-1}). Specifically, the hash data is written to an array and is intended to be in a specific index range of that array. The GPT modified source code may write out of bounds, but only the intended portion of that array is checked for algorithmic correctness of the hash function and writes outside of that array are not checked. The produced text from the compiled binaries are encoded and then decoded as utf-8 strings.

All compilation and execution was performed on Ubuntu, with \texttt{Intel(R) Xeon(R) Gold 6338 CPU @ 2.00GHz} CPUs. The compiler versions are \texttt{gcc (Ubuntu 10.5.0-1ubuntu1~22.04) 10.5.0} and \texttt{Ubuntu clang version 14.0.0-1ubuntu1.1}. All code was compiled using C standard C11. The use the compilers on the fixed hardware platforms allowed for consistent binaries to be compiled and compared. The Valgrind version used for all tests is \texttt{Valgrind-3.18.1}.

\subsubsection{Composability of Correct Function Rewrites}
\label{section:methods_GPT_output_parsing_and_testing_composability}
The last step in generating full cryptographic function variants is composing the source code functions that were determined to be correct and compilable under all optimization settings. This is done by randomly selecting a representative source code function from each of the unique meta-clusters computed by \textbf{Metric 13} in Section~\ref{section:methods_GPT_output_parsing_and_testing}. Then, all combinations of the full SHA-1 implementations are enumerated through, where we replace each of the four component functions with a re-written unique version. These re-written versions now have distinctly different source code from the original implementation for all of the component functions. These re-written versions are then processed by the same ensemble of tests that were performed on the single-function replacement tests in Section~\ref{section:methods_GPT_output_parsing_and_testing}.

\begin{table}[htbp]
\centering
\begin{tabular}{ |p{1.8cm}|p{3.5cm}|p{3.5cm}|p{3.5cm}|p{3.5cm}| } 
 \hline
 Metric & sha1\_final & sha1\_init & sha1\_update & sha1\_transform \\
 \hline
 Function Rewrites & 34,299 & 40,716 & 45,269 & 34,149 \\
 \hline
 Metric 1 & 12,431 & 8,729 & 21,893 & 12,109 \\ 
 \hline
 Metric 2$\square$ & 10 & 17 & 29 & 5 \\ 
 \hline
 Metric 3* & 10,842 & 7,672 & 19,906 & 10,214 \\ 
 \hline
 Metric 4$\square$ & 1,105 & 922 & 696 & 1,568 \\ 
 \hline
 Metric 5$\square$ & 6 & 21 & 2 & 19 \\ 
 \hline
 Metric 6* & 9,142 & 7,222 & 17,556 & 1,102  \\ 
 \hline
 Metric 7* & 1,700 & 450 & 2,350 & 9,112 \\ 
 \hline
 Metric 8$\square$ & 10 & 0 & 39 & 3 \\ 
 \hline
 Metric 9$\square$ & 99 & 29 & 156 & 209 \\ 
 \hline
 Metric 10$\square$ & 15 & 6 & 39 & 12 \\ 
 \hline
 Metric 11$\square$ & 133 & 75 & 398 & 111 \\ 
 \hline
 Metric 12* & 48-23-17-18-17-18-54-22-17-17-21-17-22 & 116-52-37-38-37-38-117-54-55-55-58-55-60 & 119-75-72-73-70-73-147-73-68-68-69-68-69 & 68-29-18-25-20-25-91-19-17-17-19-17-19 \\ 
 \hline
 Metric 13* & 13 groups:\newline 3-1014-5-1-57-1-1-1-1-1-1-1-1 & 31 groups:\newline 19-14-20-3-2-2-3-1-4-1-2-1-1-1-1-1-1-1-1-1-1-1-1-1-1-1-1-1-1-1-1 & 57 groups:\newline 426-129-32-61-2-2-3-5-7-23-5-5-7-6-10-23-10-1-1-1-1-2-8-1-10-1-1-3-2-2-4-1-2-3-1-1-1-5-1-1-1-1-1-1-2-1-1-3-2-1-1-1-1-1-1-1-1 & 8 groups\newline 1-1-1-1-1-1-1-1 \\ 
 \hline
 Metric 14* & 611 & 357 & 1,519 & 8,743 \\ 
 \hline
 Metric 15$\square$ & 0 & 13 & 1 & 0 \\ 
 \hline
 Metric 16$\square$ & 0 & 0 & 0 & 0 \\ 
 \hline
 Metric 17$\square$ & 143 & 12 & 698 & 5 \\ 
 \hline
 Metric 18$\square$ & 110 & 8 & 92 & 0 \\ 
 \hline
 Metric 19$\square$ & 1 & 0 & 0 & 0 \\ 
 \hline
 Metric 20$\square$ & 1 & 0 & 0 & 0 \\ 
 \hline
 Metric 21$\square$ & 383 & 238 & 213 & 220 \\ 
 \hline
 Metric 22$\square$ & 955 & 696 & 1,182 & 1,353 \\ 
 \hline
 Metric 23$\square$ & 0 & 3 & 2 & 36 \\ 
 \hline
 Metric 24$\square$ & 0 & 0 & 1 & 0 \\ 
 \hline
 Metric 25$\square$ & 3 & 1 & 1 & 1 \\ 
 \hline
 Metric 26$\square$ & 0 & 0 & 0 & 0 \\ 
 \hline
 Metric 27$\square$ & 1 & 0 & 2 & 0 \\ 
 \hline
 Metric 28$\square$ & 0 & 0 & 0 & 0 \\ 
 \hline
 Metric 29$\square$ & 2 & 0 & 1 & 2 \\ 
 \hline
 Metric 30$\square$ & 0 & 0 & 0 & 0 \\ 
 \hline
 Metric 31$\square$ & 611 & 234 & 401 & 225 \\ 
 \hline
 Metric 32$\square$ & 0 & 0 & 0 & 0 \\ 
 \hline
 Metric 33$\square$ & 614 & 241 & 406 & 227 \\ 
 \hline
 Metric 34$\square$ & 0 & 3 & 1 & 0 \\ 
 \hline
 Metric 35$\square$ & 0 & 10 & 0 & 13 \\ 
 \hline
 Metric 36$\square$ & 0 & 0 & 0 & 0 \\ 
 \hline
 Metric 37$\square$ & 115 & 304 & 55 & 17 \\ 
 \hline
 Metric 38$\square$ & 1 & 12 & 3 & 0 \\ 
 \hline
 Metric 39$\square$ & 658 & 683 & 207 & 1206 \\ 
 \hline
 Metric 40$\square$ & 2 & 0 & 1 & 0 \\ 
 \hline
\end{tabular}
\caption{SHA-1 C code rewriting metrics (across all GPT models, prompts, and inference temperatures). * denotes function variant metrics that are correct re-writes $\square$ denotes function variant counts that have an implementation flaw of some type causing code instability, compiler optimization instability, infinite loops, critical errors, or not correct SHA-1 implementations. }
\label{table:SHA-1_aggregate_metrics}
\end{table}

\section{Results}
\label{section:results}

The aggregated test statistics defined in Section \ref{section:methods_GPT_output_parsing_and_testing} are all reported, for each component function of the reference implementation, in Table \ref{table:SHA-1_aggregate_metrics}. These metrics include a wide range of different inference temperature settings, the $10$ different prompts, and are also aggregated from all $3$ GPT models. The most consequential metrics are the total number of \emph{attempted} function re-writes, which is $154,433$ in total, but only $55,162$ of those could be compiled with all compiler settings. Importantly, this means that more than half of the GPT outputs could not be parsed as valid C code (using the markdown-style prompts as described in Sections \ref{section:methods_GPT_implementation} and \ref{section:methods_GPT_output_parsing_and_testing}). Note that the total number of function re-writes entry would be the exact same count (and exactly equal to the total number of GPT queries made) if not for the last parsing step (step number 7 in Section \ref{section:methods_GPT_output_parsing_and_testing}), where we occasionally get \emph{extra} function re-writes from ancillary text produced by the GPT models. 

Table \ref{table:SHA-1_aggregate_metrics} shows that there are a large number of GPT function re-writes that are instances of very flawed C code. For example, Metrics 31 and 33 in the table total over 1,000 instances, and both of these are measures that show fundamental issues with the code implementation. 

The function \texttt{sha1\_init} defines several constants that initialize the SHA-1 algorithm. Metrics 12 and 13 in Table \ref{table:SHA-1_aggregate_metrics} show quantify correct function re-writes, and one of the interesting questions to ask specifically about the function re-writes of \texttt{sha1\_init} is whether any of those functions define constants that are not the same as the original implementation (which is given in Appendix \ref{section:appendix_reference_source_code_SHA-1}). The answer is that all of the constants defined in these variants were correct, and not different compared to the original source code. Many of these variants add ancillary constants, or define data-structures that are not used, but fundamentally the correct constants are always defined.

In regards to the versions that produced hash functions that were seemingly valid hash functions (characterized by Metric 39), but not correct to the reference algorithm, these examples are very likely not cryptographically secure (as in, they are very likely to have fundamental algorithmic weaknesses beyond implementation risks such as side channel attacks). The scope of this study is not to thoroughly evaluate these instances. However, these function re-write instances were unexpected results of these experiments, and they are indicators that open source GPT models can be tools for the proliferation of (incorrect) versions of important algorithms such as cryptographic algorithms. If such examples were minimally evaluated and found to have the basic requirements of hash functions, these could then be used (at fairly low cost) in new malware variants, thus obfuscating the functionality of the software. Even if these algorithms are insecure, they can be generated at scale, and therefore pose a risk to the community of cybersecurity analysts to the increase of availability of such tools. Therefore, these instances warrant future research. 

Across all source code variants and compiler settings, the binary execution fatal errors that were encountered were $10685$ \texttt{SIGSEGV} errors, $559$ \texttt{SIGABRT} errors, $9$ \texttt{SIGILL} errors, $3$ \texttt{SIGFPE} errors, and $13$ \texttt{SIGBUS} errors. Across all source code variants and compiler settings $1263$ of the compiled binaries reached an apparent infinite loop state, as determined by a timeout check of 10 seconds. 

The different hash string output lengths, as determined by Metric 40, across all of the function re-writes and compiled binaries are; $40, 0, 160, 2228264$. The vast majority of the generated hashes have length of $40$, but one function re-write produces hashes (for some compiler optimization settings) of length $0$ (meaning an empty string), one function re-write produces a hash of length $160$, and a different function re-write produces hashes of length $2228264$ (for some compiler optimization settings), a majority of these characters are zeroes. Note that the testing code (Code Listing \ref{source_code:SHA-1_reference_test_and_main}) is set up to read a specified number of indices from the array in which the hash data is computed.

\begin{figure}[htbp]
    \centering
    \includegraphics[width=0.24\textwidth]{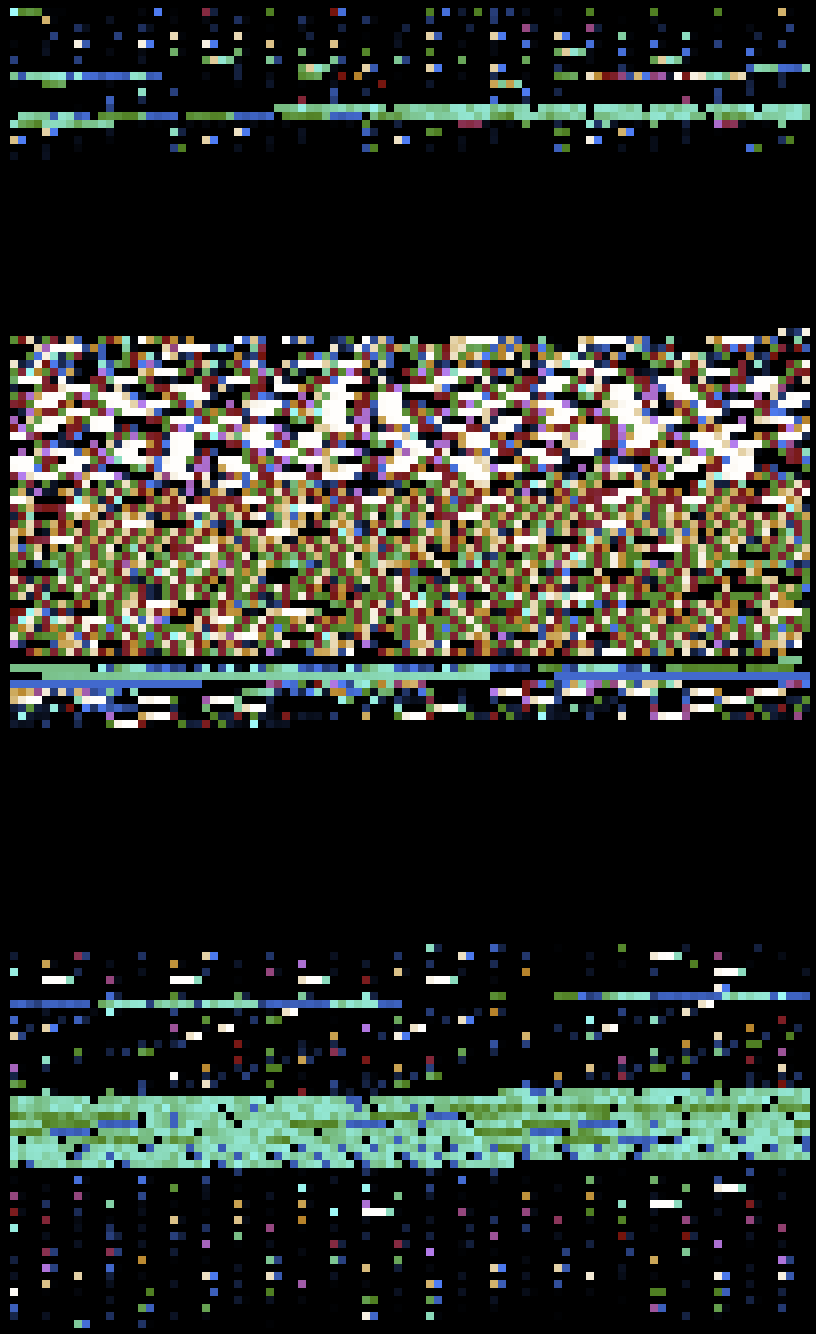}
    \includegraphics[width=0.24\textwidth]{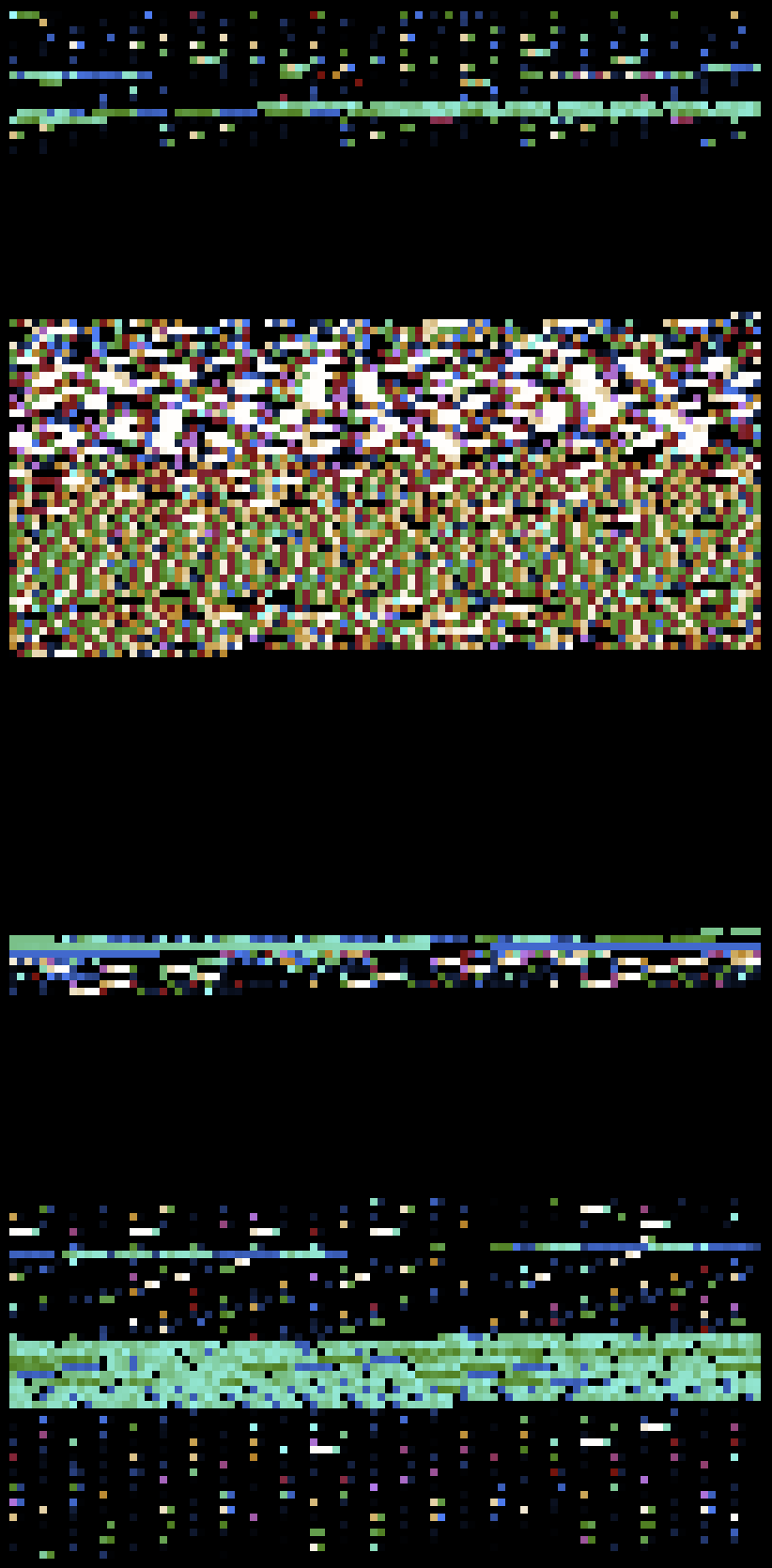}
    \includegraphics[width=0.24\textwidth]{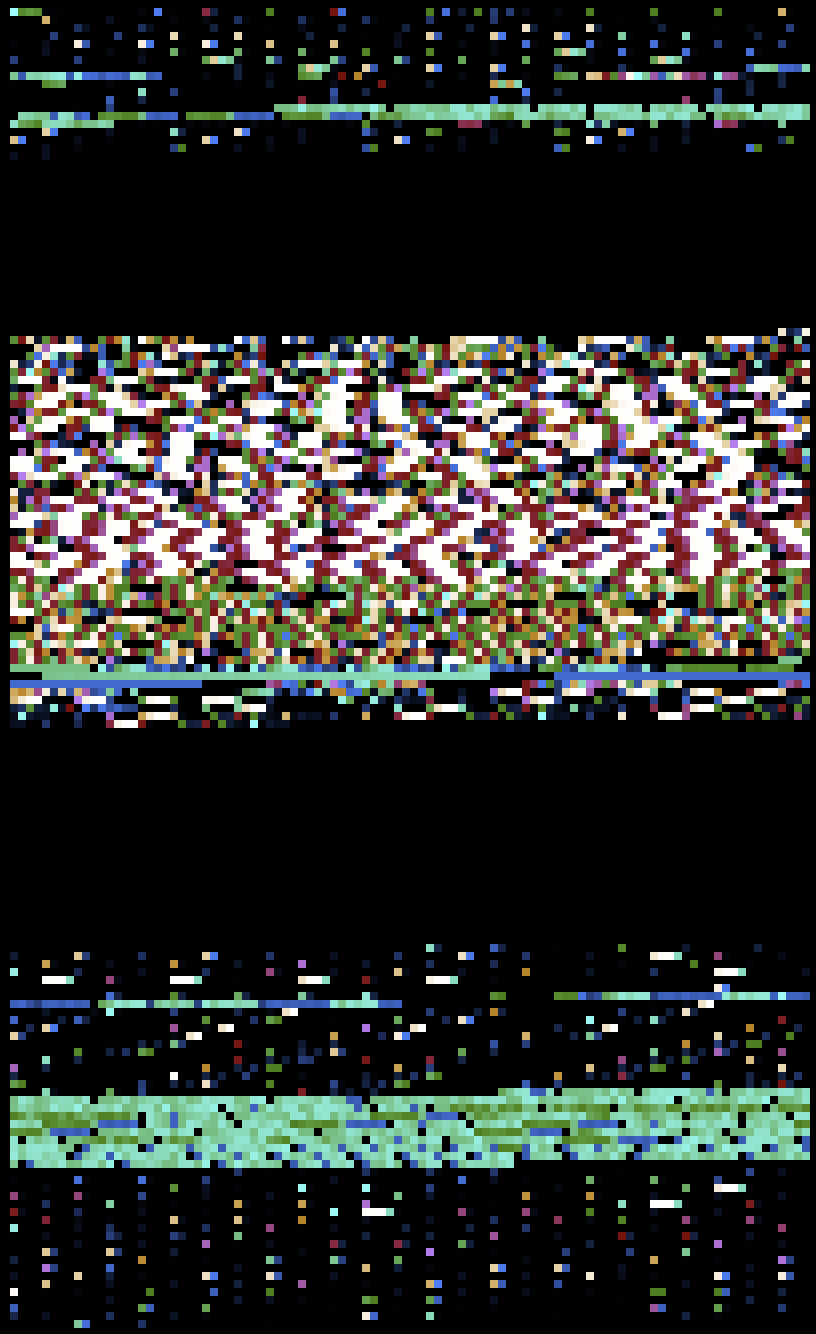}
    \includegraphics[width=0.24\textwidth]{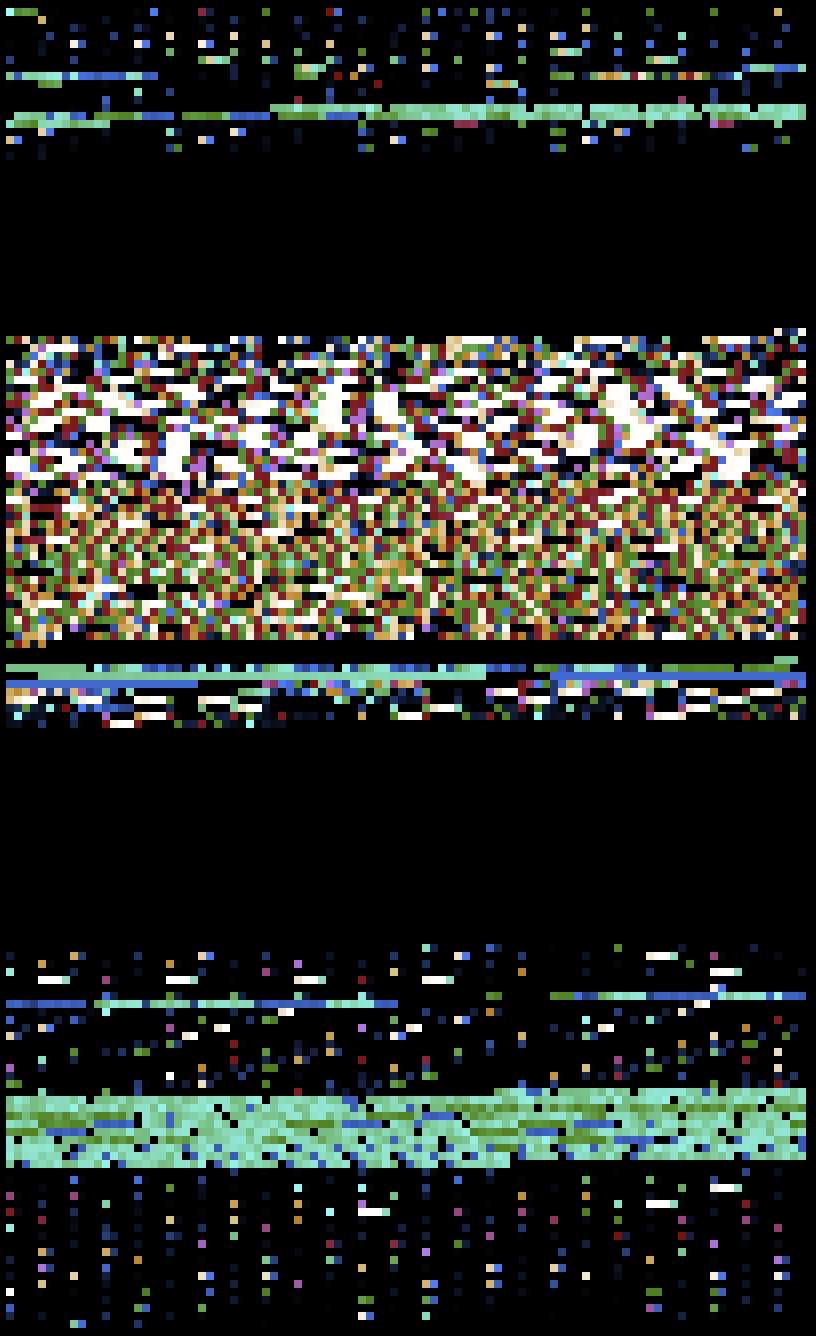}
    \caption{Visualized examples of compiled SHA-1 binaries using binocle. In order from top left to right; a single function-rewrite that is fully correct, 2 function rewrites where all test cases fail (and the outputs are not close to the correct SHA-1 hashes), and 1 function re-write that resulted in some of the test vectors producing a correct SHA-1 hash but failing for at least one test vector (and was not compiler optimization unstable). All of these example binaries were compiled using clang with an optimization level of $0$. These binaries were arbitrarily selected as representative examples. }
    \label{fig:compiled_binary_visualization}
\end{figure}

Figure \ref{fig:compiled_binary_visualization} shows compiled binary visualizations for $4$ of the compiled SHA-1 binaries. These examples include re-writes that execute SHA-1 algorithms correctly, and re-writes that are incorrect and cause the output to be incorrect. These visualizations were generated using the tool binocle \cite{binocle_github}, with consistent binary data layout dimensions of the visual window and all default visualization settings otherwise. Note that these visualizations necessarily include the entirety of the testing code (the exact syntax of which is shown in Code Listing \ref{source_code:SHA-1_reference_test_and_main}), and the only differences between each of the binaries is at most one of the four SHA-1 component functions being changed. There are clearly some variations that can be seen in these compiled binaries, but overall their structure is quite similar.

\begin{figure}[htbp]
    \centering
    \includegraphics[width=0.24\textwidth]{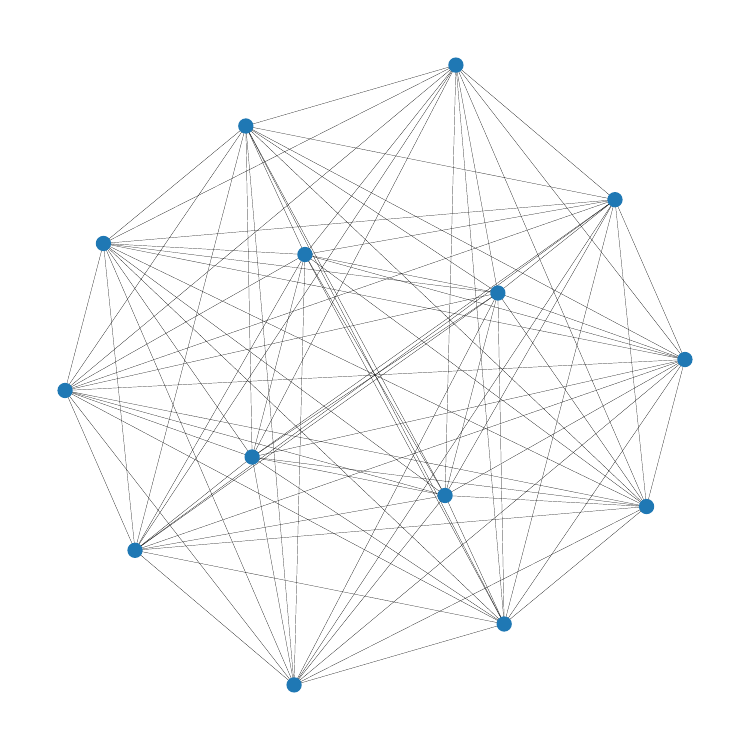}
    \includegraphics[width=0.24\textwidth]{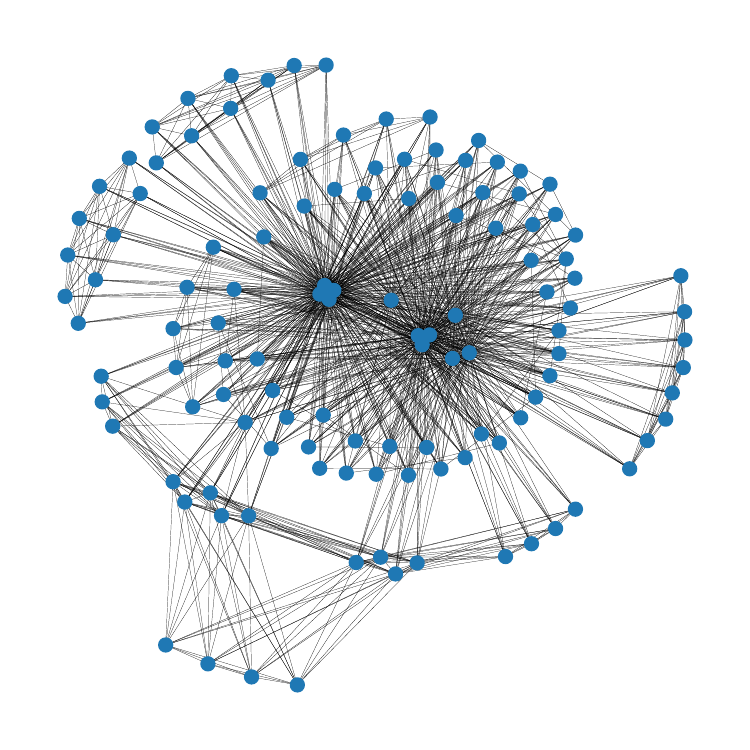}
    \includegraphics[width=0.24\textwidth]{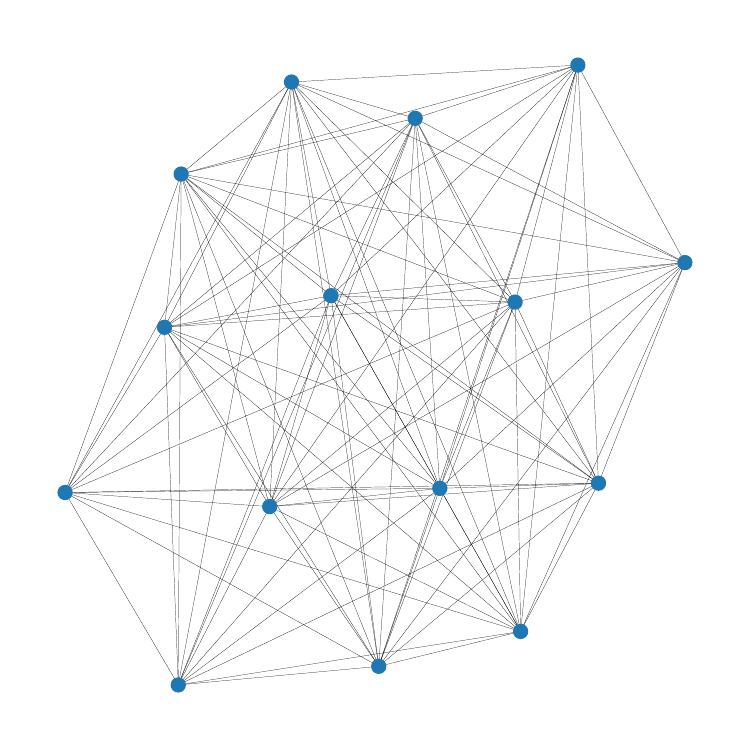}
    \includegraphics[width=0.24\textwidth]{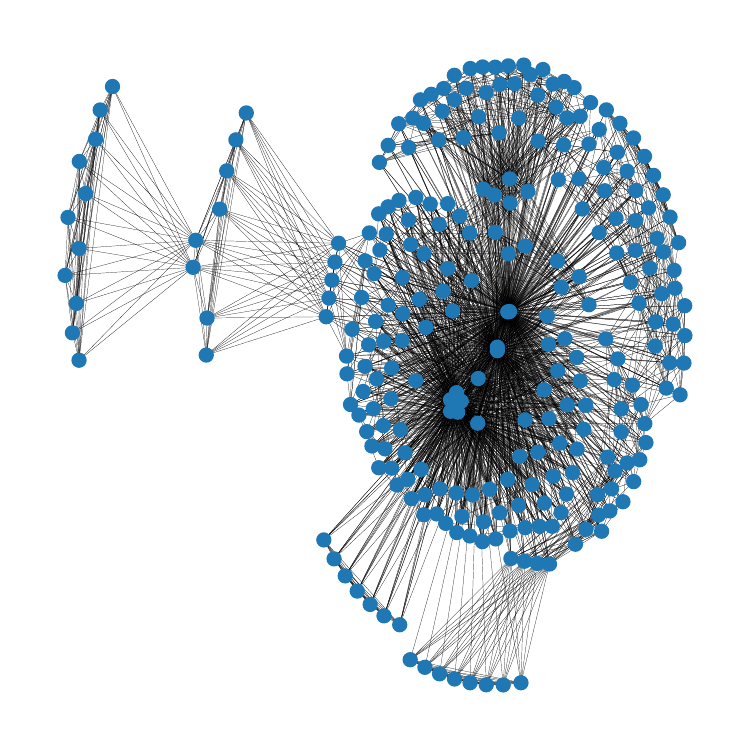}
    \includegraphics[width=0.24\textwidth]{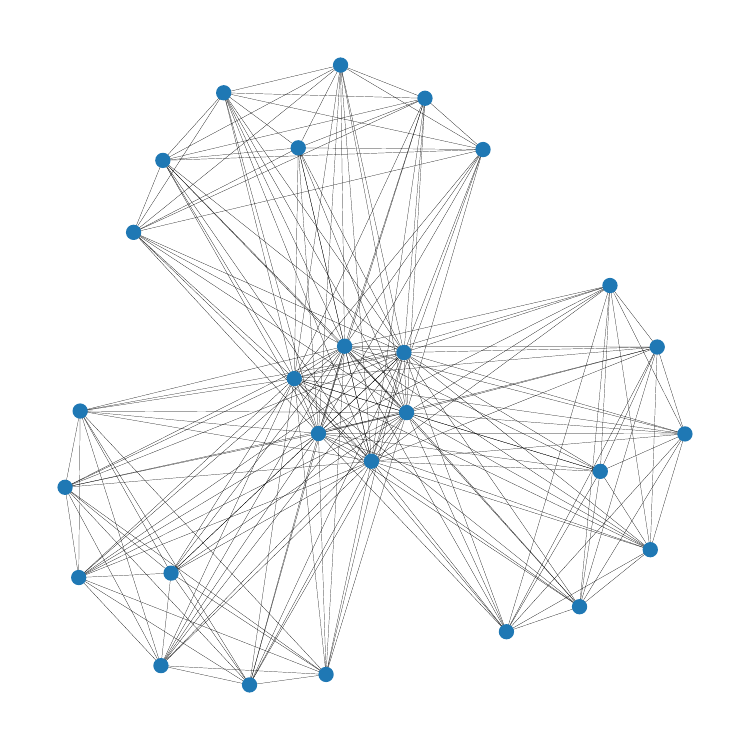}
    \includegraphics[width=0.24\textwidth]{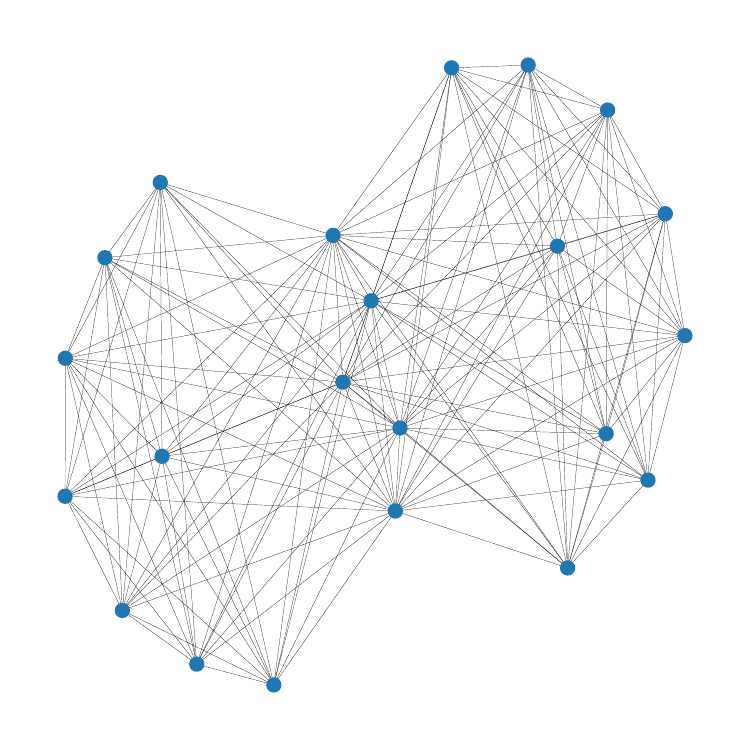}
    \includegraphics[width=0.24\textwidth]{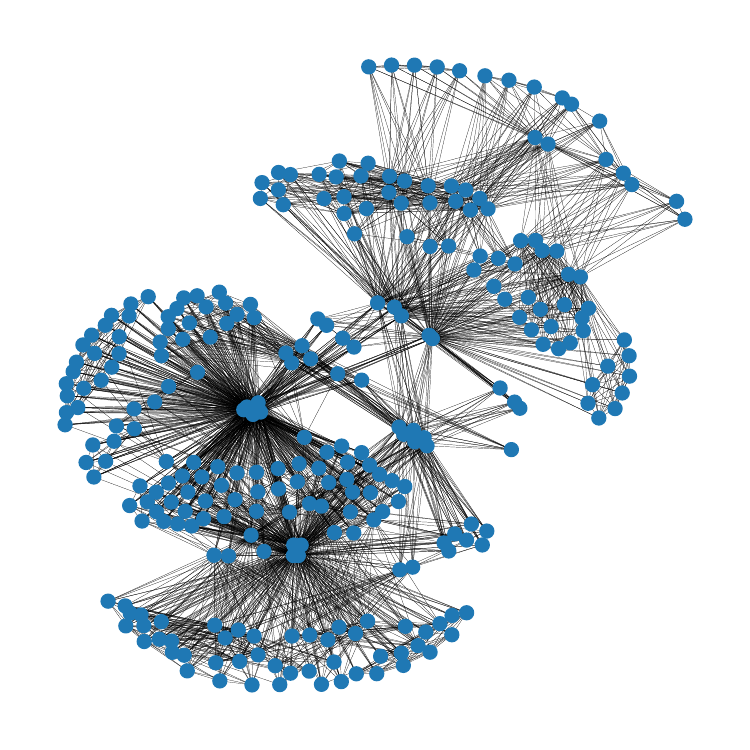}
    \includegraphics[width=0.24\textwidth]{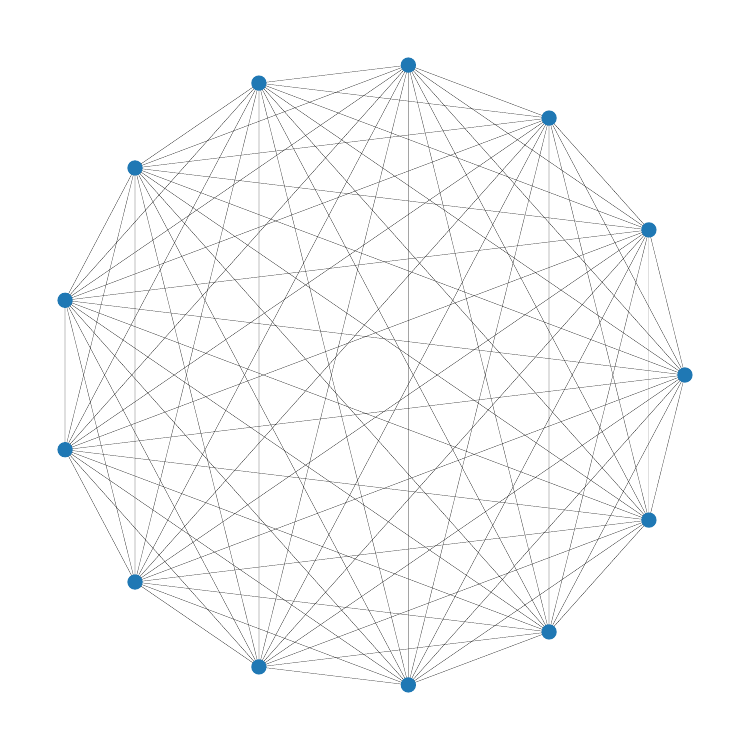}
    \includegraphics[width=0.24\textwidth]{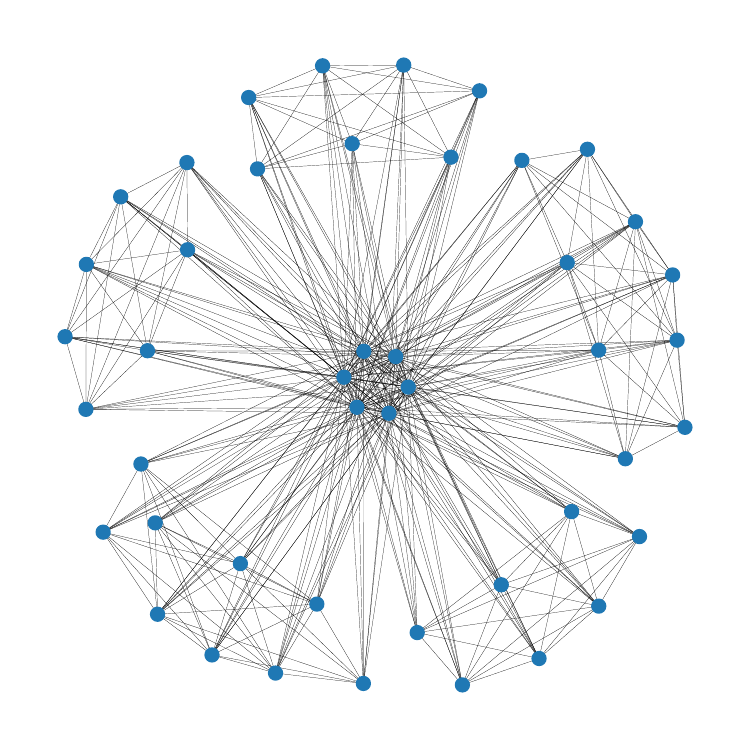}
    \includegraphics[width=0.24\textwidth]{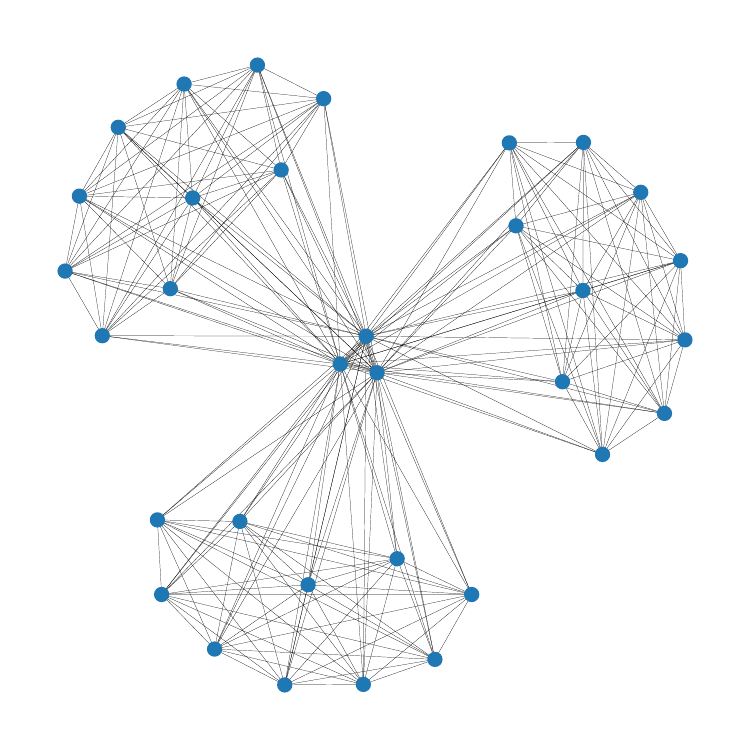}
    \includegraphics[width=0.24\textwidth]{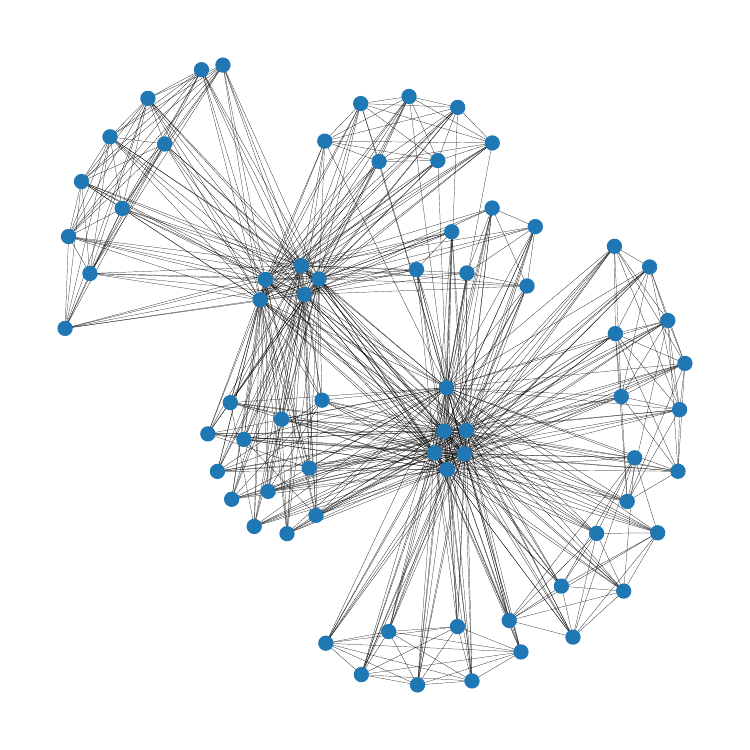}
    \includegraphics[width=0.24\textwidth]{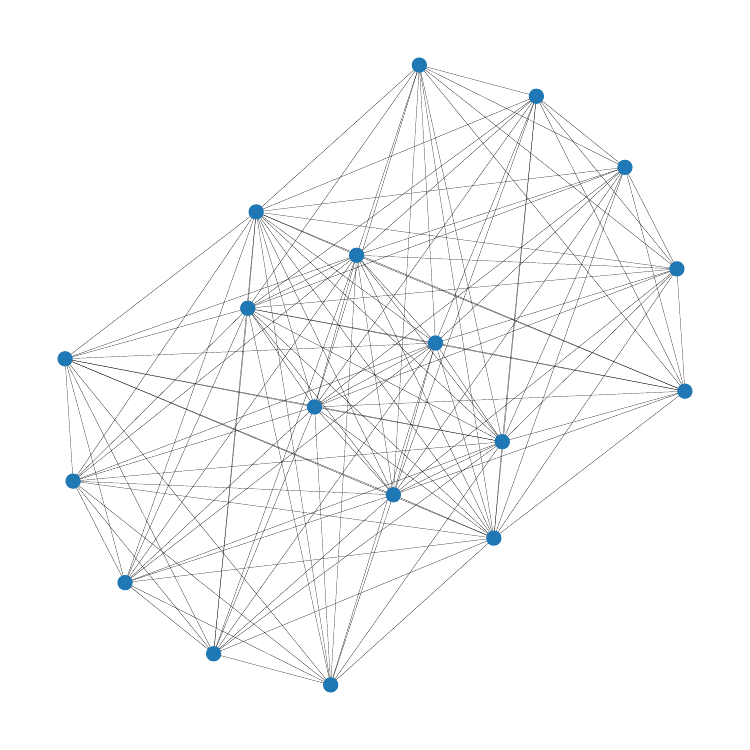}
    \includegraphics[width=0.24\textwidth]{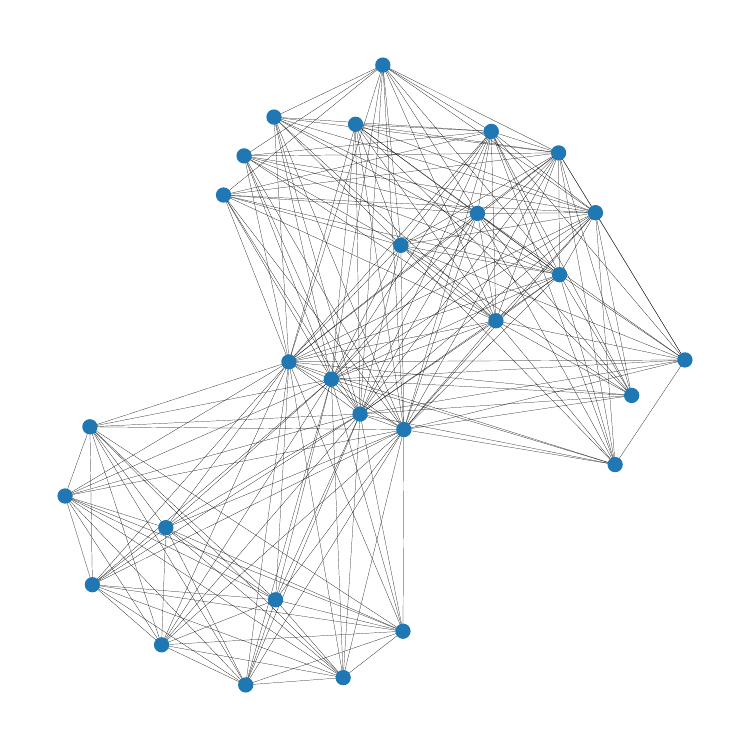}
    \includegraphics[width=0.24\textwidth]{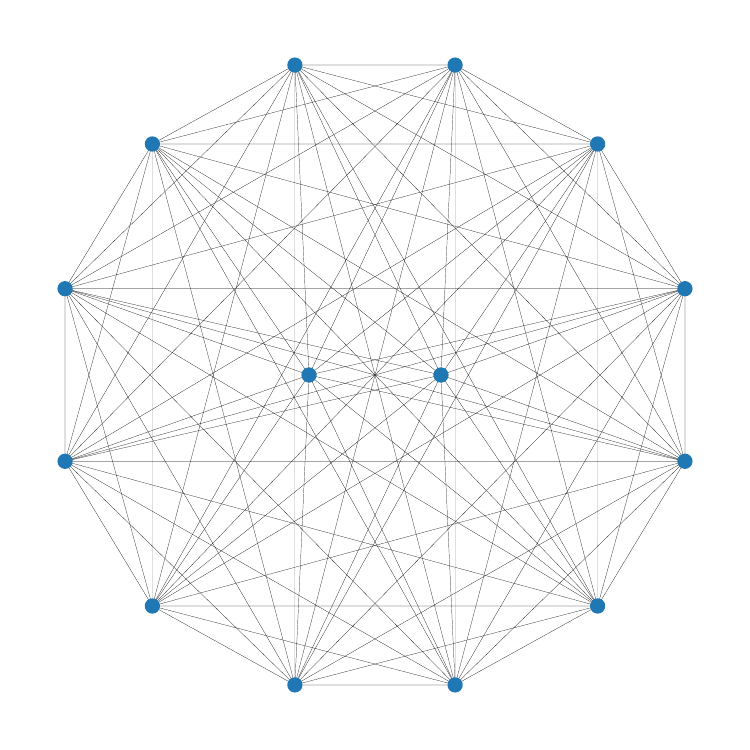}
    \includegraphics[width=0.24\textwidth]{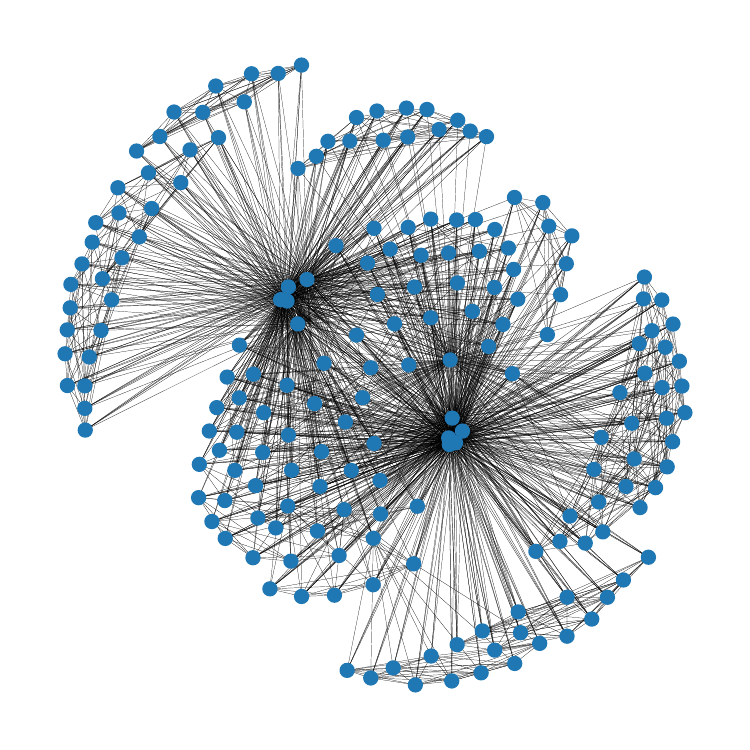}
    \includegraphics[width=0.24\textwidth]{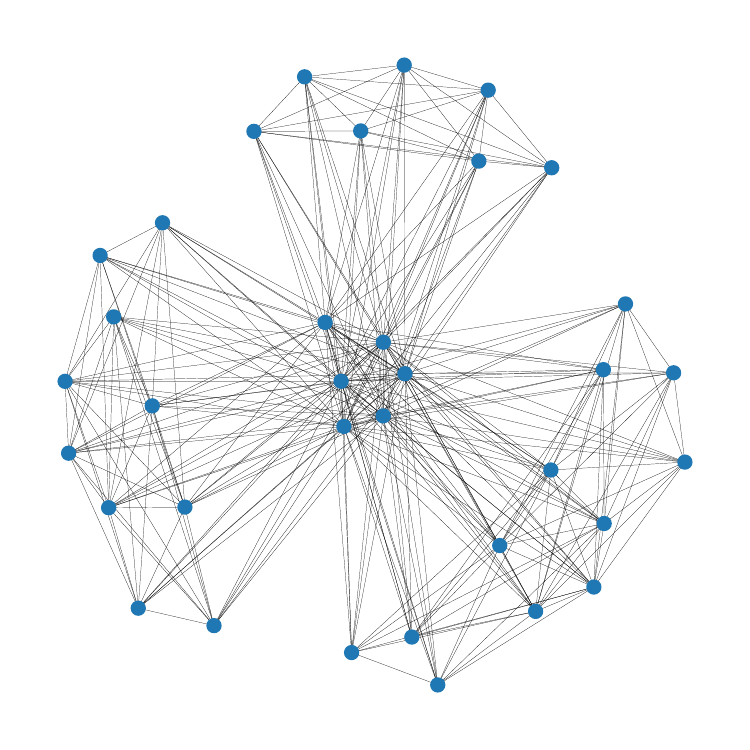}
    \caption{Graph renderings of various example connected components from the compiled binary clustering procedure; specifically for the GPT function re-writes of the SHA-1 C code where the compiled binary correctly produced SHA-1 hashes (and did not have fatal errors, or compiler optimization instability). Each node (blue) represents a tuple of a single SHA-1 component function re-write whose source code had Levenshtein character distance greater than $0$ compared to the original source code (after repeated whitespace and code comments were removed) and one of the $13$ compiler optimization settings used (either gcc or clang, with varying optimization levels). In other words, each node represents a single compiled binary that correctly executed the SHA-1 algorithm. Each edge represents the SHA-256 checksum of the compiled binary being equal for the two compiled binaries that the edge connects. These networks are not the comprehensive clustering of the correct SHA-1 rewrites, but they do represent a majority of the graphs that were produced. Notably, four of these graphs which are noticeably larger and more densely connected than the other graphs correspond to the graphs of function re-writes that are equivalent to the original source code due to the syntax changes made by the GPT models being relatively minimal. Each of these graphs are single connected components from the overall binary hashing clustering procedure, which is described in Section \ref{section:methods_GPT_output_parsing_and_testing} and the summary statistics for are shown in Table \ref{table:SHA-1_aggregate_metrics}. }
    \label{fig:connected_component_graphs_compiled_binary_hashes}
\end{figure}

Figure \ref{fig:connected_component_graphs_compiled_binary_hashes} contain renderings of several of the connected components of the graphs produced by the correct SHA-1 re-write clustering, which are summarized by Metrics 12 and 13 in Table \ref{table:SHA-1_aggregate_metrics}. Connections (e.g., paths and connected regions) of these graphs indicate that for some compiler optimization setting, an identical hash was produced for another piece of code (potentially the same code) with another compiler optimization setting. These connected graph components thus show function re-writes that are actually performing the same computation once compiled, even though the exact syntax of the code may be different. These graphs are all undirected. These graphs were generated using the Python 3 libraries Networkx \cite{hagberg2008exploring} and Matplotlib \cite{thomas_a_caswell_2021_5194481, Hunter:2007}, and drawn using the kamada-kawai layout algorithm. Note that, counter-intuitively, more SHA-1 rewrites can actually cause more connections to be revelealed by the compilers, and thus fewer unique function clusters be found by Metric 13.

\begin{figure}[htb!]
    \centering
    \includegraphics[width=0.59\textwidth]{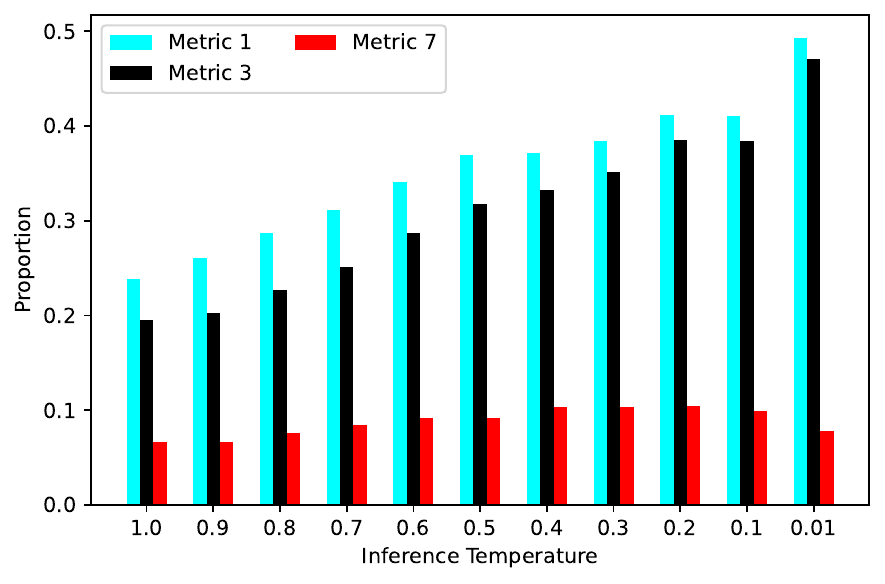}
    \caption{Correct code rewrite performance metrics as a function of inference temperature. }
    \label{fig:performance_across_temperatures}
\end{figure}

\begin{figure}[htb!]
    \centering
    \includegraphics[width=0.59\textwidth]{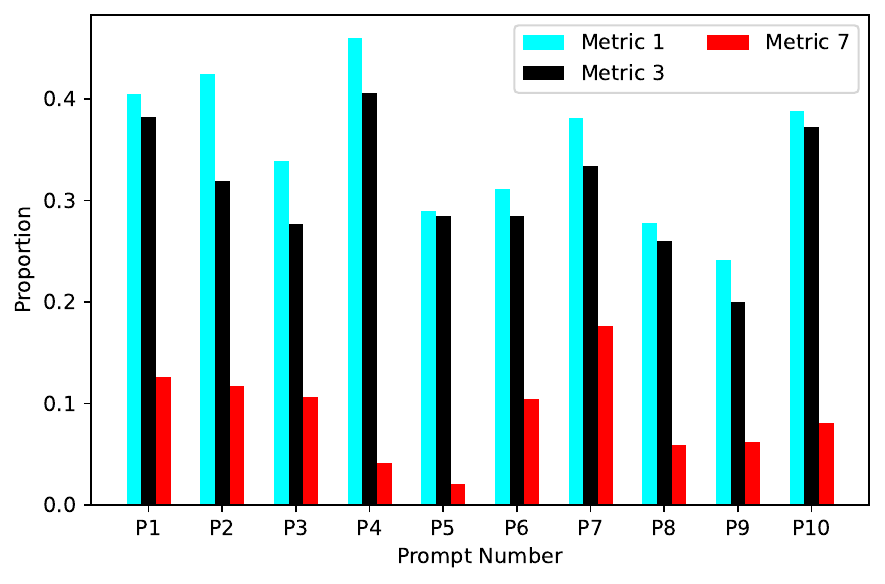}
    \caption{Correct code rewrite performance metrics as a function of the $10$ different prompts. }
    \label{fig:performance_across_prompts}
\end{figure}

\begin{figure}[htb!]
    \centering
    \includegraphics[width=0.59\textwidth]{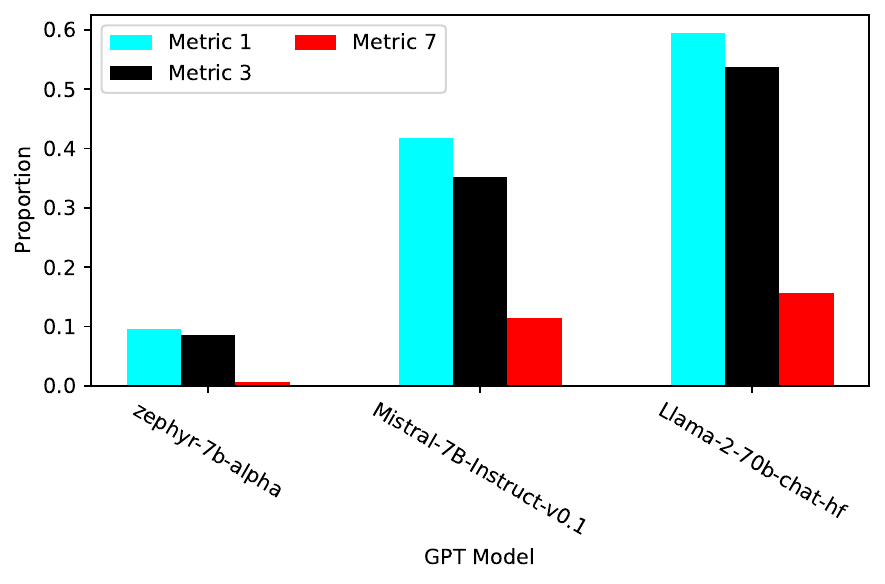}
    \caption{Correct code rewrite performance metrics as a function of the three GPT models. }
    \label{fig:performance_across_GPT_models}
\end{figure}

So as to compare the various inference temperatures, GPT models used, and prompts, we use three high level metrics (that do not depend on the clustering of the various versions produced by different compiler settings); \textbf{Metric 1} (was able to be compiled for all compiler settings), \textbf{Metric 3} (compiled binary was output-verified for all compiler settings), \textbf{Metric 7} (the compiled binary was output-verified for all compiler settings, and had source code distance greater than 0 compared to the original) as outlined in Section \ref{section:methods_GPT_output_parsing_and_testing}. For all three of these summary metrics, the larger they are the better the model and parameter choice is performing. Metric 7 is the most fundamental and important measure of code quality, and also being measurably different from the original source code. Figure \ref{fig:performance_across_temperatures} shows these three metrics as a function of inference temperature (across all $10$ prompts and $3$ GPT models). Figure \ref{fig:performance_across_prompts} shows these three metrics as a function of the $10$ different prompts. Figure \ref{fig:performance_across_GPT_models} shows these three metrics as a function of the $3$ different GPT models. 

There are several clear findings from these aggregate performance metrics. Figure \ref{fig:performance_across_temperatures} shows that the proportion of compilable functions increases consistently as the temperature gets lower. However, Metric 7 peaks at a temperature of $0.4$ to $0.2$ - this shows that some stochasticity in the token choice is required for the generated code to be different from the original code, but too high stochasticity makes the code quality decrease as well. Figure \ref{fig:performance_across_prompts} shows that the best performing prompt to maximize Metric 7 was Prompt number 7, and the worst prompt to maximize Metric 7 was Prompt number 5. Finally, Figure \ref{fig:performance_across_GPT_models} shows that \texttt{Llama-2-70b-chat-hf} performed the best.

\subsection{Function Composability Results}
\label{section:results_function_composable}
A total of $13 \cdot 32 \cdot 58 \cdot 8 = 193024$ different composed SHA-1 codebase re-writes were generated. This comes from all of the possible combinations of the unique and correct versions of the function re-writes, as determined by the compiled binary clustering of Metric 13 in Table \ref{table:SHA-1_aggregate_metrics}. Of these $193024$ versions, a vast majority are entirely correct, new, and unique variants of the SHA-1 codebase where they are all compilable, output-verified, they have no out-of-bounds writes, no integer overflows, no memory leaks, and are compiler optimization stable. Interestingly, $464$ of these function re-writes however could not be compiled, but the underlying cause was due to re-definitions of C functions (specifically, functions with names that are not in the original SHA-1 source code) with conflicting type information. These ancillary functions that were produced could be removed in an automated system and then these conflicting definition errors would be mitigated, but here we did not apply this further step. Of the versions that could be compiled, using the SHA-256 checksums of the compiled binaries it was determined that all of these variants are all unique (e.g., there are not duplicates) under all compiler optimization settings using both gcc and clang, and are not the same as the original source code. Code Listings \ref{re_written_source_code:SHA-1_reference_functions_ex1}, \ref{re_written_source_code:SHA-1_reference_functions_ex2}, \ref{re_written_source_code:SHA-1_reference_functions_ex3} are three examples of composed SHA-1 codebase re-writes that are fully correct and unique versions of the SHA-1 source code.

\subsection{Example Function Variants and Compiled Binary Outputs}
\label{section:results_examples}

This section lists a number of concrete function re-write examples produced by the GPT models that exhibit a wide range of the behaviors that are summarized in Table \ref{table:SHA-1_aggregate_metrics}. These include function re-writes that have memory leaks, that generate partially correct SHA-1 hashes, that have compiler optimization instability, among many other interesting software flaws. To conserve the total amount of space used for providing these functions, single instances will be listed in this section, and then many additional examples are given in Appendix \ref{section:appendix_fully_correct_SHA1_rewrites} and Appendix \ref{section:appendix_additional_source_code_examples}. The source code examples include syntax highlighting for C keywords, and if the compiled binary produced output, the text of the generated hashes are shown below the source code. The text of the hashes are color-coded such that black text denotes the hexadecimal character matches the corresponding character in the (correct) SHA-1 hash, and otherwise the character is red. The output of the hashes are ordered sequentially starting at test 1, and ending at test 4 (the test vectors are given in Appendix \ref{section:appendix_hash_function_test_vectors}).

\begin{lstlisting}[caption={\textbf{Fully correct SHA-1 codebase GPT model re-write example Number 1. Each of these four functions were manually selected from the correct and unique re-writes of that function across all of the GPT model outputs. Minimal formatting of the raw parsed GPT output has been applied so as to reduce the amount of whitespace. }},captionpos=b,label={re_written_source_code:SHA-1_reference_functions_ex1},language=C,style=base,keywordstyle=\color{blue}]

void sha1_init(SHA1_CTX *ctx){
    ctx->data[0] = ctx->data[1] = ctx->data[2] = ctx->data[3] = ctx->data[4] = 0x00;
    ctx->data[5] = ctx->data[6] = ctx->data[7] = ctx->data[8] = ctx->data[9] = ctx->data[10] = ctx->data[11] = ctx->data[12] = ctx->data[13] = ctx->data[14] = ctx->data[15] = 0xFF;
    ctx->data[16] = ctx->data[17] = ctx->data[18] = ctx->data[19] = ctx->data[20] = ctx->data[21] = ctx->data[22] = ctx->data[23] = ctx->data[24] = ctx->data[25] = ctx->data[26] = ctx->data[27] = ctx->data[28] = ctx->data[29] = ctx->data[30] = 0x00;
    ctx->data[31] = ctx->data[32] = ctx->data[33] = ctx->data[34] = ctx->data[35] = ctx->data[36] = ctx->data[37] = ctx->data[38] = ctx->data[39] = ctx->data[40] = ctx->data[41] = ctx->data[42] = ctx->data[43] = ctx->data[44] = ctx->data[45] = ctx->data[46] = ctx->data[47] = ctx->data[48] = ctx->data[49] = ctx->data[50] = ctx->data[51] = ctx->data[52] = ctx->data[53] = ctx->data[54] = ctx->data[55] = 0xFF;
    ctx->datalen = 0;
    ctx->bitlen = 0;
    // initialize with values defined in the original implementation
    ctx->state[0] = 0x67452301;
    ctx->state[1] = 0xEFCDAB89;
    ctx->state[2] = 0x98BADCFE;
    ctx->state[3] = 0x10325476;
    ctx->state[4] = 0xc3d2e1f0;
    ctx->k[0] = 0x5a827999;
    ctx->k[1] = 0x6ed9eba1;
    ctx->k[2] = 0x8f1bbcdc;
    ctx->k[3] = 0xca62c1d6;
}
void sha1_update(SHA1_CTX *ctx, const BYTE data[], size_t len){
	size_t i;
	for (i = 0; i < len; ++i) {
		ctx->data[ctx->datalen] = data[i];
		ctx->datalen++;
		if (ctx->datalen == 64) {
			sha1_transform(ctx, ctx->data);
			ctx->bitlen += 512;
			ctx->datalen = 0;
			ctx->k[0] = 0x5a827999;
			ctx->k[1] = 0x6ed9eba1;
			ctx->k[2] = 0x8f1bbcdc;
			ctx->k[3] = 0xca62c1d6;
		}
	}
}
void sha1_transform(SHA1_CTX *ctx, const BYTE data[]){
    WORD a = ctx->state[0];
    WORD b = ctx->state[1];
    WORD c = ctx->state[2];
    WORD d = ctx->state[3];
    WORD e = ctx->state[4];
    WORD *m = (WORD *) malloc(80 * sizeof(WORD));
    for (int i = 0, j = 0; i < 16; ++i, j += 4){
        m[i] = (data[j] << 24) + (data[j + 1] << 16) + (data[j + 2] << 8) + (data[j + 3]);
    }
    for (int i = 16; i < 80; ++i){
        m[i] = (m[i - 3] ^ m[i - 8] ^ m[i - 14] ^ m[i - 16]);
        m[i] = (m[i] << 1) | (m[i] >> 31);
    }
    for (int i = 0; i < 20; ++i){
        WORD t = ROTLEFT(a, 5) + ((b & c) ^ (~b & d)) + e + ctx->k[0] + m[i];
        e = d;
        d = c;
        c = ROTLEFT(b, 30);
        b = a;
        a = t;
    }
    for (int i = 20; i < 40; ++i){
        WORD t = ROTLEFT(a, 5) + (b ^ c ^ d) + e + ctx->k[1] + m[i];
        e = d;
        d = c;
        c = ROTLEFT(b, 30);
        b = a;
        a = t;
    }
    for (int i = 40; i < 60; ++i){
        WORD t = ROTLEFT(a, 5) + ((b & c) ^ (b & d) ^ (c & d))  + e + ctx->k[2] + m[i];
        e = d;
        d = c;
        c = ROTLEFT(b, 30);
        b = a;
        a = t;
    }
    for (int i = 60; i < 80; ++i){
        WORD t = ROTLEFT(a, 5) + (b ^ c ^ d) + e + ctx->k[3] + m[i];
        e = d;
        d = c;
        c = ROTLEFT(b, 30);
        b = a;
        a = t;
    }
    ctx->state[0] += a;
    ctx->state[1] += b;
    ctx->state[2] += c;
    ctx->state[3] += d;
    ctx->state[4] += e;
    free(m);
}
void sha1_final(SHA1_CTX *ctx, BYTE hash[]){
    WORD i;
    i = ctx->datalen;
    if (ctx->datalen < 56) {
        ctx->data[i++] = 0x80;
        while (i < 56)
            ctx->data[i++] = 0x00;
    } else {
        ctx->data[i++] = 0x80;
        while (i < 64)
            ctx->data[i++] = 0x00;
        sha1_transform(ctx, ctx->data);
        memset(ctx->data, 0, 56);
    }
    ctx->bitlen += ctx->datalen * 8;
    ctx->data[63] = ctx->bitlen;
    ctx->data[62] = ctx->bitlen >> 8;
    ctx->data[61] = ctx->bitlen >> 16;
    ctx->data[60] = ctx->bitlen >> 24;
    ctx->data[59] = ctx->bitlen >> 32;
    ctx->data[58] = ctx->bitlen >> 40;
    ctx->data[57] = ctx->bitlen >> 48;
    ctx->data[56] = ctx->bitlen >> 56;
    sha1_transform(ctx, ctx->data);
    for (i = 0; i < 4; ++i) {
        BYTE temp1 = ctx->state[0] >> (24 - i * 8);
        BYTE temp2 = ctx->state[1] >> (24 - i * 8);
        BYTE temp3 = ctx->state[2] >> (24 - i * 8);
        BYTE temp4 = ctx->state[3] >> (24 - i * 8);
        BYTE temp5 = ctx->state[4] >> (24 - i * 8);
        hash[i] = temp1 & 0x000000ff;
        hash[i + 4] = temp2 & 0x000000ff;
        hash[i + 8] = temp3 & 0x000000ff;
        hash[i + 12] = temp4 & 0x000000ff;
        hash[i + 16] = temp5 & 0x000000ff;
    }
}
\end{lstlisting}


\noindent\begin{minipage}{.49\textwidth}
\begin{lstlisting}[caption={ Example GPT function re-write of sha1\_update produced from Llama-2-70b-chat-hf with inference temperature 0.9 and prompt number 2. This function re-write is an instance of the function category found by the Metric 17 definition, in this case where the generated hashes were correct for 3 out of the 4 test vectors. The output for the four test vectors are given below the source code function.  },captionpos=b,label={source_code:correct_for_at_least_one_test_vector_but_are_incorrect_example1},language=C,style=base,keywordstyle=\color{blue}]


void sha1_update(SHA1_CTX *ctx, const BYTE data[], size_t len) {
    size_t i;
    for (i = 0; i < len; ++i) {
        ctx->data[ctx->datalen] = data[i];
        ctx->datalen = (ctx->datalen + 1) % 64;
        if (ctx->datalen == 0) {
            sha1_transform(ctx, ctx->data);
            ctx->bitlen += 512;
            ctx->datalen = 63;
        }
    }
}

Compiled Binary output:
a9993e364706816aba3e25717850c26c9cd0d89d
84983e441c3bd26ebaae4aa1f95129e5e54670f1
3@f@@1@@b@@7@@c@@8@@0@@b@@5@@4@@a@d@4@@d@@6@@7@@7@@b@@4@@d@b@b@@f@@8@@1@@d@d2@1@@b@@9@@5@@2@3@9@@1@@5@@8@@9@
04575f6b701b0333133f720bc5c1353844075b57


\end{lstlisting}

\end{minipage}

\noindent\begin{minipage}{.49\textwidth}
\begin{lstlisting}[caption={ Example GPT function re-write of sha1\_final produced from Mistral-7B-Instruct-v0.1 with inference temperature 0.5 and prompt number 4. This function re-write is an instance of the function category found by the Metric 18 definition, meaning that for some of the test vectors the generated hashes are correct, but the output changes depending on what compiler optimizations are used. The generated checksums for the four test vectors are given below the source code function, in this case for three specific compiler optimization settings (3 sets of 4 hexadecimal strings). The hash for the third test vector was correct for all gcc optimization settings and correct for clang with optimization level 0. All higher optimization levels used with clang cause the compiled binary to generate incorrect checksums. },captionpos=b,label={source_code:correct_for_at_least_one_test_vector_but_are_incorrect_optimization_unstable_example1},language=C,style=base,keywordstyle=\color{blue}]

void sha1_final(SHA1_CTX *ctx, BYTE hash[])
{
	WORD i;
	if (ctx->datalen < 56) {
		ctx->data[i++] = 0x80;
		while (i < 56)
			ctx->data[i++] = 0x00;
	}
	else {
		ctx->data[i++] = 0x80;
		while (i < 64)
			ctx->data[i++] = 0x00;
		sha1_transform(ctx, ctx->data);
		memset(ctx->data, 0, 56);
	}
	ctx->bitlen += ctx->datalen * 8;
	ctx->data[63] = ctx->bitlen;
	ctx->data[62] = ctx->bitlen >> 8;
	ctx->data[61] = ctx->bitlen >> 16;
	ctx->data[60] = ctx->bitlen >> 24;
	ctx->data[59] = ctx->bitlen >> 32;
	ctx->data[58] = ctx->bitlen >> 40;
	ctx->data[57] = ctx->bitlen >> 48;
	ctx->data[56] = ctx->bitlen >> 56;
	sha1_transform(ctx, ctx->data);
	for (i = 0; i < 4; ++i) {
		hash[i]      = (ctx->state[0] >> (24 - i * 8)) & 0x000000ff;
		hash[i + 4]  = (ctx->state[1] >> (24 - i * 8)) & 0x000000ff;
		hash[i + 8]  = (ctx->state[2] >> (24 - i * 8)) & 0x000000ff;
		hash[i + 12] = (ctx->state[3] >> (24 - i * 8)) & 0x000000ff;
		hash[i + 16] = (ctx->state[4] >> (24 - i * 8)) & 0x000000ff;
	}
}


Compiled Binary output for gcc with optimization level 0:
@f@@1@@e@@2@@4@@6@@0@@e@@a@@b@@a@@c@@e@@9@@2@@7@@9@@0@@b@@2@2@f@@f@@4@@f@@5@@1@0@2@@6@@5@@1@@4@@7@@c@@1@@9@@a@@1@@1@
@9@@9@@c@@7@@b@@c@@2@@1@@b@@9@@a@b@3@@8@@1@e@9@@8@@1@@0@@9@@9@@b@@f@@a@@8@@f@@2@@0@@5@@2@@b@@3@@3@@f@@a@7@6@@2@@9@
34aa973cd4c4daa4f61eeb2bdbad27316534016f
@9@@9@@c@7@b@@c@@2@@1@@b@@9@@a@b@3@@8@@1@@e@@9@@8@@1@@0@@9@@9@@b@@f@@a@@8@@f@@2@@0@5@2@@b@@3@@3@@f@@a@@7@@6@@2@@9@

Compiled Binary output for clang with optimization level 1:
@6@@a@@c@@4@@0@@a@@b@@e@@a@@5@@f@@9@@5@@3@@3@@2@@7@@2@@9@@8@@3@@a@@d@@b@@a@@4@@2@@c@@4@@b@@7@@1@@4@@e@@e@@f@@2@@7@@a@@7@
@2@@8@@c@@0@@e@@b@@8@@a@@5@@7@@0@@c@@f@@b@@d@@7@@3@@7@@1@@d@@9@@d@@f@@8@@7@@2@@6@@3@@9@9@0@@d@@9@@6@@d@@c@@d@@3@@d@@a@
@8@@8@@b@@b@@5@@e@@1@@d@@e@@3@@a@@9@@9@@f@@c@@b@@9@@5@@9@@7@e@a@@5@@8@@2@@c@@b@@a@@c@@f@@d@@6@@7@@a@@b@@7@@3@@0@@0@@3@
@d@@9@@3@@e@@6@@5@@a@@c@7@3@@c@@f@@1@@5@@1@@4@@f@3@5@@9@@d@@5@@d@@6@@9@@0@@9@@e@@4@@7@@c@@5@@c@@5@@6@@b@@9@@1@@4@@a@

Compiled Binary output for clang with optimization level z:
@9@9@f@@b@@1@@5@@b@@4@4@4@@9@@b@@f@@9@@c@@5@@8@@4@@6@@9@@3@@a@7@6@@c@@1@@a@0@0@@8@@1@@1@@4@@0@@a@@f@@2@@1@@6@@6@
@e@@1@@c@@6@@6@@3@@9@@d@@2@@2@@2@@7@@4@@b@@7@@6@@0@@7@@3@@b@4@e@@e@@2@@2@@7@@6@@7@@4@@1@@1@@0@@0@@9@@1@@b@@9@@b@@b@@f@
@6@@2@@7@@0@@d@@8@@a@c@7@@f@c@1@@8@@3@@4@@b@@a@@8@@b@@1@@c@b@6@@a@@7@@8@a@0@@6@@a@@4@@c@@7@@0@@9@4@b@@c@@5@@6@
@e@@1@@c@@6@@6@@3@@9@@d@@2@@2@@2@@7@@4@@b@@7@@6@@0@@7@3@b@@4@@e@@e@@2@@2@@7@@6@@7@@4@@1@@1@@0@@0@@9@@1@@b@@9@b@b@@f@


\end{lstlisting}

\end{minipage}

\noindent\begin{minipage}{.49\textwidth}
\begin{lstlisting}[caption={ This function re-write is an example where the algorithmic correctness of the compiled binary is dependent on the compiler optimization setting that is used (this is categorized by Metric 5). Specifically, using optimization level 0 with both gcc and clang result in the binary correctly producing SHA-1 hashes, but any higher optimization levels cause the binary to produce incorrect SHA-1 hashes. The incorrect hashes are the same for all higher levels of gcc optimization (optimization levels 1 and fast are shown as examples), and the incorrect hashes for the compiled clang binaries are different from the incorrect gcc compiled binaries (optimization levels 1 and fast are also shown for clang as examples). Notably, visually the incorrect hashes seem to have reasonably high entropy and do not show obvious low-entropy correlations. This function re-write was generated by Mistral-7B-Instruct-v0.1 with inference temperature 0.4 with prompt number 4.  },captionpos=b,label={source_code:alg_correctness_compiler_optimization_unstable1},language=C,style=base,keywordstyle=\color{blue}]

void sha1_init(SHA1_CTX *ctx){
	ctx->datalen = 0;
	ctx->bitlen = 0;
	ctx->state[0] = 0x67452301;
	ctx->state[1] = 0xEFCDAB89;
	ctx->state[2] = 0x98BADCFE;
	ctx->state[3] = 0x10325476;
	ctx->state[4] = 0xc3d2e1f0;
	ctx->k[0] = 0x5a827999;
	ctx->k[1] = 0x6ed9eba1;
	ctx->k[2] = 0x8f1bbcdc;
	ctx->k[3] = 0xca62c1d6;
	ctx->state[0] = ctx->state[0] >> 32;
	ctx->state[1] = ctx->state[1] >> 32;
	ctx->state[2] = ctx->state[2] >> 32;
	ctx->state[3] = ctx->state[3] >> 32;
	ctx->state[4] = ctx->state[4] >> 32;
	ctx->k[0] = ctx->k[0] >> 32;
	ctx->k[1] = ctx->k[1] >> 32;
	ctx->k[2] = ctx->k[2] >> 32;
	ctx->k[3] = ctx->k[3] >> 32;
}

gcc with optimization level 0:
a9993e364706816aba3e25717850c26c9cd0d89d
84983e441c3bd26ebaae4aa1f95129e5e54670f1
34aa973cd4c4daa4f61eeb2bdbad27316534016f
04575f6b701b0333133f720bc5c1353844075b57

gcc with optimization level 1:
@8@@7@@5@@d@@9@@1@@7@@8@@1@@2@@f@@d@@0@@4@@9@@0@@8@@7@@4@@7@@5@@1@@2@@8@@2@@2@@4@@c@c@6@6@6@@d@@6@@6@@3@d@b@@0@@7@
@7@@c@@4@@4@@9@@d@@2@@d@@f@@a@@4@@4@@3@@4@@b@@5@@7@@2@a@9@4@8@@1@@a@@1@@8@@8@@f@@6@@c@@a@@4@@4@@0@@5@@5@@8@@a@@1@@4@
@a@@a@@b@@c@@7@@9@@8@@1@@1@@c@@6@@7@@2@@3@@2@@8@@1@@f@1@2@@0@@9@2@c@@f@@3@@f@@8@2@3@@6@@4@@b@5@9@@a@0@f@@8@@3@
@4@@3@@b@@8@@a@@7@@c@@d@@c@@2@@f@@6@0@c@@8@@a@@e@3@c@@b@@9@@6@@8@@3@@e@@2@@d@1@4@@6@@5@@6@@f@@9@@e@@b@5@8@@f@@a@

gcc with optimization level fast:
@8@@7@@5@@d@@9@@1@@7@@8@@1@@2@@f@@d@@0@@4@@9@@0@@8@@7@@4@@7@@5@@1@@2@@8@@2@@2@@4@@c@c@6@6@6@@d@@6@@6@@3@d@b@@0@@7@
@7@@c@@4@@4@@9@@d@@2@@d@@f@@a@@4@@4@@3@@4@@b@@5@@7@@2@a@9@4@8@@1@@a@@1@@8@@8@@f@@6@@c@@a@@4@@4@@0@@5@@5@@8@@a@@1@@4@
@a@@a@@b@@c@@7@@9@@8@@1@@1@@c@@6@@7@@2@@3@@2@@8@@1@@f@1@2@@0@@9@2@c@@f@@3@@f@@8@2@3@@6@@4@@b@5@9@@a@0@f@@8@@3@
@4@@3@@b@@8@@a@@7@@c@@d@@c@@2@@f@@6@0@c@@8@@a@@e@3@c@@b@@9@@6@@8@@3@@e@@2@@d@1@4@@6@@5@@6@@f@@9@@e@@b@5@8@@f@@a@

clang with optimization level 0:
a9993e364706816aba3e25717850c26c9cd0d89d
84983e441c3bd26ebaae4aa1f95129e5e54670f1
34aa973cd4c4daa4f61eeb2bdbad27316534016f
04575f6b701b0333133f720bc5c1353844075b57

clang with optimization level 1:
@7@@8@@c@@3@@9@@4@@b@@7@4@9@@4@@f@@5@@0@@7@@e@@8@@1@@2@@5@@5@@1@@8@17@a@@7@@a@c@5@@c@c@f@@3@@4@@4@@7@@6@@b@@6@
@a@@8@@5@8@c@@6@@b@@f@@c@@1@@5@@f@@8@@3@@5@@b@@5@@1@@b@@b@@2@@3@@1@@3@@4@9@9@@6@@9@@5@@9@@9@@c@@4@@7@@5@@0@@f@@6@@0@
@d@4@f@@e@@7@@b@@f@@e@@e@@5@@b@@6@@6@@6@@1@@5@f@b@@2@@5@@7@@e@@7@@d@@6@@c@@d@@5@@7@@c@@e@@7@@b@@6@@9@@6@@d@@4@@9@f
@1@@b@@8@@5@@3@@7@@b@@7@@2@@9@@0@@c@@6@@5@@5@@d@@6@@4@@c@@b@@a@@6@@7@@d@@d@@3@@9@@5@@1@@d@@b@@e@@f@@c@@e@@e@@3@@5@@1@@3@

clang with optimization level fast:
@7@@8@@c@@3@@9@@4@@b@@7@4@9@@4@@f@@5@@0@@7@@e@@8@@1@@2@@5@@5@@1@@8@17@a@@7@@a@c@5@@c@c@f@@3@@4@@4@@7@@6@@b@@6@
@a@@8@@5@8@c@@6@@b@@f@@c@@1@@5@@f@@8@@3@@5@@b@@5@@1@@b@@b@@2@@3@@1@@3@@4@9@9@@6@@9@@5@@9@@9@@c@@4@@7@@5@@0@@f@@6@@0@
@d@4@f@@e@@7@@b@@f@@e@@e@@5@@b@@6@@6@@6@@1@@5@f@b@@2@@5@@7@@e@@7@@d@@6@@c@@d@@5@@7@@c@@e@@7@@b@@6@@9@@6@@d@@4@@9@f
@1@@b@@8@@5@@3@@7@@b@@7@@2@@9@@0@@c@@6@@5@@5@@d@@6@@4@@c@@b@@a@@6@@7@@d@@d@@3@@9@@5@@1@@d@@b@@e@@f@@c@@e@@e@@3@@5@@1@@3@

\end{lstlisting}
\end{minipage}
\noindent\begin{minipage}{.49\textwidth}
\begin{lstlisting}[caption={ Example function re-write, that does not produce correct SHA-1 hashes, and where at least one of the generated hashes is very close to the correct SHA-1 hash, in this case by being incorrect by only two hexadecimal characters. This function re-write was generated by Mistral-7B-Instruct-v0.1 with inference temperature 0.9 with prompt number 2. },captionpos=b,label={source_code:hash_off_by_less_than_5_characters_example1},language=C,style=base,keywordstyle=\color{blue}]

void sha1_final(SHA1_CTX *ctx, BYTE *hash){
	WORD i;
	i = ctx->datalen;
	if (ctx->datalen < 56) {
		ctx->data[i++] = 0x80;
		while (i < 56)
			ctx->data[i++] = 0x00;
	}
	else {
		ctx->data[i++] = 0x80;
		while (i < 64)
			ctx->data[i++] = 0x00;
		sha1_transform(ctx, ctx->data);
		memset(ctx->data, 0, 56);
	}
	ctx->bitlen += ctx->datalen * 8;
	ctx->data[63] = ctx->bitlen;
	ctx->data[62] = ctx->bitlen >> 8;
	ctx->data[61] = ctx->bitlen >> 16;
	ctx->data[60] = ctx->bitlen >> 24;
	ctx->data[59] = ctx->bitlen >> 32;
	ctx->data[58] = ctx->bitlen >> 40;
	ctx->data[57] = ctx->bitlen >> 48;
	ctx->data[56] = ctx->bitlen >> 56;
	sha1_transform(ctx, ctx->data);
	for (i = 0; i < 4; ++i) {
		hash[i]      = (ctx->state[0] >> (24 - i * 8)) & 0x0000007f;
		hash[i + 4]  = (ctx->state[1] >> (24 - i * 8)) & 0x0000007f;
		hash[i + 8]  = (ctx->state[2] >> (24 - i * 8)) & 0x0000007f;
		hash[i + 12] = (ctx->state[3] >> (24 - i * 8)) & 0x0000007f;
		hash[i + 16] = (ctx->state[4] >> (24 - i * 8)) & 0x0000007f;
	}
}

Compiled Binary output:
@2@9@1@93e364706@0@16a@3@a3e25717850@4@26c@1@c@5@0@5@8@1@d
@0@4@1@83e441c3b@5@26e@3@a@2@e4a@2@1@7@95129@6@5@6@54670@7@1
34@2@a@1@73c@5@4@4@4@5@a@2@4@7@61e@6@b2b@5@b@2@d27316534016f
04575f6b701b0333133f720b@4@5@4@1353844075b57
\end{lstlisting}

\end{minipage}

\begin{lstlisting}[caption={ Example function re-write of sha1\_transform where for some compiler optimization settings the compiled binary, when executed, results in a fatal error, but for at least one other optimization setting the compiled binary correctly produces SHA-1 hashes. For 6 of the compiler optimization settings (specifically gcc with levels 0, 1, 2, 3, fast and clang with level 0), the compiled binary threw a Signals.SIGSEGV fatal error. The binaries compiled using gcc with optimization level s, and clang with optimization levels 1, 2, s, z correctly generated SHA-1 hashes. This was produced by Llama-2-70b-chat-hf with prompt number 2 and inference temperature 0.5. },captionpos=b,label={source_code:optimization_unstable_between_fatal_error_and_being_SHA1_correct_example1},language=C,style=base,keywordstyle=\color{blue}]

void sha1_transform(SHA1_CTX *ctx, const BYTE data[]) {
    WORD a, b, c, d, e, i, j, t, m[80];
    // Initialize variables
    a = ctx->state[0];
    b = ctx->state[1];
    c = ctx->state[2];
    d = ctx->state[3];
    e = ctx->state[4];
    // Process input data
    for (i = 0; i < 16; ++i) {
        m[i] = (data[j] << 24) + (data[j + 1] << 16) + (data[j + 2] << 8) + (data[j + 3]);
        j += 4;
    }
    for (; i < 80; ++i) {
        m[i] = (m[i - 3] ^ m[i - 8] ^ m[i - 14] ^ m[i - 16]);
        m[i] = (m[i] << 1) | (m[i] >> 31);
    }
    // Perform transformations
    for (i = 0; i < 20; ++i) {
        t = ROTLEFT(a, 5) + ((b & c) ^ (~b & d)) + e + ctx->k[0] + m[i];
        e = d;
        d = c;
        c = ROTLEFT(b, 30);
        b = a;
        a = t;
    }
    for (; i < 40; ++i) {
        t = ROTLEFT(a, 5) + (b ^ c ^ d) + e + ctx->k[1] + m[i];
        e = d;
        d = c;
        c = ROTLEFT(b, 30);
        b = a;
        a = t;
    }
    for (; i < 60; ++i) {
        t = ROTLEFT(a, 5) + ((b & c) ^ (b & d) ^ (c & d))  + e + ctx->k[2] + m[i];
        e = d;
        d = c;
        c = ROTLEFT(b, 30);
        b = a;
        a = t;
    }
    for (; i < 80; ++i) {
        t = ROTLEFT(a, 5) + (b ^ c ^ d) + e + ctx->k[3] + m[i];
        e = d;
        d = c;
        c = ROTLEFT(b, 30);
        b = a;
        a = t;
    }
    // Update state
    ctx->state[0] += a;
    ctx->state[1] += b;
    ctx->state[2] += c;
    ctx->state[3] += d;
    ctx->state[4] += e;
}

Compiled Binary output for gcc with optimization level s:
a9993e364706816aba3e25717850c26c9cd0d89d
84983e441c3bd26ebaae4aa1f95129e5e54670f1
34aa973cd4c4daa4f61eeb2bdbad27316534016f
04575f6b701b0333133f720bc5c1353844075b57

Compiled Binary output for clang with optimization level 1:
a9993e364706816aba3e25717850c26c9cd0d89d
84983e441c3bd26ebaae4aa1f95129e5e54670f1
34aa973cd4c4daa4f61eeb2bdbad27316534016f
04575f6b701b0333133f720bc5c1353844075b57

Compiled Binary output for clang with optimization level 2:
a9993e364706816aba3e25717850c26c9cd0d89d
84983e441c3bd26ebaae4aa1f95129e5e54670f1
34aa973cd4c4daa4f61eeb2bdbad27316534016f
04575f6b701b0333133f720bc5c1353844075b57

Compiled Binary output for clang with optimization level 3:
@d@@a@@4@9@6@@8@@e@@b@@2@@e@@3@@7@@7@@c@@1@@f@@8@@8@@4@e@8@@f@@5@@2@@8@@3@5@2@@4@@b@@e@@b@@e@@7@@4@@e@@b@@d@@b@d
@3@@1@@3@@9@@0@@8@@d@@8@@9@@e@@0@@4@@c@@b@@0@@b@@2@@c@@0@@b@@c@@8@@e@@9@@6@@d@@e@12@a@@a@@a@@4@@7@@3@@a@@8@@d@@b@@e@
@c@@2@@2@@c@@c@@a@@1@@0@@b@@a@@a@@8@@4@@1@@0@@7@@9@@e@@0@@0@@d@@5@@b@@4@@e@b@5@d@5@@3@@9@@d@@1@@d@@8@@5@@e@@6@@d@@1@
@2@@1@@c@@6@@3@@9@@8@@4@@2@@6@@9@@9@@5@@4@@9@@f@@b@@c@@f@@7@@5@@5@@c@@8@@0@@c@@6@1@f@@0@@e@@7@4@2@@2@@8@@a@@8@@0@@a@

Compiled Binary output for clang with optimization level s:
a9993e364706816aba3e25717850c26c9cd0d89d
84983e441c3bd26ebaae4aa1f95129e5e54670f1
34aa973cd4c4daa4f61eeb2bdbad27316534016f
04575f6b701b0333133f720bc5c1353844075b57

Compiled Binary output for clang with optimization level fast:
@d@@a@@4@9@6@@8@@e@@b@@2@@e@@3@@7@@7@@c@@1@@f@@8@@8@@4@e@8@@f@@5@@2@@8@@3@5@2@@4@@b@@e@@b@@e@@7@@4@@e@@b@@d@@b@d
@3@@1@@3@@9@@0@@8@@d@@8@@9@@e@@0@@4@@c@@b@@0@@b@@2@@c@@0@@b@@c@@8@@e@@9@@6@@d@@e@12@a@@a@@a@@4@@7@@3@@a@@8@@d@@b@@e@
@c@@2@@2@@c@@c@@a@@1@@0@@b@@a@@a@@8@@4@@1@@0@@7@@9@@e@@0@@0@@d@@5@@b@@4@@e@b@5@d@5@@3@@9@@d@@1@@d@@8@@5@@e@@6@@d@@1@
@2@@1@@c@@6@@3@@9@@8@@4@@2@@6@@9@@9@@5@@4@@9@@f@@b@@c@@f@@7@@5@@5@@c@@8@@0@@c@@6@1@f@@0@@e@@7@4@2@@2@@8@@a@@8@@0@@a@

Compiled Binary output for clang with optimization level z:
a9993e364706816aba3e25717850c26c9cd0d89d
84983e441c3bd26ebaae4aa1f95129e5e54670f1
34aa973cd4c4daa4f61eeb2bdbad27316534016f
04575f6b701b0333133f720bc5c1353844075b57



\end{lstlisting}


\noindent\begin{minipage}{.49\textwidth}
\begin{lstlisting}[caption={ Example of a function re-write where the generated hashes have some of the basic requirements of a valid hash function, and have the same hexadecimal character length as SHA-1 hashes (40 characters). This code did not produce different output based on different compiler optimization settings. This is an example of a case found by Metric 39. This function re-write was generated by Mistral-7B-Instruct-v0.1 with inference temperature 0.6 with prompt number 4. Additionally, this function re-write had at least one detected out-of-bounds error by the automated memory sanitizer check. },captionpos=b,label={source_code:potentially_good_hash_but_not_correct_SHA1_compiler_stable_example1},language=C,style=base,keywordstyle=\color{blue}]

void sha1_init(SHA1_CTX *ctx)
{
	ctx->datalen = 0;
	ctx->bitlen = 0;
	ctx->state[0] = 0x67452301;
	ctx->state[1] = 0xEFCDAB89;
	ctx->state[2] = 0x98BADCFE;
	ctx->state[3] = 0x10325476;
	ctx->state[4] = 0xc3d2e1f0;
	ctx->k[0] = 0x5a827999;
	ctx->k[1] = 0x6ed9eba1;
	ctx->k[2] = 0x8f1bbcdc;
	ctx->k[3] = 0xca62c1d6;
	ctx->state[0] ^= ctx->k[0];
	ctx->state[1] ^= ctx->k[1];
	ctx->state[2] ^= ctx->k[2];
	ctx->state[3] ^= ctx->k[3];
	ctx->state[4] ^= ctx->k[4];
}

Compiled Binary output:
@4@@2@@7@@5@@e@@c@@a@@f@@3@@5@0@9@@7@1@0@@1@@5@@e@@d@@6@@3@@3@7@6@@c@@2@@9@@d@@8@@d@@8@@7@@8@@3@@1@@8@@7@@b@@b@@1@
@3@4@a@@4@@a@@1@@1@@7@@3@@f@@5@@3@d@9@@d@@7@@e@@2@@9@@1@@5@@d@@e@@d@@8@@3@@6@@5@@4@@e@@f@@a@@c@@8@@3@@9@@3@@b@@2@@c@
@a@@5@@0@@d@@2@@6@@6@@b@@6@4@b@@e@@8@@9@@1@@7@@a@@2@@6@@9@@d@@7@@5@@5@@6@@8@@6@@1@@e@@3@@4@@a@@5@@b@@0@@b@@b@@f@@f@@6@
@c@@2@@1@@4@@4@@3@@d@@8@7@8@@5@@8@@3@@e@@4@@9@@8@@6@@4@@a@7@c@@b@@4@@9@@2@@0@@a@3@0@@5@@c@@7@4@4@@a@@6@@a@@b@@9@


\end{lstlisting}

\end{minipage}

\noindent\begin{minipage}{.49\textwidth}
\begin{lstlisting}[caption={Incorrect SHA-1 function implementation of sha1\_final which causes a detected memory leak using Valgrind (for a binary compiled using optimization level 0 for either gcc or clang or both). Note that the compiled binary output for this example function does not change when different compiler optimization levels are applied. All of these output hash digests are very far away from the correct SHA-1 hash digests. This example was produced by Mistral-7B-Instruct-v0.1 using inference temperature of 0.6 and prompt number 4. },captionpos=b,label={code:mem_leak_example1},language=C,style=base,keywordstyle=\color{blue}]

void sha1_final(SHA1_CTX *ctx, BYTE hash[]){
    WORD i;
    BYTE *padded_data = malloc(ctx->datalen + 56);
    for (i = 0; i < ctx->datalen; ++i) {
        padded_data[i] = ctx->data[i];
    }
    for (i = ctx->datalen; i < 56; ++i) {
        padded_data[i] = 0x00;
    }
    ctx->bitlen += ctx->datalen * 8;
    padded_data[ctx->datalen] = ctx->bitlen;
    padded_data[ctx->datalen + 1] = ctx->bitlen >> 8;
    padded_data[ctx->datalen + 2] = ctx->bitlen >> 16;
    padded_data[ctx->datalen + 3] = ctx->bitlen >> 24;
    padded_data[ctx->datalen + 4] = ctx->bitlen >> 32;
    padded_data[ctx->datalen + 5] = ctx->bitlen >> 40;
    padded_data[ctx->datalen + 6] = ctx->bitlen >> 48;
    padded_data[ctx->datalen + 7] = ctx->bitlen >> 56;
    sha1_transform(ctx, padded_data);
    memset(padded_data, 0, ctx->datalen + 56);
    for (i = 0; i < 4; ++i) {
        hash[i]      = (ctx->state[0] >> (24 - i * 8)) & 0x000000ff;
        hash[i + 4]  = (ctx->state[1] >> (24 - i * 8)) & 0x000000ff;
        hash[i + 8]  = (ctx->state[2] >> (24 - i * 8)) & 0x000000ff;
        hash[i + 12] = (ctx->state[3] >> (24 - i * 8)) & 0x000000ff;
        hash[i + 16] = (ctx->state[4] >> (24 - i * 8)) & 0x000000ff;
    }
}

Compiled binary output:
@1@@2@@4@@2@@a@@4@@6@@d@@f@@f@@b@@f@@d@1@a@a@1@@f@@5@@0@@c@@d@@0@@c@@0@@a@@9@@c@@e@@c@@5@@f@@c@@2@@9@@6@@2@@6@@0@@a@
@c@@a@@1@@d@@c@@8@@9@@a@@b@@7@@0@@6@@c@@4@@8@@5@@0@@5@@2@@6@@6@@1@@c@@6@@6@@8@@4@@7@@f@@c@@b@@d@@7@@c@@1@@7@@b@@d@@e@@8@
@8@@9@@2@@0@@1@@4@@6@@5@@3@@d@@4@@d@@b@@d@@6@@5@@1@@8@@4@@c@@2@@d@2@8@@2@@9@@6@@3@@4@@9@@9@@5@@0@@c@@9@@9@@a@@b@@5@f
@a@@e@@6@@1@@f@@5@@a@@6@@b@@c@@b@@d@0@6@@f@@e@@7@@1@@c@@1@@3@@6@@c@@f@@a@@7@@8@@5@@b@@9@@e@@c@@1@@0@@d@@c@@4@@0@@3@@e@


\end{lstlisting}
\end{minipage}

Code Listing \ref{re_written_source_code:SHA-1_reference_functions_ex1} shows a complete SHA-1 re-written codebase example that is fully correct, and all of the component functions are distinct from the original source code (as determined by the compiled binary hashing clustering). In each of these functions, the code functions are selected arbitrarily from the entirety of the correct (and unique) function re-writes to serve as representative, and interesting, examples of the GPT code-writing results. Code Listings \ref{re_written_source_code:SHA-1_reference_functions_ex2} and \ref{re_written_source_code:SHA-1_reference_functions_ex3} in Appendix \ref{section:appendix_fully_correct_SHA1_rewrites} shows two more example SHA-1 codebases that were completely re-written and unique.

Code Listing \ref{source_code:correct_for_at_least_one_test_vector_but_are_incorrect_example1} shows a specific function re-write where three of the output checksums are correct, but one of them is not correct. Code Listing \ref{source_code:correct_for_at_least_one_test_vector_but_are_incorrect_optimization_unstable_example1} shows a similar function re-write, except in that case the generated output varied based on the compiler optimization settings. Code Listing \ref{source_code:alg_correctness_compiler_optimization_unstable1} shows an example of a function re-write which is correct for all 4 test vectors, if the code is compiled using specific optimization settings, and otherwise the output checksums are not at all correct. 

Code Listing \ref{source_code:hash_off_by_less_than_5_characters_example1} shows a function re-write where the output checksums are not correct (e.g. not SHA-1 hashes), but are incorrect by only a few hexadecimal characters. 

Code Listing \ref{source_code:optimization_unstable_between_fatal_error_and_being_SHA1_correct_example1} shows a function re-write that is compiler optimization unstable where for some settings the compiled binary correct produces SHA-1 hashes, for other settings it crashes in a fatal error, and still for other settings does produce output but the output is not correct SHA-1 hexadecimal hashes. 

Code Listing \ref{source_code:potentially_good_hash_but_not_correct_SHA1_compiler_stable_example1} shows an example function re-write that produces incorrect checksums, but that have the interesting property of appearing to be good hashes (e.g. no apparent dependence on the input, and reasonably high entropy hexadecimal strings). Note that this specific example is not necessarily a good hash function -- the notable thing is that the function re-write was quite minimal, and yet the compiled binary produced hexadecimal strings that were not obviously terrible checksums. 

Code Listing \ref{code:mem_leak_example1} shows a function re-write example where the compiled binary had a detected memory leak by Valgrind. Interestingly, the compiled binary was able to be executed, and hexadecimal output was produced, however it was not correct SHA-1 hashes.

\section{Discussion and Conclusion}
\label{section:conclusion}

This study shows that open source GPT models, without customized fine-tuning, can be used to construct correct algorithmic invariant implementations of a cryptographic hash function, in particular SHA-1. Code Listings \ref{re_written_source_code:SHA-1_reference_functions_ex1}, \ref{re_written_source_code:SHA-1_reference_functions_ex2}, \ref{re_written_source_code:SHA-1_reference_functions_ex3} explicitly show three examples of re-written SHA-1 source code that maintain the same correct functionality of the original C source code. However, the success rate of function re-writes in terms of algorithmic correctness, or even being compilable, is quite low. Meaning that to assess a current GPT model for code re-writing capabilities, one needs to produce a large distribution of function re-writes and thoroughly test the characteristics of that output. 

Fundamentally, using GPT models to produce source code, or as shown in this study to produce source code re-writes, is a cybersecurity risk for the integrity and stability of software development. The suite of tests that have been applied to these SHA-1 function re-writes has shown that GPT re-written functions can contain critical software flaws, some of which could be hard to detect without proper software validation and testing. GPT models have been shown to be incredibly effective at producing these SHA-1 re-writes at scale, but this capability of GPT models should serve as a research tool for testing interesting and useful versions of software, not as a solution for writing correct software. The tendency for GPT produced source code to contain bugs has been observed in several previous studies \cite{sandoval2023lost, pearce2021asleep, spiess2024calibration, bhatt2023purple}, meaning the findings of this study are consistent with the existing literature. The remarkable finding here is that these faulty code implementations were generated as a byproduct of GPT models with prompts requesting \emph{accurate} code function re-writes, as opposed to other studies where the aim was to induce the generation of faulty code from GPT models \cite{wu2023deceptprompt}. However, the other very notable finding of this study is the dramatic variety of code re-writes that were produced. While many of these re-writes were incorrect or contained serious software flaws, this suggests that GPT models offer a unique capability of producing large amounts of highly variable source code - which could be used for studying properties of computer code (namely, security properties), or fuzzing computer code. The code re-writing procedure described in this study should also be applied to implementations of other cryptographic algorithms to determine how well the GPT models perform at rewriting potentially even more complex code than this SHA-1 implementation.

This type of generative machine learning code re-writing could also be tested on malware source code, and will likely be used for this purpose in the wild. In the case of malware, the measurement of correctness is more difficult to capture - and potentially more dangerous in the case of undefined behavior. However, this study is an interesting first step in this direction - especially because cryptographic algorithms are commonly used in malware \cite{begovic2023cryptographic, 7958617, 8960303, 6461007, 8392623, 10.1145/2382196.2382217, cryptoeprint:2022/1699, 5403858}, for example when encrypting web traffic for command and control, or for disruption in the case of ransomware. Meaning that even GPT model re-writes of relatively small functions could further propagate malware variants.

The prompting of GPT models unintentionally producing hash function implementations that fit the basic properties of hash functions, but are not correct to the actual algorithm (in this case SHA-1), is also very notable (Code Listings \ref{source_code:potentially_good_hash_but_not_correct_SHA1_compiler_stable_example1}, \ref{source_code:potentially_good_hash_but_not_correct_SHA1_compiler_stable_example2}, \ref{source_code:potentially_good_hash_but_not_correct_SHA1_compiler_stable_example3}, \ref{source_code:potentially_good_hash_but_not_correct_SHA1_compiler_stable_example4}, \ref{source_code:potentially_good_hash_but_not_correct_SHA1_compiler_stable_example5}, \ref{source_code:potentially_good_hash_but_not_correct_SHA1_compiler_stable_example6}, \ref{source_code:potentially_good_hash_but_not_correct_SHA1_compiler_stable_example7}, \ref{source_code:potentially_good_hash_but_not_correct_SHA1_compiler_stable_example8} are explicit examples of these instances, but there are other instances of this occurring in other function re-write examples namely where the output is compiler optimization unstable). Fundamentally, being able to easily produce hash function implementations using generative transformer models that are valid hash functions (but not necessarily secure \emph{cryptographic} hash functions), will make malware analysis more challenging. Not only does this change the source code, and therefore signatures of the compiled binaries, but it also makes automated cryptographic algorithm detection in binaries \cite{7958617, 8960303, 6461007, 8392623, 10.1145/2382196.2382217, cryptoeprint:2022/1699, 5403858} significantly more challenging since these rely on finding known cryptographic constants used in standard cryptographic algorithms. Future work should analyze how secure these incorrect (e.g., non-SHA-1) hash function implementations are, for example using cryptographic fuzzing tools \cite{10.1145/3628160, jin2022differential, ammann2024dy}. More detailed analysis of the GPT rewritten code, specifically the compiled binaries, can be performed in future research - namely things such as memory usage, compute time efficiency, and detecting other potential security flaws such as side channel attack susceptibility.

The $10$ prompts that were used in this study are selected to be reasonable hand-crafted prompts that attempt in, different ways, to extract useful source code. However, GPT prompts are effectively another hyperparameter that can be tuned to get good performance, and good natural language prompts may not be very obvious. There are a couple of proposed techniques \cite{yang2023large, khattab2023dspy} for automatically constructing good natural language prompts, and it is likely that such methods could be used to get even better performance for GPT code re-writing. 

Future research on source code re-writing using GPT models should be focused on domains where producing a large amount of different source code is useful. The challenge in utilizing GPT models for this is testing for correctness (although, perhaps there are situations where the source code does not need to be an exactly correct computation invariant of the original source code). The clearest case where this could be used is in pre-computing a large number of versions of known malware (where the source code, or compile-able disassembled code, is known) and then computing signatures of those binaries that are used for standard dynamic or static malware detection (such as fuzzy hashing \cite{191669, 9177636}). This set of signatures of the artificially created versions of the malware could then be used to detect those versions of the malware in the wild if a developer were to ever produce those versions. This capability is especially promising since it can be done in an entirely automated system -- this does not require human developers to produce different versions of source code. This pre-computation of malware signatures would also help preemptively combat the likely future of GPT produced malware. 

Another possible application for producing a large number of correct variants of source code is in optimizing the source code for a specific purpose - such as speed, or reduced memory usage. Because GPT models can generate a wide variety of code reasonably fast, they could be prompted to generate re-written and novel versions of code that have some desired characteristic such as executing faster. As seen in this study, it is likely that such code re-writes will not always be correct or adhere to the given prompt, but for a sufficiently large number of samples of GPT produced code, some versions could have the required properties. 

Lastly, another aspect of this study which could be expanded on is the correctness analysis of the function re-writes which passed all of the SHA-1 tests. In principle, a cryptographic implementation passing these tests means that with very high likelihood the implementation is correct. However, we have also seen the numerous ways in which the GPT produced code can be incorrect, and therefore it is plausible, if unlikely, that the functions we have found to be re-written correctly actually contain interesting edge-case flaws.

\section{Acknowledgments}
\label{section:acknowledgments}

Sandia National Laboratories is a multi-mission laboratory managed and operated by National Technology \& Engineering Solutions of Sandia, LLC (NTESS), a wholly owned subsidiary of Honeywell International Inc., for the U.S. Department of Energy’s National Nuclear Security Administration (DOE/NNSA) under contract DE-NA0003525. This written work is authored by an employee of NTESS. The employee, not NTESS, owns the right, title and interest in and to the written work and is responsible for its contents. Any subjective views or opinions that might be expressed in the written work do not necessarily represent the views of the U.S. Government. The publisher acknowledges that the U.S. Government retains a non-exclusive, paid-up, irrevocable, world-wide license to publish or reproduce the published form of this written work or allow others to do so, for U.S. Government purposes. The DOE will provide public access to results of federally sponsored research in accordance with the DOE Public Access Plan.

\clearpage

\setlength\bibitemsep{0pt}
\printbibliography

\appendix
\section{Hash Function Test Vectors}
\label{section:appendix_hash_function_test_vectors}

The four test vectors that are tested against:

\begin{itemize}[noitemsep]
    \item abc
    \item abcdbcdecdefdefgefghfghighijhijkijkljklmklmnlmnomnopnopq
    \item 1,000,000 repeats of the character a
    \item 55555555555555555555555555555555555555555555555555555555
\end{itemize}

The correct SHA-1 hash digests, in hexadecimal, of these four test vectors is the following:

\begin{itemize}[noitemsep]
    \item a9993e364706816aba3e25717850c26c9cd0d89d
    \item 84983e441c3bd26ebaae4aa1f95129e5e54670f1
    \item 34aa973cd4c4daa4f61eeb2bdbad27316534016f
    \item 04575f6b701b0333133f720bc5c1353844075b57
\end{itemize}

These are the same test vectors defined in the original SHA-1 standard \cite{SHA1_NIST_specifications}, but also with an additional test vector that we manually added to test against. The additional test vector is motivated by having a test vector with a character length equal to the second test vector to specifically check if some of the instances where the hash outputs were correct for some test vectors but not other were due to input character length similarities (specifically having the same character length as the second test vector).

\section{Source Code Reference Function Implementations for SHA-1}
\label{section:appendix_reference_source_code_SHA-1}

Code Listing~\ref{source_code:SHA-1_reference_functions} shows the reference source code implementation for the 4 functions that implement the SHA-1 algorithm. This implementation is based on the original NIST implementation, defined in ref. \cite{SHA1_NIST_specifications}. Note that this is from ref.~\cite{cryptographic_algorithms_Github}, and the comments from the original were removed for these versions which were used in the prompts to the GPT models. Code Listing \ref{source_code:SHA-1_reference_header} defines the fixed library imports and the macro that are concatenated with the re-written source code functions in order to generate the complete source code. 

Code listing~\ref{source_code:SHA-1_reference_test_and_main} is adapted from \cite{cryptographic_algorithms_Github}, and includes one additional test vector that we manually added that has the same character length as one of the original test vectors. Importantly, the C code implementations passing these tests is an indication that the underlying source code is very likely a correct implementation of SHA-1. However, it is not certain that the implementation is correct -- and in particular many more tests vectors would need to be applied to the source code in order to better verify its correctness. 

Of these source codes, only the individual functions in Code Listing~\ref{source_code:SHA-1_reference_functions} are re-written and evaluated using the GPT models, since these contain the SHA-1 algorithmic syntax.

\begin{lstlisting}[caption={\textbf{The 4 C functions that comprise the SHA-1 algorithm reference implementation. Note that this version is only the C syntax of the functions which are given as part of the GPT prompts, and does not include comments that were in the original version and does not include macros or library importants. }},captionpos=b,label={source_code:SHA-1_reference_functions},language=C,style=base,keywordstyle=\color{blue}]

#include <stdlib.h>
#include <memory.h>
#include "sha1.h"

#define ROTLEFT(a, b) ((a << b) | (a >> (32 - b)))

void sha1_transform(SHA1_CTX *ctx, const BYTE data[]){
	WORD a, b, c, d, e, i, j, t, m[80];
	for (i = 0, j = 0; i < 16; ++i, j += 4)
		m[i] = (data[j] << 24) + (data[j + 1] << 16) + (data[j + 2] << 8) + (data[j + 3]);
	for ( ; i < 80; ++i) {
		m[i] = (m[i - 3] ^ m[i - 8] ^ m[i - 14] ^ m[i - 16]);
		m[i] = (m[i] << 1) | (m[i] >> 31);
	}
	a = ctx->state[0];
	b = ctx->state[1];
	c = ctx->state[2];
	d = ctx->state[3];
	e = ctx->state[4];
	for (i = 0; i < 20; ++i) {
		t = ROTLEFT(a, 5) + ((b & c) ^ (~b & d)) + e + ctx->k[0] + m[i];
		e = d;
		d = c;
		c = ROTLEFT(b, 30);
		b = a;
		a = t;
	}
	for ( ; i < 40; ++i) {
		t = ROTLEFT(a, 5) + (b ^ c ^ d) + e + ctx->k[1] + m[i];
		e = d;
		d = c;
		c = ROTLEFT(b, 30);
		b = a;
		a = t;
	}
	for ( ; i < 60; ++i) {
		t = ROTLEFT(a, 5) + ((b & c) ^ (b & d) ^ (c & d))  + e + ctx->k[2] + m[i];
		e = d;
		d = c;
		c = ROTLEFT(b, 30);
		b = a;
		a = t;
	}
	for ( ; i < 80; ++i) {
		t = ROTLEFT(a, 5) + (b ^ c ^ d) + e + ctx->k[3] + m[i];
		e = d;
		d = c;
		c = ROTLEFT(b, 30);
		b = a;
		a = t;
	}
	ctx->state[0] += a;
	ctx->state[1] += b;
	ctx->state[2] += c;
	ctx->state[3] += d;
	ctx->state[4] += e;
}
void sha1_init(SHA1_CTX *ctx){
	ctx->datalen = 0;
	ctx->bitlen = 0;
	ctx->state[0] = 0x67452301;
	ctx->state[1] = 0xEFCDAB89;
	ctx->state[2] = 0x98BADCFE;
	ctx->state[3] = 0x10325476;
	ctx->state[4] = 0xc3d2e1f0;
	ctx->k[0] = 0x5a827999;
	ctx->k[1] = 0x6ed9eba1;
	ctx->k[2] = 0x8f1bbcdc;
	ctx->k[3] = 0xca62c1d6;
}
void sha1_update(SHA1_CTX *ctx, const BYTE data[], size_t len){
	size_t i;
	for (i = 0; i < len; ++i) {
		ctx->data[ctx->datalen] = data[i];
		ctx->datalen++;
		if (ctx->datalen == 64) {
			sha1_transform(ctx, ctx->data);
			ctx->bitlen += 512;
			ctx->datalen = 0;
		}
	}
}
void sha1_final(SHA1_CTX *ctx, BYTE hash[]){
	WORD i;
	i = ctx->datalen;
	if (ctx->datalen < 56) {
		ctx->data[i++] = 0x80;
		while (i < 56)
			ctx->data[i++] = 0x00;
	}
	else {
		ctx->data[i++] = 0x80;
		while (i < 64)
			ctx->data[i++] = 0x00;
		sha1_transform(ctx, ctx->data);
		memset(ctx->data, 0, 56);
	}
	ctx->bitlen += ctx->datalen * 8;
	ctx->data[63] = ctx->bitlen;
	ctx->data[62] = ctx->bitlen >> 8;
	ctx->data[61] = ctx->bitlen >> 16;
	ctx->data[60] = ctx->bitlen >> 24;
	ctx->data[59] = ctx->bitlen >> 32;
	ctx->data[58] = ctx->bitlen >> 40;
	ctx->data[57] = ctx->bitlen >> 48;
	ctx->data[56] = ctx->bitlen >> 56;
	sha1_transform(ctx, ctx->data);
	for (i = 0; i < 4; ++i) {
		hash[i]      = (ctx->state[0] >> (24 - i * 8)) & 0x000000ff;
		hash[i + 4]  = (ctx->state[1] >> (24 - i * 8)) & 0x000000ff;
		hash[i + 8]  = (ctx->state[2] >> (24 - i * 8)) & 0x000000ff;
		hash[i + 12] = (ctx->state[3] >> (24 - i * 8)) & 0x000000ff;
		hash[i + 16] = (ctx->state[4] >> (24 - i * 8)) & 0x000000ff;
	}
}

\end{lstlisting}

\begin{lstlisting}[caption={ \textbf{Reference SHA-1 implementation header file }},captionpos=b,label={source_code:SHA-1_reference_header_file},language=C,style=base,keywordstyle=\color{blue}]
#ifndef sha1_H
#define sha1_H

/*************************** HEADER FILES ***************************/
#include <stddef.h>

/****************************** MACROS ******************************/
#define sha1_BLOCK_SIZE 20              // SHA-1 outputs a 20 byte digest

/**************************** DATA TYPES ****************************/
typedef unsigned char BYTE;             // 8-bit byte
typedef unsigned int  WORD;             // 32-bit word, change to "long" for 16-bit machines

typedef struct {
	BYTE data[64];
	WORD datalen;
	unsigned long long bitlen;
	WORD state[5];
	WORD k[4];
} SHA-1_CTX;

/*********************** FUNCTION DECLARATIONS **********************/
void SHA-1_init(sha1_CTX *ctx);
void SHA-1_update(sha1_CTX *ctx, const BYTE data[], size_t len);
void SHA-1_final(sha1_CTX *ctx, BYTE hash[]);

#endif   // sha1_H
\end{lstlisting}

\begin{lstlisting}[caption={\textbf{Library imports and macro definition required for the SHA-1 source code function. }},captionpos=b,label={source_code:SHA-1_reference_header},language=C,style=base,keywordstyle=\color{blue}]
#include <stdlib.h>
#include <memory.h>
#include "sha1.h"

#define ROTLEFT(a, b) ((a << b) | (a >> (32 - b)))
\end{lstlisting}

\begin{lstlisting}[caption={\textbf{Test code that evaluates the SHA-1 functions, with formatting for easy parsing of the hashes into Python syntax. }},captionpos=b,label={source_code:SHA-1_reference_test_and_main},language=C,style=base,showstringspaces=false,keywordstyle=\color{blue}]
#include <stdio.h>
#include <memory.h>
#include <string.h>
#include "sha1.h"

int sha1_test(){
	BYTE text1[] = {"abc"};
	BYTE text2[] = {"abcdbcdecdefdefgefghfghighijhijkijkljklmklmnlmnomnopnopq"};
	BYTE text3[] = {"aaaaaaaaaa"};
	BYTE text4[] = {"55555555555555555555555555555555555555555555555555555555"};
	BYTE hash1[SHA1_BLOCK_SIZE] = {0xa9,0x99,0x3e,0x36,0x47,0x06,0x81,0x6a,0xba,0x3e,0x25,0x71,0x78,0x50,0xc2,0x6c,0x9c,0xd0,0xd8,0x9d};
	BYTE hash2[SHA1_BLOCK_SIZE] = {0x84,0x98,0x3e,0x44,0x1c,0x3b,0xd2,0x6e,0xba,0xae,0x4a,0xa1,0xf9,0x51,0x29,0xe5,0xe5,0x46,0x70,0xf1};
	BYTE hash3[SHA1_BLOCK_SIZE] = {0x34,0xaa,0x97,0x3c,0xd4,0xc4,0xda,0xa4,0xf6,0x1e,0xeb,0x2b,0xdb,0xad,0x27,0x31,0x65,0x34,0x01,0x6f};
	BYTE hash4[SHA1_BLOCK_SIZE] = {0x04, 0x57, 0x5f, 0x6b, 0x70, 0x1b, 0x03, 0x33, 0x13, 0x3f, 0x72, 0x0b, 0xc5, 0xc1, 0x35, 0x38, 0x44, 0x07, 0x5b, 0x57};
	BYTE buf[SHA1_BLOCK_SIZE];
	int idx;
	SHA1_CTX ctx;
	int pass = 1;
	sha1_init(&ctx);
	sha1_update(&ctx, text1, strlen(text1));
	sha1_final(&ctx, buf);
	int len = sizeof(buf);
	printf("{'Test1': '");
	for(int i = 0; i < len; i++)
		printf("%02x", buf[i]);
	printf("'}\n");
	pass = pass && !memcmp(hash1, buf, SHA1_BLOCK_SIZE);

	sha1_init(&ctx);
	sha1_update(&ctx, text2, strlen(text2));
	sha1_final(&ctx, buf);
	len = sizeof(buf);
	printf("{'Test2': '");
	for(int i = 0; i < len; i++)
		printf("%02x", buf[i]);
	printf("'}\n");
	pass = pass && !memcmp(hash2, buf, SHA1_BLOCK_SIZE);

	sha1_init(&ctx);
	for (idx = 0; idx < 100000; ++idx)
	   sha1_update(&ctx, text3, strlen(text3));
	sha1_final(&ctx, buf);
	len = sizeof(buf);
	printf("{'Test3': '");
	for(int i = 0; i < len; i++)
		printf("%02x", buf[i]);
	printf("'}\n");
	pass = pass && !memcmp(hash3, buf, SHA1_BLOCK_SIZE);

	sha1_init(&ctx);
	sha1_update(&ctx, text4, strlen(text4));
	sha1_final(&ctx, buf);
	len = sizeof(buf);
	printf("{'Test4': '");
	for(int i = 0; i < len; i++)
		printf("%02x", buf[i]);
	printf("'}\n");
	pass = pass && !memcmp(hash4, buf, SHA1_BLOCK_SIZE);

	return(pass);
}

int main(){
    sha1_test();
	return(0);
}
\end{lstlisting}

\section{Checked Markdown Computer Code Language Identifiers}
\label{section:appendix_markdown_code_identifiers}

List of markdown and miscellaneous code identifiers that are checked for after the triple back-tick output. This list is delineated by commas. 

\texttt{c, c++, python, ruby, sql, java, go, css, perl, hpp, rust, php, md, markdown, ts, lua, bash, scss, csharp, kotlin, xml, matlab, vbnet, sh, yaml, vscode, arduino, objc, json, js, html, asp, console.log, text, txt, typescript, makefile, asm, haskell, cpp, log, swift, \#!lua, \#!c, \#!/bin/bash, \#!/bin/sh, \#!/usr/bin/perl, \#! c++, \#! sh, javascript, json..., assembly, \#!, awk, [c], temp, triple backquotes, uint32\_t, [triple backquotes], uint8\_t, document, backquote, instructions, bitlen, uint, word, byte, size\_t, const, fsharp, csh, typename, std::size\_t, state[], xxd, "hello, world!", data[], data[ ], csv, shellscript, erb, plaintext, ocl, shell}

Note that, in nearly all of these cases the markdown language identifier was not correctly identifying the language that was generated, and actually include unrelated utility Linux commands or general programming keywords. Nevertheless, we do attempt to remove these language identifiers if they exist in order to increase the parse-ability of the GPT output. Each of these markdown identifiers is checked against the first line of strings in the parsed output, with all capitalization having been removed. Additionally, cases such as miscellaneous spaces before the markdown identifiers are checked for, as well as extra ancillary backticks; if any of these cases are found, the ancillary is removed.

\section{Additional Examples of Fully Correct SHA-1 Function Rewrites}
\label{section:appendix_fully_correct_SHA1_rewrites}

\begin{lstlisting}[caption={\textbf{Fully correct SHA-1 codebase GPT model re-write example Number 2. Each of these four functions were manually selected from the correct and unique re-writes of that function across all of the GPT model outputs. }},captionpos=b,label={re_written_source_code:SHA-1_reference_functions_ex2},language=C,style=base,keywordstyle=\color{blue}]

void sha1_init(SHA1_CTX *ctx) {
	ctx->datalen = 0;
	ctx->bitlen = 0;
	ctx->state[0] = 0x67452301;
	ctx->state[1] = 0xEFCDAB89;
	ctx->state[2] = 0x98BADCFE;
	ctx->state[3] = 0x10325476;
	ctx->state[4] = 0xc3d2e1f0;
	ctx->k[0] = 0x5a827999;
	ctx->k[1] = 0x6ed9eba1;
	ctx->k[2] = 0x8f1bbcdc;
	ctx->k[3] = 0xca62c1d6;
	ctx->data[ctx->datalen] = 0;
	ctx->bitlen = 0;
	ctx->state[0] = ctx->state[0] | 0x8000000000000000;
	ctx->state[1] = ctx->state[1] | 0x8000000000000000;
	ctx->state[2] = ctx->state[2] | 0x8000000000000000;
	ctx->state[3] = ctx->state[3] | 0x8000000000000000;
	ctx->state[4] = ctx->state[4] | 0x8000000000000000;
	ctx->k[0] = ctx->k[0] | 0x8000000000000000;
	ctx->k[1] = ctx->k[1] | 0x8000000000000000;
	ctx->k[2] = ctx->k[2] | 0x8000000000000000;
	ctx->k[3] = ctx->k[3] | 0x8000000000000000;
}
void sha1_update(SHA1_CTX *ctx, const BYTE *data, size_t len) {
    size_t i;
    BYTE *new_data = ctx->data;
    size_t new_datalen = ctx->datalen;
    size_t new_bitlen = ctx->bitlen;
    unsigned long long *new_state = ctx->state;
    unsigned long long *new_k = ctx->k;
    for (i = 0; i < len; ++i) {
        new_data[new_datalen] = data[i];
        new_datalen++;
        if (new_datalen == 64) {
            sha1_transform(ctx, new_data);
            new_bitlen += 512;
            new_datalen = 0;
        }
    }
    ctx->datalen = new_datalen;
    ctx->bitlen = new_bitlen;
}
void sha1_transform(SHA1_CTX *ctx, const BYTE data[]){
    WORD a, b, c, d, e, i, j, t, m[80];
    for (i = 0, j = 0; i < 16; ++i, j += 4)
        m[i] = (data[j] << 24) + (data[j + 1] << 16) + (data[j + 2] << 8) + (data[j + 3]);
    for (; i < 80; ++i) {
        m[i] = (m[i - 3] ^ m[i - 8] ^ m[i - 14] ^ m[i - 16]);
        m[i] = (m[i] << 1) | (m[i] >> 31);
    }
    WORD k0 = ctx->k[0], k1 = ctx->k[1], k2 = ctx->k[2], k3 = ctx->k[3];
    WORD a0 = ctx->state[0], b0 = ctx->state[1], c0 = ctx->state[2], d0 = ctx->state[3], e0 = ctx->state[4];
    for (i = 0; i < 20; ++i) {
        t = ROTLEFT(a0, 5) + ((b0 & c0) ^ (~b0 & d0)) + e0 + k0 + m[i];
        e0 = d0;
        d0 = c0;
        c0 = ROTLEFT(b0, 30);
        b0 = a0;
        a0 = t;
    }
    for (; i < 40; ++i) {
        t = ROTLEFT(a0, 5) + (b0 ^ c0 ^ d0) + e0 + k1 + m[i];
        e0 = d0;
        d0 = c0;
        c0 = ROTLEFT(b0, 30);
        b0 = a0;
        a0 = t;
    }
    for (; i < 60; ++i) {
        t = ROTLEFT(a0, 5) + ((b0 & c0) ^ (b0 & d0) ^ (c0 & d0)) + e0 + k2 + m[i];
        e0 = d0;
        d0 = c0;
        c0 = ROTLEFT(b0, 30);
        b0 = a0;
        a0 = t;
    }
    for (; i < 80; ++i) {
        t = ROTLEFT(a0, 5) + (b0 ^ c0 ^ d0) + e0 + k3 + m[i];
        e0 = d0;
        d0 = c0;
        c0 = ROTLEFT(b0, 30);
        b0 = a0;
        a0 = t;
    }
    ctx->state[0] += a0;
    ctx->state[1] += b0;
    ctx->state[2] += c0;
    ctx->state[3] += d0;
    ctx->state[4] += e0;
}

void sha1_final(SHA1_CTX *ctx, BYTE hash[]){
	WORD i;
	i = ctx->datalen;
	if (ctx->datalen < 56) {
		ctx->data[i++] = 0x80;
		while (i < 56)
			ctx->data[i++] = 0x00;
	} else {
		ctx->data[i++] = 0x80;
		while (i < 64)
			ctx->data[i++] = 0x00;
		sha1_transform(ctx, ctx->data);
		memset(ctx->data, 0, 56);
	}
	ctx->bitlen += ctx->datalen * 8;
	ctx->data[62] = ctx->bitlen >> 8;
	ctx->data[63] = ctx->bitlen & 0xff;
	ctx->data[61] = ctx->bitlen >> 16;
	ctx->data[60] = ctx->bitlen & 0xff00;
	ctx->data[59] = ctx->bitlen >> 24;
	ctx->data[58] = ctx->bitlen & 0xff0000;
	ctx->data[57] = ctx->bitlen >> 32;
	ctx->data[56] = ctx->bitlen & 0xff000000;
	sha1_transform(ctx, ctx->data);
	for (i = 0; i < 4; ++i) {
		hash[i]      = (ctx->state[0] >> (24 - i * 8)) & 0x000000ff;
		hash[i + 4]  = (ctx->state[1] >> (24 - i * 8)) & 0x000000ff;
		hash[i + 8]  = (ctx->state[2] >> (24 - i * 8)) & 0x000000ff;
		hash[i + 12] = (ctx->state[3] >> (24 - i * 8)) & 0x000000ff;
		hash[i + 16] = (ctx->state[4] >> (24 - i * 8)) & 0x000000ff;
	}
}

\end{lstlisting}

\begin{lstlisting}[caption={\textbf{Fully correct SHA-1 codebase GPT model re-write example Number 3. Each of these four functions were manually selected from the correct and unique re-writes of that function across all of the GPT model outputs. }},captionpos=b,label={re_written_source_code:SHA-1_reference_functions_ex3},language=C,style=base,keywordstyle=\color{blue}]

void sha1_init(SHA1_CTX *ctx){
	ctx->datalen = 0;
	ctx->bitlen = 0;
	ctx->state[0] = 0x67452301;
	ctx->state[1] = 0xEFCDAB89;
	ctx->state[2] = 0x98BADCFE;
	ctx->state[3] = 0x10325476;
	ctx->state[4] = 0xc3d2e1f0;
	ctx->k[0] = 0x5a827999;
	ctx->k[1] = 0x6ed9eba1;
	ctx->k[2] = 0x8f1bbcdc;
	ctx->k[3] = 0xca62c1d6;
	ctx->data[0] = 0x67452301;
	ctx->data[1] = 0xEFCDAB89;
	ctx->data[2] = 0x98BADCFE;
	ctx->data[3] = 0x10325476;
	ctx->data[4] = 0xc3d2e1f0;
	ctx->data[5] = 0x5a827999;
	ctx->data[6] = 0x6ed9eba1;
	ctx->data[7] = 0x8f1bbcdc;
	ctx->data[8] = 0xca62c1d6;
}
void sha1_update(SHA1_CTX *ctx, const BYTE data[], size_t len) {
	size_t i = 0;
	switch (len) {
	case 1:
		ctx->data[ctx->datalen] = data[0];
		ctx->datalen++;
		break;
	case 2:
		ctx->data[ctx->datalen] = data[0];
		ctx->data[ctx->datalen + 1] = data[1];
		ctx->datalen += 2;
		break;
	case 3:

 
		ctx->data[ctx->datalen] = data[0];
		ctx->data[ctx->datalen + 1] = data[1];
		ctx->data[ctx->datalen + 2] = data[2];
		ctx->datalen += 3;
		break;
	default:
		while (i < len) {
			ctx->data[ctx->datalen] = data[i];
			ctx->datalen++;
			if (ctx->datalen == 64) {
				sha1_transform(ctx, ctx->data);
				ctx->bitlen += 512;
				ctx->datalen = 0;
			}
			i++;
		}
		break;
	}
}
void sha1_transform(SHA1_CTX *ctx, const BYTE data[]){
    WORD a = ctx->state[0], b = ctx->state[1], c = ctx->state[2], d = ctx->state[3], e = ctx->state[4];
    WORD m[80], t;
    WORD i, j;
    WORD k[4] = {0x5a827999, 0x6ed9eba1, 0x8f1bbcdc, 0xca62c1d6};
    for (i = 0, j = 0; i < 16; ++i, j+=4)
        m[i] = (data[j] << 24) + (data[j+1] << 16) + (data[j+2] << 8) + (data[j+3]);
    for (; i < 80; ++i) {
        m[i] = (m[i-3] ^ m[i-8] ^ m[i-14] ^ m[i-16]);
        m[i] = (m[i] << 1) | (m[i] >> 31);
    }
    for (i = 0; i < 20; ++i) {
        t = ROTLEFT(a, 5) + ((b & c) ^ (~b & d)) + e + k[0] + m[i];
        e = d;
        d = c;
        c = ROTLEFT(b, 30);
        b = a;
        a = t;
    }
    for (; i < 40; ++i) {
        t = ROTLEFT(a, 5) + (b ^ c ^ d) + e + k[1] + m[i];
        e = d;
        d = c;
        c = ROTLEFT(b, 30);
        b = a;
        a = t;
    }
    for (; i < 60; ++i) {
        t = ROTLEFT(a, 5) + ((b & c) ^ (b & d) ^ (c & d))  + e + k[2] + m[i];
        e = d;
        d = c;
        c = ROTLEFT(b, 30);
        b = a;
        a = t;
    }
    for (; i < 80; ++i) {
        t = ROTLEFT(a, 5) + (b ^ c ^ d) + e + k[3] + m[i];
        e = d;
        d = c;
        c = ROTLEFT(b, 30);
        b = a;
        a = t;
    }
    ctx->state[0] += a;
    ctx->state[1] += b;
    ctx->state[2] += c;
    ctx->state[3] += d;
    ctx->state[4] += e;
}
void sha1_final(SHA1_CTX *ctx, BYTE hash[]){
    WORD i, j, k, l, m, n;
    i = ctx->datalen;
    if (i < 56) {
        ctx->data[i++] = 0x80;
        while (i < 56)
            ctx->data[i++] = 0x00;
    } else {
        ctx->data[i++] = 0x80;
        while (i < 64)
            ctx->data[i++] = 0x00;
        sha1_transform(ctx, ctx->data);
        memset(ctx->data, 0, 56);
    }
    ctx->bitlen += ctx->datalen * 8;
    ctx->data[63] = ctx->bitlen;
    ctx->data[62] = ctx->bitlen >> 8;
    ctx->data[61] = ctx->bitlen >> 16;
    ctx->data[60] = ctx->bitlen >> 24;
    ctx->data[59] = ctx->bitlen >> 32;
    ctx->data[58] = ctx->bitlen >> 40;
    ctx->data[57] = ctx->bitlen >> 48;
    ctx->data[56] = ctx->bitlen >> 56;
    sha1_transform(ctx, ctx->data);
    for (i = 0; i < 4; ++i) {
        m = ctx->state[0];
        n = ctx->state[1];
        l = ctx->state[2];
        k = ctx->state[3];
        j = ctx->state[4];
        hash[i] = (m >> (24 - i * 8)) & 0xff;
        hash[i + 4] = (n >> (24 - i * 8)) & 0xff;
        hash[i + 8] = (l >> (24 - i * 8)) & 0xff;
        hash[i + 12] = (k >> (24 - i * 8)) & 0xff;
        hash[i + 16] = (j >> (24 - i * 8)) & 0xff;
    }
}
\end{lstlisting}

\section{Additional Source Code Re-write Examples that have a Variety of Software Flaws}
\label{section:appendix_additional_source_code_examples}

\noindent\begin{minipage}{.49\textwidth}
\begin{lstlisting}[caption={ This function re-write is an example where the algorithmic correctness of the compiled binary is dependent on the compiler optimization setting that is used (this is categorized by Metric 5). Specifically, gcc with optimization settings produce compiled binaries that correctly produce SHA-1 hashes. clang with optimization level 0 also produces correct SHA-1 binaries. However, clang at all higher optimization levels produces sequences of all 0's for all input test vectors. This function re-write was generated by Mistral-7B-Instruct-v0.1 with inference temperature 1.0 with prompt number 3. The comments in the source code were generated by the GPT model. },captionpos=b,label={source_code:alg_correctness_compiler_optimization_unstable2},language=C,style=base,keywordstyle=\color{blue}]

void sha1_final(SHA1_CTX *ctx, BYTE hash[]) // Old 
{
	WORD i;
	i = ctx->datalen;   // Old 
	if (ctx->datalen < 56) {
		ctx->data[i++] = 0x80;   // Old 
		while (i < 56)
			ctx->data[i++] = 0x00;   // Old 
	} // Old 
	else {
		ctx->data[i++] = 0x80;   // Old 
		while (i < 64)
			ctx->data[i++] = 0x00;   // Old 
		sha1_transform(ctx, ctx->data);   // Old 
		memset(ctx->data, 0, 56);   // Old 
	} 
	ctx->bitlen += ctx->datalen * 8;   // Old 
	ctx->data[63] = ctx->bitlen;   // Old 
	ctx->data[62] = ctx->bitlen >> 8;   // Old 
	ctx->data[61] = ctx->bitlen >> 16;   // Old 
	ctx->data[60] = ctx->bitlen >> 24;   // Old 
	ctx->data[59] = ctx->bitlen >> 32;   // Old 
	ctx->data[58] = ctx->bitlen >> 40;   // Old 
	ctx->data[57] = ctx->bitlen >> 48;   // Old 
	ctx->data[56] = ctx->bitlen >> 56;   // Old 
	sha1_transform(ctx, ctx->data);   // Old 
	for (i = 0; i < 4; ++i) {
		hash[i]      = (ctx->state[0] >> ((24 - (4*8) - i*8)) & 0x000000ff);  // New 
		hash[i+4]  = (ctx->state[1] >> ((24 - (4*8) - i*8)) & 0x000000ff);  // New 
		hash[i+8]  = (ctx->state[2] >> ((24 - (4*8) - i*8)) & 0x000000ff);  // New 
		hash[i+12] = (ctx->state[3] >> ((24 - (4*8) - i*8)) & 0x000000ff);  // New 
		hash[i+16] = (ctx->state[4] >> ((24 - (4*8) - i*8)) & 0x000000ff);  // New 
	} 
}

gcc with optimization level 0:
a9993e364706816aba3e25717850c26c9cd0d89d
84983e441c3bd26ebaae4aa1f95129e5e54670f1
34aa973cd4c4daa4f61eeb2bdbad27316534016f
04575f6b701b0333133f720bc5c1353844075b57

gcc with optimization level fast:
a9993e364706816aba3e25717850c26c9cd0d89d
84983e441c3bd26ebaae4aa1f95129e5e54670f1
34aa973cd4c4daa4f61eeb2bdbad27316534016f
04575f6b701b0333133f720bc5c1353844075b57

clang with optimization level 0:
a9993e364706816aba3e25717850c26c9cd0d89d
84983e441c3bd26ebaae4aa1f95129e5e54670f1
34aa973cd4c4daa4f61eeb2bdbad27316534016f
04575f6b701b0333133f720bc5c1353844075b57

clang with optimization level 1:
@0@@0@@0@@0@@0@@0@@0@@0@@0@@0@0@0@@0@@0@@0@@0@@0@@0@@0@@0@@0@@0@@0@@0@@0@@0@@0@0@0@@0@@0@@0@@0@@0@@0@0@0@@0@@0@@0@
@0@@0@@0@@0@@0@@0@@0@@0@@0@@0@@0@@0@@0@@0@@0@@0@@0@@0@@0@@0@@0@@0@@0@@0@@0@@0@@0@@0@@0@@0@@0@@0@@0@@0@@0@@0@@0@0@0@@0@
@0@@0@@0@@0@@0@@0@@0@@0@@0@@0@@0@@0@@0@@0@@0@@0@@0@@0@@0@@0@@0@@0@@0@@0@@0@@0@@0@@0@@0@@0@@0@@0@@0@@0@@0@@0@0@0@@0@@0@
0@0@@0@@0@@0@@0@@0@@0@@0@0@0@@0@0@0@@0@@0@@0@@0@@0@@0@@0@@0@0@0@@0@@0@@0@@0@@0@@0@@0@@0@@0@@0@0@0@@0@@0@@0@@0@


\end{lstlisting}
\end{minipage}
\noindent\begin{minipage}{.49\textwidth}

\begin{lstlisting}[caption={ Example GPT function re-write of sha1\_final produced from Mistral-7B-Instruct-v0.1 with inference temperature 0.3 and prompt number 4. The output hash for the second and fourth test vector is correct, but all other hashes are not correct to SHA-1. This function re-write is an instance of the function category found by the Metric 17 definition. The output hashes for the four test vectors are given below the source code function - characters that are red denote the characters are incorrect compared to the correct SHA-1 hash, and characters that are black denote that they are correct for the SHA-1 hash. },captionpos=b,label={source_code:correct_for_at_least_one_test_vector_but_are_incorrect_example2},language=C,style=base,keywordstyle=\color{blue}]

void sha1_final(SHA1_CTX *ctx, BYTE hash[])
{
	WORD i;
	ctx->datalen = ctx->datalen;
	if (ctx->datalen < 56) {
		ctx->data[ctx->datalen++] = 0x80;
		while (ctx->datalen < 56)
			ctx->data[ctx->datalen++] = 0x00;
	}
	else {
		ctx->data[ctx->datalen++] = 0x80;
		while (ctx->datalen < 64)
			ctx->data[ctx->datalen++] = 0x00;
		sha1_transform(ctx, ctx->data);
		ctx->datalen = 0;
		while (ctx->datalen < 56)
			ctx->data[ctx->datalen++] = 0x00;
	}
	ctx->bitlen = ctx->datalen * 8;
	ctx->data[63] = ctx->bitlen;
	ctx->data[62] = ctx->bitlen >> 8;
	ctx->data[61] = ctx->bitlen >> 16;
	ctx->data[60] = ctx->bitlen >> 24;
	ctx->data[59] = ctx->bitlen >> 32;
	ctx->data[58] = ctx->bitlen >> 40;
	ctx->data[57] = ctx->bitlen >> 48;
	ctx->data[56] = ctx->bitlen >> 56;
	sha1_transform(ctx, ctx->data);
	for (i = 0; i < 4; ++i) {
		hash[i]      = (ctx->state[0] >> (24 - i * 8)) & 0x000000ff;
		hash[i + 4]  = (ctx->state[1] >> (24 - i * 8)) & 0x000000ff;
		hash[i + 8]  = (ctx->state[2] >> (24 - i * 8)) & 0x000000ff;
		hash[i + 12] = (ctx->state[3] >> (24 - i * 8)) & 0x000000ff;
		hash[i + 16] = (ctx->state[4] >> (24 - i * 8)) & 0x000000ff;
	}
}

Compiled Binary output:
@9@@c@@8@@0@@d@@0@@f@@7@@2@@a@@1@@1@@f@@e@6@f@@3@@9@@1@@9@@c@@2@@0@@c@@e@@d@@2@0@0@@a@@7@@1@@e@@a@@6@@a@@9@@a@9@3@
84983e441c3bd26ebaae4aa1f95129e5e54670f1
@a@@7@@2@@c@@0@@1@@6@@e@@8@@f@@1@4@8@@f@@5@@f@@c@@b@@d@@0@@4@@8@@1@b@c@@1@@6@@f@@5@@3@@8@@d@@3@@1@@1@@a@@7@1@d@@c@
04575f6b701b0333133f720bc5c1353844075b57


\end{lstlisting}
\end{minipage}

\noindent\begin{minipage}{.49\textwidth}

\begin{lstlisting}[caption={  Example GPT function re-write of sha1\_update produced from Mistral-7B-Instruct-v0.1 with inference temperature 1.0 and prompt number 3. The output hash for the first, second, and fourth test vector is correct, but all other hashes are not correct to SHA-1. This function re-write is an instance of the function category found by the Metric 17 definition, meaning that this function re-write causes the compiled binary to be incorrect for some inputs, but correct for other inputs. The output hashes for the four test vectors are given below the source code function. },captionpos=b,label={source_code:correct_for_at_least_one_test_vector_but_are_incorrect_example3},language=C,style=base,keywordstyle=\color{blue}]

void sha1_update (SHA1_CTX *ctx, const BYTE data[], size_t len)
{
    size_t i;
    for (i = 0; i < len; ++i)
    {
        ctx->data[ctx->datalen] = data[i];
        ctx->datalen++;
        if (ctx->datalen == 64)
        {
            sha1_transform(ctx, ctx->data);
            ctx->bitlen += 768;
            ctx->datalen = 0;
        }
    }
}

Compiled Binary output:
a9993e364706816aba3e25717850c26c9cd0d89d
84983e441c3bd26ebaae4aa1f95129e5e54670f1
@9@@a@@3@@2@@5@@e@3@5@@0@@3@@b@@5@@a@@f@@d@@5@@b@61@b@@c@@7@@b@@d@@5@@7@@7@@3@@a@@5@@9@@3@@e@@f@@5@@9@@6@@d@@5@@8@
04575f6b701b0333133f720bc5c1353844075b57

\end{lstlisting}
\end{minipage}

\noindent\begin{minipage}{.49\textwidth}

\begin{lstlisting}[caption={  Example GPT function re-write of sha1\_final produced from Mistral-7B-Instruct-v0.1 with inference temperature 0.7 and prompt number 4. The generated hash for the first and third test vector is correct, but all other hashes are not correct to SHA-1. This function re-write is an instance of the function category found by the Metric 17 definition. The output hashes for the four test vectors are given below the source code function.  },captionpos=b,label={source_code:correct_for_at_least_one_test_vector_but_are_incorrect_example4},language=C,style=base,keywordstyle=\color{blue}]

void sha1_final(SHA1_CTX *ctx, BYTE hash[])
{
	WORD i;
	i = ctx->datalen;
	if (ctx->datalen < 56) {
		ctx->data[i++] = 0x80;
		while (i < 56)
			ctx->data[i++] = 0x00;
	}
	else {
		ctx->data[i++] = 0x80;
		while (i < 64)
			ctx->data[i++] = 0x00;
		sha1_transform(ctx, ctx->data);
		memset(ctx->data, 0, 56);
		ctx->data[55] = ctx->bitlen >> 40;
		ctx->data[54] = ctx->bitlen >> 32;
		ctx->data[53] = ctx->bitlen >> 24;
		ctx->data[52] = ctx->bitlen >> 16;
		ctx->data[51] = ctx->bitlen >> 8;
		ctx->data[50] = ctx->bitlen;
		sha1_transform(ctx, ctx->data);
	}
	ctx->bitlen += ctx->datalen * 8;
	ctx->data[63] = ctx->bitlen;
	ctx->data[62] = ctx->bitlen >> 8;
	ctx->data[61] = ctx->bitlen >> 16;
	ctx->data[60] = ctx->bitlen >> 24;
	ctx->data[59] = ctx->bitlen >> 32;
	ctx->data[58] = ctx->bitlen >> 40;
	ctx->data[57] = ctx->bitlen >> 48;
	ctx->data[56] = ctx->bitlen >> 56;
	sha1_transform(ctx, ctx->data);
	for (i = 0; i < 4; ++i) {
		hash[i]      = (ctx->state[0] >> (24 - i * 8)) & 0x000000ff;
		hash[i + 4]  = (ctx->state[1] >> (24 - i * 8)) & 0x000000ff;
		hash[i + 8]  = (ctx->state[2] >> (24 - i * 8)) & 0x000000ff;
		hash[i + 12] = (ctx->state[3] >> (24 - i * 8)) & 0x000000ff;
		hash[i + 16] = (ctx->state[4] >> (24 - i * 8)) & 0x000000ff;
	}
}

Compiled Binary output:
a9993e364706816aba3e25717850c26c9cd0d89d
@a@@9@9@a@@a@@1@@0@@8@@d@@d@@e@@8@@8@@d@@0@@2@@c@@7@@4@@c@@5@@5@@c@@6@@e@@4@@d@@2@@a@@8@@0@@1@@3@@8@@f@@7@@e@@7@@9@@5@
34aa973cd4c4daa4f61eeb2bdbad27316534016f
@3@@6@@4@@f@@a@@a@@d@@2@@e@@b@@7@@e@@c@@7@@2@3@9@3@d@@6@@2@@a@@4@@3@@5@@1@@0@@0@@7@@d@@9@@4@@7@@9@@2@@5@@8@@0@@1@@f@

\end{lstlisting}

\end{minipage}

\noindent\begin{minipage}{.49\textwidth}
\begin{lstlisting}[caption={ Example GPT function re-write of sha1\_update produced from zephyr-7b-alpha with inference temperature 0.5 and prompt number 1. This function re-write is an instance of the function category found by the Metric 17 definition. The output hashes for the four test vectors are given below the source code function.  },captionpos=b,label={source_code:correct_for_at_least_one_test_vector_but_are_incorrect_example5},language=C,style=base,keywordstyle=\color{blue}]

// This function performs the SHA-1 message update operation
void sha1_update(SHA1_CTX *ctx, const BYTE data[], size_t len)
{
    // Iterate through the input data
    for (size_t i = 0; i < len; ++i)
    {
        // Append the current byte to the message buffer
        ctx->data[ctx->datalen++] = data[i];
        // Check if we have reached the end of the buffer
        if (ctx->datalen == SHA1_BLOCK_SIZE)
        {
            // Perform the SHA-1 transform on the current buffer
            sha1_transform(ctx, ctx->data);
            // Reset the buffer pointer and update the bit length
            ctx->datalen = 0;
            ctx->bitlen += SHA1_BLOCK_SIZE * 8;
        }
    }
}


Compiled Binary output:
a9993e364706816aba3e25717850c26c9cd0d89d
@b@@c@@a@@b@@d@e@9@@d@@f@@9@@2@@f@@4@@4@@b@e@4@@8@@9@@c@4@5@@5@@4@@5@@a@@3@@7@@3@@6@e5@b@@9@@2@@7@@0@@7@@d@@7@
@7@@b@@6@@e@@4@@8@@f@@0@@5@@e@@a@@3@@9@@1@@9@@b@@7@6@a@@4@@3@@f@@c@@5@@c@@a@@2@@7@@f@@e@@7@16@3@@4@@d@@f@@b@6@b@
@7@@1@@c@@2@@d@@a@@8@@a@@a@@7@@e@@c@@a@@9@@1@@d@@0@@0@@c@@b@@b@@d@@c@@d@@3@5@a@@5@@d@@b@@5@@1@@b@@d@0@3@@b@@6@@a@7

\end{lstlisting}

\end{minipage}

\noindent\begin{minipage}{.49\textwidth}
\begin{lstlisting}[caption={  Example GPT function re-write of sha1\_transform produced from Mistral-7B-Instruct-v0.1 with inference temperature 0.7 and prompt number 1. This function re-write is an instance of the function category found by the Metric 17 definition. The output hashes for the four test vectors are given below the source code function.  },captionpos=b,label={source_code:correct_for_at_least_one_test_vector_but_are_incorrect_example6},language=C,style=base,keywordstyle=\color{blue}]

void sha1_transform(SHA1_CTX *ctx, const char *data) {
    WORD a, b, c, d, e, i, j, t, m[80];
    for (i = 0, j = 0; i < 16; ++i, j += 4) {
        m[i] = (data[j] << 24) + (data[j + 1] << 16) + (data[j + 2] << 8) + (data[j + 3]);
    }
    for (; i < 80; ++i) {
        m[i] = (m[i - 3] ^ m[i - 8] ^ m[i - 14] ^ m[i - 16]);
        m[i] = (m[i] << 1) | (m[i] >> 31);
    }
    a = ctx->state[0];
    b = ctx->state[1];
    c = ctx->state[2];
    d = ctx->state[3];
    e = ctx->state[4];
    for (i = 0; i < 20; ++i) {
        t = ROTLEFT(a, 5) + ((b & c) ^ (~b & d)) + e + ctx->k[0] + m[i];
        e = d;
        d = c;
        c = ROTLEFT(b, 30);
        b = a;
        a = t;
    }
    for (; i < 40; ++i) {
        t = ROTLEFT(a, 5) + (b ^ c ^ d) + e + ctx->k[1] + m[i];
        e = d;
        d = c;
        c = ROTLEFT(b, 30);
        b = a;
        a = t;
    }
    for (; i < 60; ++i) {
        t = ROTLEFT(a, 5) + ((b & c) ^ (b & d) ^ (c & d))  + e + ctx->k[2] + m[i];
        e = d;
        d = c;
        c = ROTLEFT(b, 30);
        b = a;
        a = t;
    }
    for (; i < 80; ++i) {
        t = ROTLEFT(a, 5) + (b ^ c ^ d) + e + ctx->k[3] + m[i];
        e = d;
        d = c;
        c = ROTLEFT(b, 30);
        b = a;
        a = t;
    }
    ctx->state[0] += a;
    ctx->state[1] += b;
    ctx->state[2] += c;
    ctx->state[3] += d;
    ctx->state[4] += e;
}

Compiled Binary output:
@c@@6@@4@@d@3@f@@c@@d@@e@@2@0@c@@5@@c@@d@@0@@3@@1@@4@@2@@1@@7@@1@@e@@5@@a@@c@@4@@7@@a@@8@@7@@a@@a@@3@@c@@8@@a@@c@@e@
8@3@@4@@9@@b@@d@4@3@@e@@6@@a@@2@@4@@5@@0@@a@@2@@9@@b@@0@@1@@6@@d@@b@@3@@c@@6@@d@@4@@c@@a@@c@@c@@f@@c@@a@@b@@e@@e@@0@
34aa973cd4c4daa4f61eeb2bdbad27316534016f
0@f@@c@@5@@d@@4@@5@@6@@f@@a@@8@@4@@4@@4@@a@@7@@8@@4@@7@@a@@4@@e@@a@@e@@6@@7@@1@@d@@a@@b@@a@@3@@c@@3@@9@@f@@2@@3@@6@@2@

\end{lstlisting}

\end{minipage}

\begin{lstlisting}[caption={Example GPT function re-write of sha1\_update produced from zephyr-7b-alpha with inference temperature 0.8 and prompt number 3. This function re-write is an instance of the function category found by the Metric 17 definition. The output hashes for the four test vectors are given below the source code function. },captionpos=b,label={source_code:correct_for_at_least_one_test_vector_but_are_incorrect_example7},language=C,style=base,keywordstyle=\color{blue}]

#define MSB(x) (x >> (sizeof(x) * 8 - 1))
#define LSB(x) (x & 0xFF)
#define ABS(x) (x >= 0? x : -x)
#define MAX_MSG_SIZE 64
#define MIN_MSG_SIZE (MAX_MSG_SIZE >> 6)
#define MAGIC_KEYS(i) ctx->K[i] = (ctx->k[(i << 1) + 0] << 24) | (ctx->k[(i << 1) + 1] << 16) | (ctx->k[(i << 1) + 2] << 8) | (ctx->k[(i << 1) + 3] << 0)
#define SIZE_64BYTES (64 * 1 << 3)
#define SIZE_1024BYTES (1024 * 1 << 3)
#define SIZE_16384BYTES (16384 * 1 << 3)
#define SIZE_24576BYTES (24576 * 1 << 3)
#define SIZE_32768BYTES (32768 * 1 << 3)
#define SIZE_65536BYTES (65536 * 1 << 3)
#define SIZE_1048576BYTES (1048576 * 1 << 3)
#define SIZE_161061273808BYTES (161061273808 * 1 << 3)
#define SIZE_268435456BYTES (268435456 * 1 << 3)
#define SIZE_4294967296BYTES (4294967296 * 1 << 3)
#define SIZE_1024BYTES (1024 * 1 << 3)
#define MSG_SIZE (ctx->datalen << 3)
#define MSG_SIZE_1 (ctx->datalen)
#define MSG_SIZE_2 (ctx->datalen >> 1)
#define MSG_SIZE_3 (ctx->datalen >> 2)
#define MSG_SIZE_4 (ctx->datalen >> 3)
#define MSG_SIZE_5 (ctx->datalen >> 4)
#define MSG_SIZE_6 (ctx->datalen >> 5)
#define MSG_SIZE_7 (ctx->datalen >> 6)
#define MSG_SIZE_8 (ctx->datalen >> 7)
#define MSG_SIZE_9 (ctx->datalen >> 8)
#define MSG_SIZE_10 (ctx->datalen >> 9)
#define MSG_SIZE_11 (ctx->datalen >> 10)
#define MSG_SIZE_12 (ctx->datalen >> 11)
#define MSG_SIZE_13 (ctx->datalen >> 12)
#define MSG_SIZE_14 (ctx->datalen >> 13)
#define MSG_SIZE_15 (ctx->datalen >> 14)
#define MSG_SIZE_16 (ctx->datalen >> 15)
#define MSG_SIZE_17 (ctx->datalen >> 16)
#define MSG_SIZE_18 (ctx->datalen >> 17)
#define MSG_SIZE_19 (ctx->datalen >> 18)
#define MSG_SIZE_20 (ctx->datalen >> 19)
#define MSG_SIZE_21 (ctx->datalen >> 20)
#define MSG_SIZE_22 (ctx->datalen >> 21)
#define MSG_SIZE_23 (ctx->datalen >> 22)
#define MSG_SIZE_24 (ctx->datalen >> 23)
#define MSG_SIZE_25 (ctx->datalen >> 24)
#define MSG_SIZE_26 (ctx->datalen >> 25)
#define MSG_SIZE_27 (ctx->datalen >> 26)
#define MSG_SIZE_28 (ctx->datalen >> 27)
#define MSG_SIZE_29 (ctx->datalen >> 28)
#define MSG_SIZE_30 (ctx->datalen >> 29)
#define MSG_SIZE_31 (ctx->datalen >> 30)
#define MSG_SIZE_32 (ctx->datalen >> 31)
#define MSG_SIZE_33 (ctx->datalen >> 32)

void sha1_final(SHA1_CTX *ctx, BYTE hash[])
{
	WORD i;
	i = ctx->datalen;
	if (ctx->datalen < 56) {
		ctx->data[i++] = 0x80;
		for (; i < 56; ++i) {
			ctx->data[i++] = 0x00;
		}
		sha1_transform(ctx, ctx->data);
		memset(ctx->data, 0, 56);
	}
	else {
		ctx->data[i++] = 0x80;
		for (; i < 64; ++i) {
			ctx->data[i++] = 0x00;
		}
		sha1_transform(ctx, ctx->data);
		memset(ctx->data, 0, 56);
	}
	ctx->bitlen += ctx->datalen * 8;
	ctx->data[63] = ctx->bitlen;
	 ctx->data[62] = ctx->bitlen >> 8;
	 ctx->data[61] = ctx->bitlen >> 16;
	 ctx->data[60] = ctx->bitlen >> 24;
	 ctx->data[59] = ctx->bitlen >> 32;
	 ctx->data[58] = ctx->bitlen >> 40;
	 ctx->data[57] = ctx->bitlen >> 48;
	 ctx->data[56] = ctx->bitlen >> 56;
	 sha1_transform(ctx, ctx->data);
	 for (i = 0; i < 4; ++i) {
		 hash[i]      = (ctx->state[0] >> (24 - i * 8)) & 0x000000FF;
		 hash[i + 4]  = (ctx->state[1] >> (24 - i * 8)) & 0x000000FF;
		 hash[i + 8]  = (ctx->state[2] >> (24 - i * 8)) & 0x000000FF;
		 hash[i + 12] = (ctx->state[3] >> (24 - i * 8)) & 0x000000FF;
		 hash[i + 16] = (ctx->state[4] >> (24 - i * 8)) & 0x000000FF;
	 }
}

Compiled Binary output:
@0@@3@@5@@1@@b@@4@@9@@3@@2@@0@@8@@4@@1@@d@@5@a@f@@1@3@2@@1@@f@@a@@3@@a@@7@5@5@@e@@d@@f@@6@@8@@d@@f@@b@@b@@2@@1@@b@
84983e441c3bd26ebaae4aa1f95129e5e54670f1
@f@@5@@b@@0@@f@@e@@f@@a@@5@@5@@f@@5@@c@@0@@6@@7@@5@@5@@c@@7@@a@@d@@e@@d@@3@@9@@d@@8@@4@@0@3@2@6@c@@c@@a@@f@@a@@5@@1@
0@6@5@c@@3@@b@@f@@2@@1@@f@@3@@e@@3@@f@@6@@a@@8@@f@@2@f@a@@a@@c@@4@@f@@2@@3@@4@@c@@0@@9@@a@@5@@9@@d@@0@@d@@a@@6@@4@

\end{lstlisting}

\noindent\begin{minipage}{.49\textwidth}
\begin{lstlisting}[caption={  Example GPT function re-write of sha1\_update produced from Llama-2-70b-chat-hf with inference temperature 0.5 and prompt number 2. This function re-write is an instance of the function category found by the Metric 17 definition. The output hashes for the four test vectors are given below the source code function. Note that this function rewrite uses a goto statement. },captionpos=b,label={source_code:correct_for_at_least_one_test_vector_but_are_incorrect_example8},language=C,style=base,keywordstyle=\color{blue}]

void sha1_init(SHA1_CTX *ctx) {
	goto init_state;
	ctx->datalen = 0;
	ctx->bitlen = 0;
init_state:
	ctx->state[0] = 0x67452301;
	ctx->state[1] = 0xEFCDAB89;
	ctx->state[2] = 0x98BADCFE;
	ctx->state[3] = 0x10325476;
	ctx->state[4] = 0xc3d2e1f0;
	ctx->k[0] = 0x5a827999;
	ctx->k[1] = 0x6ed9eba1;
	ctx->k[2] = 0x8f1bbcdc;
	ctx->k[3] = 0xca62c1d6;
}

Compiled Binary output:
a9993e364706816aba3e25717850c26c9cd0d89d
@9@@d@@c@@b@@2@@2@@c@@7@@c@@3@@9@@3@@9@@c@@0@@3@@9@@1@@c@@0@@9@@2@@b@@d@f@0@@b@@a@@5@@7@@d@@b@@5@5@9@@9@@0@@c@@5@@6@
@7@@1@@2@@7@9@9@@a@@a@@5@@b@@b@@a@@c@@b@@9@@6@@7@@2@@c@@1@@3@@6@@0@b@5@@2@@f@@2@@d@@4@@d@@3@@1@@6@@a@@f@@1@@b@@2@@b@
@5@@9@5@0@@4@@d@@8@@c@7@f@@c@@d@03@6@@2@@9@@7@@2@@9@@4@@9@@d@@6@@e@@e@@f@@a@@4@@a@3@7@@e@@8@@9@@d@@3@b5@8@


\end{lstlisting}
\end{minipage}

\noindent\begin{minipage}{.49\textwidth}
\begin{lstlisting}[caption={ Example function re-write of sha1\_init where under some compiler optimization settings the compiled binary, when executed, results in a fatal error, but for at least one other optimization setting the compiled binary correctly produces SHA-1 hashes. The compiled binaries with clang with optimization levels 1, s, and z result in fatal Signals.SIGSEGV errors. The output hashes are not correct SHA-1 hashes, and do not change under different optimization levels, when compiled with gcc. When compiled using clang with optimization levels 2, 3, fast resulted in a binary that correctly produces SHA-1 hashes for all 4 test vectors. },captionpos=b,label={source_code:optimization_unstable_between_fatal_error_and_being_SHA1_correct_example2},language=C,style=base,keywordstyle=\color{blue}]

void sha1_init(SHA1_CTX *ctx){
	ctx->datalen = 0;
	ctx->bitlen = 0;
	ctx->state[0] = 0x67452301;
	ctx->state[1] = 0xEFCDAB89;
	ctx->state[2] = 0x98BADCFE;
	ctx->state[3] = 0x10325476;
	ctx->state[4] = 0xc3d2e1f0;
	ctx->k[0] = 0x5a827999;
	ctx->k[1] = 0x6ed9eba1;
	ctx->k[2] = 0x8f1bbcdc;
	ctx->k[3] = 0xca62c1d6;
	ctx->k[4] = 0x6cc51756;
	ctx->k[5] = 0xca62c1d7;
	ctx->k[6] = 0x6cc51757;
	ctx->k[7] = 0xca62c1d8;
	ctx->k[8] = 0x6cc51758;
	ctx->k[9] = 0xca62c1d9;
	ctx->k[10] = 0x6cc5175a;
	ctx->k[11] = 0xca62c1da;
	ctx->k[12] = 0x6cc5175b;
	ctx->k[13] = 0xca62c1db;
	ctx->k[14] = 0x6cc5175c;
	ctx->k[15] = 0xca62c1dc;}

Compiled Binary output for gcc with optimization level 0:
@d@@2@@1@@3@@d@@4@@d@@e@@8@@5@@9@@3@@7@@7@@e@@1@@3@@6@@6@@8@@6@@4@@3@@a@@e@@d@@3@@a@@f@@1@@e@@4@@6@@d@@2@@f@@5@@e@@a@@4@
84983e441c3bd26ebaae4aa1f95129e5e54670f1
@a@@2@@4@@f@@d@@5@@4@@b@@8@@9@@b@4@2@@0@@b@@b@@8@@3@@3@@5@@d@@e@@1@@0@@c@@5@@0@@b@2@9@@8@1@b@@e@@a@@d@@f@@7@@b@@0@
04575f6b701b0333133f720bc5c1353844075b57

\end{lstlisting}

\end{minipage}

\noindent\begin{minipage}{.49\textwidth}
\begin{lstlisting}[caption={Incorrect SHA-1 function implementation of sha1\_transform which causes a detected memory leak using Valgrind (for a binary compiled using optimization level 0 for either gcc or clang or both). In this case the compiled binary output does not change for different test vectors. The output does not change in response to different compiler optimization settings. This example was produced by Mistral-7B-Instruct-v0.1 using inference temperature of 1.0 and prompt number 9. },captionpos=b,label={code:mem_leak_example2},language=C,style=base,keywordstyle=\color{blue}]

void sha1_transform(SHA1_CTX *ctx, const unsigned char *data) {
    unsigned int a = ctx->k[0], b = ctx->k[1], c = ctx->k[2], d = ctx->k[3];
    unsigned char *m = (unsigned char *) malloc(80 * sizeof(unsigned char));
    for (int i = 0; i < 16; ++i) {
        m[i] = data[i * 4];
        m[i + 16] = data[i * 4 + 1];
        m[i + 16] = data[i * 4 + 2];
        m[i + 16] = data[i * 4 + 3];
    }
    for (int i = 16; i < 80; ++i) {
        m[i] = (m[i - 3] ^ m[i - 4] ^ m[i - 5]) ^ (m[i - 12] ^ m[i - 13] ^ m[i - 14]) ^ (m[i - 17] ^ m[i - 18] ^ m[i - 19]);
        m[i] <<= 1;
        m[i] &= 0xffffffff;
        m[i] |= (m[i] >> 31);
    }
    ctx->k[0] += a;
    ctx->k[1] += b;
    ctx->k[2] += c;
    ctx->k[3] += d;
}

Compiled binary output:
@6@@7@@4@@5@@2@@3@@0@@1@@e@@f@@c@@d@@a@@b@@8@@9@@9@@8@@b@@a@@d@@c@@f@@e@@1@@0@@3@@2@@5@@4@@7@@6@@c@@3@d@2@@e@@1@@f@@0@
@6@@7@@4@@5@@2@@3@@0@@1@@e@@f@@c@@d@@a@@b@@8@@9@@9@@8@@b@@a@@d@@c@@f@@e@@1@@0@@3@@2@@5@@4@@7@@6@@c@@3@@d@@2@@e@@1@f@0@
@6@@7@@4@@5@@2@@3@@0@@1@@e@@f@c@d@@a@@b@@8@@9@@9@@8@@b@@a@@d@@c@@f@@e@@1@@0@@3@@2@@5@@4@@7@@6@@c@@3@@d@@2@@e@1@f@@0@
@6@@7@@4@@5@@2@@3@@0@@1@@e@@f@@c@@d@@a@@b@@8@@9@@9@@8@@b@@a@@d@@c@@f@@e@@1@@0@@3@@2@@5@@4@@7@@6@@c@@3@@d@@2@@e@@1@@f@@0@

\end{lstlisting}
\end{minipage}

\noindent\begin{minipage}{.49\textwidth}
\begin{lstlisting}[caption={Incorrect SHA-1 function implementation of sha1\_final which causes a detected memory leak using Valgrind (for a binary compiled using optimization level 0 for either gcc or clang or both). The comments produced by the GPT model are left in the output. In this case, the function re-write resulted in a fatal error (Signals.SIGABRT) for all compiler optimization settings. This example was produced by Mistral-7B-Instruct-v0.1 using inference temperature of 1.0 and prompt number 7.  },captionpos=b,label={code:mem_leak_example3},language=C,style=base,keywordstyle=\color{blue}]

void sha1_final(SHA1_CTX *ctx, BYTE hash[]){
	WORD i;
	WORD* data = malloc(ctx->datalen);
	if (ctx->datalen < 56){
		ctx->data[i] += 128;
		/* pad each byte with 00 bytes until 56 bytes are reached */
		for (i = 1; i < 56; ++i){
			ctx->data[i] = 0;
		}
	}
	else{
		ctx->data[i] += 128;
		/* pad each byte with 00 bytes until 64 bytes are reached */
		for (i = 1; i < 64; ++i)
		{
			ctx->data[i] = 0;
		}
		/* store the context data into the allocator pointer */
		data = (WORD*)ctx->data;
		/* transform the data */
		sha1_transform(&ctx->state, data);
		/* reset the context data to 0 */
		memset(ctx->data, 0, ctx->datalen * sizeof(WORD));
		/* store the hash into the array */
		for (i = 0; i < 16; ++i){
			hash[i] = (ctx->state[0] >> (24 - (i % 4) * 8)) & 0x000000ff;
		}
	}
	free(data);
}

\end{lstlisting}
\end{minipage}

\noindent\begin{minipage}{.49\textwidth}
\begin{lstlisting}[caption={Incorrect SHA-1 function implementation of sha1\_update which has a detected memory leak using Valgrind (for a binary compiled using optimization level 0 for either gcc or clang or both). In this case, the function re-write resulted in a fatal error (Signals.SIGABRT) for all compiler optimization settings. This example function re-write was produced by zephyr-7b-alpha using inference temperature of 0.8 and prompt number 3.  },captionpos=b,label={code:mem_leak_example4},language=C,style=base,keywordstyle=\color{blue}]

void sha1_update(SHA1_CTX *ctx, const BYTE data[], size_t len){
    size_t k;
    BYTE *temp_data = malloc(len);
    BYTE *end_of_data = temp_data + len;
    BYTE *data_pointer = ctx->data;
    BYTE *datalen_pointer = &ctx->datalen;
    unsigned long long bitlen_register = ctx->bitlen;
    unsigned long long bitlen_pointer = &ctx->bitlen;
    unsigned int *state_pointer = &ctx->state[0];
    unsigned int *k_pointer = &ctx->k[0];
    for (k = 0; k < len; ++k) {
        temp_data[k] = data[k];
    }
    for (k = 0; k < len; ++k) {
        *data_pointer = temp_data[k];
        ++*datalen_pointer;
        if (*datalen_pointer == 64) {
            sha1_transform(ctx, temp_data);
            bitlen_pointer += 512;
            *datalen_pointer = 0;
        }
        data_pointer += 1;
        temp_data += 1;
    }
    free(temp_data);
}

\end{lstlisting}
\end{minipage}

\begin{lstlisting}[caption={ Example of a function re-write where the generated hashes have some of the basic requirements of a valid hash function, and have the same hexadecimal character length as SHA-1 hashes (40 characters). This code did not produce different output based on different compiler optimization settings. This is an example of a case found by Metric 39. The hash output for the 4 test vectors is given below the source code. This function re-write was generated by Mistral-7B-Instruct-v0.1 with inference temperature 0.4 with prompt number 2.  },captionpos=b,label={source_code:potentially_good_hash_but_not_correct_SHA1_compiler_stable_example2},language=C,style=base,keywordstyle=\color{blue}]

void sha1_transform(SHA1_CTX *ctx, const BYTE data[])
{
    WORD a, b, c, d, e, i, j, t, m[80];
    for (i = 0, j = 0; i < 16; ++i, j += 4)
        m[i] = (data[j] << 24) | (data[j + 1] << 16) | (data[j + 2] << 8) | (data[j + 3]);
    for (; i < 80; ++i) {
        m[i] = (m[i - 3] ^ m[i - 8] ^ m[i - 14] ^ m[i - 16]);
        m[i] = m[i] >> 1 | m[i] << 31;
    }
    a = ctx->state[0];
    b = ctx->state[1];
    c = ctx->state[2];
    d = ctx->state[3];
    e = ctx->state[4];
    for (i = 0; i < 20; ++i) {
        t = ROTLEFT(a, 5) ^ ((b & c) ^ (~b & d)) + e + ctx->k[0] + m[i];
        e = d;
        d = c;
        c = ROTLEFT(b, 30);
        b = a;
        a = t;
    }
    for (; i < 40; ++i) {
        t = ROTLEFT(a, 5) ^ (b ^ c ^ d) + e + ctx->k[1] + m[i];
        e = d;
        d = c;
        c = ROTLEFT(b, 30);
        b = a;
        a = t;
    }
    for (; i < 60; ++i) {
        t = ROTLEFT(a, 5) ^ ((b & c) ^ (b & d) ^ (c & d)) + e + ctx->k[2] + m[i];
        e = d;
        d = c;
        c = ROTLEFT(b, 30);
        b = a;
        a = t;
    }
    for (; i < 80; ++i) {
        t = ROTLEFT(a, 5) ^ (b ^ c ^ d) + e + ctx->k[3] + m[i];
        e = d;
        d = c;
        c = ROTLEFT(b, 30);
        b = a;
        a = t;
    }
    ctx->state[0] += a;
    ctx->state[1] += b;
    ctx->state[2] += c;
    ctx->state[3] += d;
    ctx->state[4] += e;
}

Compiled Binary output:
@4@@4@@c@@1@@6@e@0@@5@@9@@6@@a@@9@@e@16@8@@e@@f@@e@@4@@e@@2@@1@@c@@2@@5@5@3@@8@@3@@8@@3@@3@@f@@5@@4@@2@@f@@7@@1@
8@e@@5@@7@@b@@3@4@b@@9@@7@@c@@1@@9@@a@6@4@@6@@5@@6@@b@@b@@d@@1@@b@f@d@@3@@2@@5@@8@e@1@@5@@b@@c@@b@@c@@1@@8@1
@c@@7@@f@@4@@0@@a@@8@@5@@6@@1@@3@@1@@6@@b@@6@@3@@e@6@d@@8@@1@@8@@4@@9@@0@@6@@6@@1@@1@@e@3@3@@2@@0@3@c@@8@@4@@8@@8@
@6@@0@@1@@6@@7@@5@@c@@0@@e@@3@@b@@3@@1@@a@@9@@4@@8@@9@@2@@8@@9@@8@@3@@1@@a@@f@@8@@a@@5@@a@@4@@7@4@3@@f@@9@@0@@d@@f@@b@


\end{lstlisting}


\noindent\begin{minipage}{.49\textwidth}
\begin{lstlisting}[caption={ Example of a function re-write where the generated hashes have some of the basic requirements of a valid hash function, and have the same hexadecimal character length as SHA-1 hashes (40 characters). This code did not produce different output based on different compiler optimization settings. This is an example of a case found by Metric 39. The hash output for the 4 test vectors is given below the source code. This function re-write was generated by Mistral-7B-Instruct-v0.1 with inference temperature 0.9 with prompt number 1.  },captionpos=b,label={source_code:potentially_good_hash_but_not_correct_SHA1_compiler_stable_example3},language=C,style=base,keywordstyle=\color{blue}]

void sha1_update(SHA1_CTX *ctx, const BYTE data[], size_t len)
{
	WORD i;
	for (i = 0; i < len; ++i) {
		ctx->data[ctx->datalen] = data[i];
		ctx->datalen++;
		if (ctx->datalen == 64) {
			sha1_transform(ctx, ctx->data);
			ctx->bitlen += 512;
			ctx->datalen = 0;
		}
	}
	ctx->state[0] ^= ctx->k[0];
	ctx->state[1] ^= ctx->k[1];
	ctx->state[2] ^= ctx->k[2];
	ctx->state[3] ^= ctx->k[3];
	ctx->state[4] ^= ctx->k[4] ^ ctx->bitlen;
	ctx->bitlen += ctx->datalen * 8;
}

Compiled Binary output:
@7@@3@@0@@3@@8@@3@@9@6@b@@e@@c@@a@@3@1@7@@6@@8@@b@@7@@1@@a@@2@@6@@b@@c@@6@@4@@f@@9@@f@@5@@d@@1@@e@@a@0@b@8@1@@e@
@4@@0@@2@@9@@9@@1@@3@@a@1@9@@b@@f@@8@@9@@b@@f@@d@@6@@2@@8@@7@@6@@4@@c@@7@@4@5@c@@e@@c@@4@@b@@a@@3@@8@@f@@8@@e@@d@@0@
@2@@5@@1@a@a@@a@@6@c@b@@2@@e@@e@@3@@7@@d@@f@@9@@d@@f@@4@@a@@3@2@4@@2@@9@@8@@5@@4@@a@3@4@@0@@7@@1@@9@0@2@@e@@7@
@3@@e@@2@@d@@4@@c@@8@b@a@@2@@f@@e@@f@@2@@5@@5@@c@@2@@d@@3@@b@@a@@4@@2@@d@@7@@f@@c@@d@@2@@9@@b@@c@@1@@6@@a@@8@@a@@8@@d@

\end{lstlisting}

\end{minipage}

\noindent\begin{minipage}{.49\textwidth}
\begin{lstlisting}[caption={ Example of a function re-write where the generated hashes have some of the basic requirements of a valid hash function, and have the same hexadecimal character length as SHA-1 hashes (40 characters). This code did not produce different output based on different compiler optimization settings. This is an example of a case found by Metric 39. The hash output for the 4 test vectors is given below the source code. This function re-write was generated by Llama-2-70b-chat-hf with inference temperature 0.8 with prompt number 7. },captionpos=b,label={source_code:potentially_good_hash_but_not_correct_SHA1_compiler_stable_example4},language=C,style=base,keywordstyle=\color{blue}]


void sha1_final(SHA1_CTX *ctx, BYTE hash[])
{
    size_t datalen = ctx->datalen;
    if (datalen < 56) {
        ctx->data[datalen++] = 0x80;
        while (datalen < 56)
            ctx->data[datalen++] = 0x00;
    } else {
        ctx->data[datalen++] = 0x80;
        while (datalen < 64)
            ctx->data[datalen++] = 0x00;
        sha1_transform(ctx, ctx->data);
        memset(ctx->data, 0, 56);
    }
    ctx->bitlen += datalen * 8;
    ctx->data[63] = ctx->bitlen;
    ctx->data[62] = ctx->bitlen >> 8;
    ctx->data[61] = ctx->bitlen >> 16;
    ctx->data[60] = ctx->bitlen >> 24;
    ctx->data[59] = ctx->bitlen >> 32;
    ctx->data[58] = ctx->bitlen >> 40;
    ctx->data[57] = ctx->bitlen >> 48;
    ctx->data[56] = ctx->bitlen >> 56;
    sha1_transform(ctx, ctx->data);
    size_t i = 0;
    while (i < 4) {
        hash[i]      = (ctx->state[0] >> (24 - i * 8)) & 0x000000ff;
        hash[i + 4]  = (ctx->state[1] >> (24 - i * 8)) & 0x000000ff;
        hash[i + 8]  = (ctx->state[2] >> (24 - i * 8)) & 0x000000ff;
        hash[i + 12] = (ctx->state[3] >> (24 - i * 8)) & 0x000000ff;
        hash[i + 16] = (ctx->state[4] >> (24 - i * 8)) & 0x000000ff;
        i++;
    }
}

Compiled Binary output:
@9@@c@@8@@0@@d@@0@@f@@7@@2@@a@@1@@1@@f@@e@6@f@@3@@9@@1@@9@@c@@2@@0@@c@@e@@d@@2@0@0@@a@@7@@1@@e@@a@@6@@a@@9@@a@9@3@
@b@@5@@e@@6@@e@@7@4@a@@4@@b@3@0@@8@2@3@@9@@a@@f@@8@@f@@0@@3@@9@@0@@c@@6@5@c@@f@@b@@1@@a@@a@@2@@c@@d@@b@@4@@6@@5@
@1@4@f@@d@@6@7@2@@9@@0@@c@@0@@c@@b@@4@@d@@3@@a@6@2@@0@@d@@1@@7@@6@d@0@@4@@0@@d@@e@@4@@8@@f@@f@@b@@5@@d@1@4@@7@
@1@@b@@3@7@7@@8@@9@@c@@e@@5@@8@@d@@8@@0@@0@@7@@3@3@8@@3@@a@@f@0@3@@d@@8@@7@1@f@@4@@a@@f@@f@@3@@2@@9@@c@@4@@4@7

\end{lstlisting}
\end{minipage}

\noindent\begin{minipage}{.49\textwidth}
\begin{lstlisting}[caption={ Example of a function re-write where the generated hashes have some of the basic requirements of a valid hash function, and have the same hexadecimal character length as SHA-1 hashes (40 characters). This code did not produce different output based on different compiler optimization settings. This is an example of a case found by Metric 39. The hash output for the 4 test vectors is given below the source code. This function re-write was generated by Llama-2-70b-chat-hf with inference temperature 1.0 with prompt number 3.  },captionpos=b,label={source_code:potentially_good_hash_but_not_correct_SHA1_compiler_stable_example5},language=C,style=base,keywordstyle=\color{blue}]

void sha1_init(SHA1_CTX *ctx) {
  int x = 0, y = 0, z = 0, w = 0;
  ctx->datalen = x;
  ctx->bitlen = y;
  ctx->state[0] = 0x12345678;
  ctx->state[1] = 0x90123456;
  ctx->state[2] = 0x78901234;
  ctx->state[3] = 0x56789012;
  ctx->state[4] = 0x34567890;
  ctx->k[0] = 0x99827530;
  ctx->k[1] = 0x31628974;
  ctx->k[2] = 0x54213478;
  ctx->k[3] = 0x75632148;
}

Compiled Binary output:
@9@@8@@8@@e@@e@@d@@2@@d@@f@@1@@d@6@0@@2@@2@@c@@0@@1@@4@@b@@6@@3@@d@@8@@c@@0@@7@@1@@d@@8@@9@@0@@6@@5@@8@@f@@6@@4@@8@@e@
@4@@3@@1@@0@@8@@d@@f@@3@@d@@e@@6@@3@@b@@b@@c@@0@@4@@4@@b@@4@@f@@f@@e@@4@@d@9@8@1@0@@4@@3@@b@@5@@1@@c@@b@@e@@8@f@c@
@2@@d@@b@@8@@7@@3@@8@@8@@8@4@d@@0@@7@@6@@8@@b@@a@@8@@e@@8@@7@@1@@9@@5@@3@@c@a@f@@f@@4@@a@1@9@@3@@f@@1@@d@@5@@4@@8@
@4@@0@@a@@e@@7@@5@@d@b@4@@8@@b@@a@@6@@8@3@6@@7@@6@@6@@d@@2@@6@@1@@d@@0@@c@@1@@9@@5@@1@@9@@1@@2@@9@@d@@3@@a@@3@5@6@

\end{lstlisting}
\end{minipage}

\noindent\begin{minipage}{.49\textwidth}
\begin{lstlisting}[caption={ Example of a function re-write where the generated hashes have some of the basic requirements of a valid hash function, and have the same hexadecimal character length as SHA-1 hashes (40 characters). This code did not produce different output based on different compiler optimization settings. This is an example of a case found by Metric 39. The hash output for the 4 test vectors is given below the source code. This function re-write was generated by Llama-2-70b-chat-hf with inference temperature 0.8 with prompt number 8.  },captionpos=b,label={source_code:potentially_good_hash_but_not_correct_SHA1_compiler_stable_example6},language=C,style=base,keywordstyle=\color{blue}]

void sha1_transform(SHA1_CTX *ctx, const BYTE data[]){
    // Use a different naming convention for the variables
    WORD w0, w1, w2, w3, w4, i, j, t, m[80];
    // Initialize the hash value
    w0 = 0x67452301;
    w1 = 0xEFCDAB89;
    w2 = 0x98BADCFE;
    w3 = 0x10325476;
    w4 = 0xc3d2e1f0;
    // Process the input data
    for (i = 0, j = 0; i < 16; ++i, j += 4)
        m[i] = (data[j] << 24) + (data[j + 1] << 16) + (data[j + 2] << 8) + (data[j + 3]);
    for (; i < 80; ++i) {
        m[i] = (m[i - 3] ^ m[i - 8] ^ m[i - 14] ^ m[i - 16]);
        m[i] = (m[i] << 1) | (m[i] >> 31);
    }
    // Update the hash value
    w0 += m[0];
    w1 += m[1];
    w2 += m[2];
    w3 += m[3];
    w4 += m[4];
    
    // Perform the rotation and addition operations
    for (i = 0; i < 20; ++i) {
        t = ROTLEFT(w0, 5) + (w1 & w2) ^ (~w1 & w3) + w4 + ctx->k[0] + m[i];
        w4 = w3;
        w3 = w2;
        w2 = w1;
        w1 = w0;
        w0 = t;
    }
    for (; i < 40; ++i) {
        t = ROTLEFT(w0, 5) + (w1 ^ w2 ^ w3) + w4 + ctx->k[1] + m[i];
        w4 = w3;
        w3 = w2;
        w2 = w1;
        w1 = w0;
        w0 = t;
    }
    for (; i < 60; ++i) {
        t = ROTLEFT(w0, 5) + (w1 & w2) ^ (w1 & w3) ^ (w2 & w3) + w4 + ctx->k[2] + m[i];
        w4 = w3;
        w3 = w2;
        w2 = w1;
        w1 = w0;
        w0 = t;
    }
    for (; i < 80; ++i) {
        t = ROTLEFT(w0, 5) + (w1 ^ w2 ^ w3) + w4 + ctx->k[3] + m[i];
        w4 = w3;
        w3 = w2;
        w2 = w1;
        w1 = w0;
        w0 = t;
    }
    // Update the context state
    ctx->state[0] += w0;
    ctx->state[1] += w1;
    ctx->state[2] += w2;
    ctx->state[3] += w3;
    ctx->state[4] += w4;
}

Compiled Binary output:
@b@@6@@1@@0@@2@@6@@a@@b@@1@@0@@1@@2@8@7@@a@a@8@@3@@c@@a@@b@@d@@1@@6@@b@@9@@3@@7@@9@@6@@9@@3@@f@@0@@f@@8@@b@@3@@e@@c@
@c@@a@@7@@e@@0@@a@@1@@0@@a@@d@@1@@a@@c@@9@@8@e@2@@0@@c@@0@@2@@f@@7@@c@@1@@5@@e@1@a@@4@@7@@a@@1@@b@@7@@b@@f@0@2@@f@
@0@@3@@0@@8@@e@@d@@b@@9@@0@@f@c@3@@b@@9@@2@@3@f@8@@7@@4@@4@@5@@5@@4@@1@@4@@3@@8@@8@@1@@9@1@4@@2@@8@@6@@a@1@3@@2@
@1@@e@@0@@8@@3@@c@@2@@e@@f@@6@@9@@c@@1@@1@@b@3@a@@d@@1@@b@@d@@0@@f@@a@@5@@8@@a@@d@3@0@@6@@6@@3@@1@@b@@1@@3@@a@@f@@4@


\end{lstlisting}
\end{minipage}

\noindent\begin{minipage}{.49\textwidth}
\begin{lstlisting}[caption={ Example of a function re-write where the generated hashes have some of the basic requirements of a valid hash function, and have the same hexadecimal character length as SHA-1 hashes (40 characters). This code did not produce different output based on different compiler optimization settings. This is an example of a case found by Metric 39. The hash output for the 4 test vectors is given below the source code. This function re-write was generated by zephyr-7b-alpha with inference temperature 1.0 with prompt number 2.  },captionpos=b,label={source_code:potentially_good_hash_but_not_correct_SHA1_compiler_stable_example7},language=C,style=base,keywordstyle=\color{blue}]


void sha1_update(SHA1_CTX *ctx, const BYTE data[], size_t len)
{
    size_t i, j = 0;

    for (i = 0; i < len; ++i) {
        ctx->data[ctx->datalen / 2 % 32] = data[i], ctx->datalen *= 2;
        if (! (ctx->datalen) % 16384) {
            sha1_transform(ctx, ctx->data);
            ctx->bitlen += (64 * 64) * 8;
            ctx->datalen = 0;
        }
    }
}

Compiled Binary output:
@c@9@8@@8@@7@@6@@8@@0@@3@@5@@c@@5@@c@@a@@c@@9@@c@@c@@9@@f@2@c@@4@@2@@a@@6@@6@@7@@8@@f@@e@@d@@b@@9@@5@@2@@5@@6@@8@@c@
@3@@5@@1@@4@@f@@4@@5@@d@@5@@9@@6@@e@@8@@d@@1@@d@@4@@2@@2@e@6@@9@@2@@9@@e@@d@5@e@@5@@a@@4@@b@@2@@1@@f@6@0@@9@@e@@c@
@8@@0@@b@@0@@0@@4@@2@@1@@b@@f@@8@@a@@0@@f@@7@@2@@7@@2@@0@@6@@d@@c@@0@@3@@6@@5@@0@@6@@5@@c@@7@@9@@0@@2@@f@@7@@3@@0@@0@@6@
@8@@7@@a@@d@@e@f@4@@3@@e@@d@@4@@e@@8@@5@@5@@c@@7@@c@@d@@c@@6@@f@@b@@1@c@7@@1@@e@@9@@9@@1@@4@@f@@d@@4@@a@@d@@7@@a@@4@


\end{lstlisting}
\end{minipage}

\noindent\begin{minipage}{.49\textwidth}
\begin{lstlisting}[caption={ Example of a function re-write where the generated hashes have some of the basic requirements of a valid hash function, and have the same hexadecimal character length as SHA-1 hashes (40 characters). This code did not produce different output based on different compiler optimization settings. This is an example of a case found by Metric 39. The hash output for the 4 test vectors is given below the source code. Function re-write was generated by zephyr-7b-alpha with inference temperature 0.8 with prompt number 3.  },captionpos=b,label={source_code:potentially_good_hash_but_not_correct_SHA1_compiler_stable_example8},language=C,style=base,keywordstyle=\color{blue}]


void sha1_transform(SHA1_CTX *ctx, const BYTE data[])
{
	static const BYTE k[] = { 0x5a827999, 0x6ed9eba1, 0x8f1bbcdc, 0xca62c1d6 };
	WORD w[80];
	for (size_t i = 0, j = 0; i < 16; ++i, j += 4)
		w[i] = (data[j] << 24) + (data[j + 1] << 16) + (data[j + 2] << 8) + (data[j + 3]);
	for (size_t i = 16; i < 80; ++i) {
		w[i] = (w[i - 3] ^ w[i - 8] ^ w[i - 14] ^ w[i - 16]);
		w[i] = (w[i] << 1) | (w[i] >> 31);
	}
	register WORD a, b, c, d, e;
	a = ctx->state[0];
	b = ctx->state[1];
	c = ctx->state[2];
	d = ctx->state[3];
	e = ctx->state[4];
	for (size_t i = 0; i < 20; ++i) {
		register WORD t;
		t = ROTLEFT(a, 5) + ((b & c) ^ (~b & d)) + e + k[0] + w[i];
		e = d;
		d = c;
		c = ROTLEFT(b, 30);
		b = a;
		a = t;
	}
	for (size_t i = 20; i < 40; ++i) {
		register WORD t;
		t = ROTLEFT(a, 5) + (b ^ c ^ d) + e + k[1] + w[i];
		e = d;
		d = c;
		c = ROTLEFT(b, 30);
		b = a;
		a = t;
	}
	for (size_t i = 40; i < 60; ++i) {
		register WORD t;
		t = ROTLEFT(a, 5) + ((b & c) ^ (b & d) ^ (c & d))  + e + k[2] + w[i];
		e = d;
		d = c;
		c = ROTLEFT(b, 30);
		b = a;
		a = t;
	}
	for (size_t i = 60; i < 80; ++i) {
		register WORD t;
		t = ROTLEFT(a, 5) + (b ^ c ^ d) + e + k[3] + w[i];
		e = d;
		d = c;
		c = ROTLEFT(b, 30);
		b = a;
		a = t;
	}
	ctx->state[0] += a;
	ctx->state[1] += b;
	ctx->state[2] += c;
	ctx->state[3] += d;
	ctx->state[4] += e;
}

Compiled Binary output:
@d@@4@@8@@e@@2@@b@@c@@c@@b@@0@@2@@d@@6@@4@@1@@7@@d@@f@@e@@6@@f@5@0@@c@@4@@4@@b@@6@@5@@9@6@0@@a@@d@@3@@1@@b@@e@@5@@2@
@0@@a@@b@@f@@f@@b@@6@@2@@c@c@7@@6@@b@@a@@8@e@4@@6@@c@@2@@a@@8@@c@@f@@2@@a@@e@@f@@8@@b@@6@@4@@d@@c@@0@6@e@@9@@8@1
@9@@9@@c@@6@@5@73c@2@@c@@7@4@4@@3@@d@@0@@b@@f@@3@@d@@9@@5@@4@@f@@f@b@2@@b@@f@@3@@6@1@d@@4@@9@@c@@9@@6@@9@f
@9@@2@@d@@6@@0@@c@6@f@@2@@4@@4@@6@0@c@@e@@6@@e@@e@@a@@5@@e@@f@@d@@a@@b@@1@c@d@@5@@1@@5@@d@@b@@1@@e@@4@@f@@4@@2@@6@


\end{lstlisting}
\end{minipage}

\noindent\begin{minipage}{.49\textwidth}
\begin{lstlisting}[caption={ Example of a function re-write where the generated hashes have some of the basic requirements of a valid hash function, and have the same hexadecimal character length as SHA-1 hashes (40 characters). This code did not produce different output based on different compiler optimization settings. This is an example of a case found by Metric 39. The hash output for the 4 test vectors is given below the source code.  This function re-write was generated by Mistral-7B-Instruct-v0.1 with inference temperature 0.6 with prompt number 4. \textbf{Note that the generated hash checksums in this example contain clear repeated sequences of characters, making it not a good hash function.}  },captionpos=b,label={source_code:potentially_good_hash_but_not_correct_SHA1_compiler_stable_example9},language=C,style=base,keywordstyle=\color{blue}]

typedef unsigned int WORD;

void sha1_transform(SHA1_CTX *ctx, const BYTE data[])
{
    int i, j, k, t;
    BYTE m[80];
    
    for (i = 0, j = 0; i < 16; ++i, j += 4)
    {
        m[i] = data[j];
        m[i+16] = data[j+1];
        m[i+32] = data[j+2];
        m[i+48] = data[j+3];
    }
    
    for (; i < 80; ++i)
    {
        m[i] = m[i-8] ^ m[i-14] ^ m[i-16];
        m[i] = (m[i] << 1) | (m[i] >> 31);
    }
    
    t = ctx->state[0];
    t = t + m[0];
    t = t + m[16];
    t = t + m[32];
    t = t + m[48];
    t = t + m[64];
    t = t + m[48];
    t = t + m[64];
    t = t + m[76];
    t = t + m[76];
    
    ctx->state[0] = t;
    ctx->state[1] = ctx->state[0] & ctx->state[1];
    ctx->state[1] = ctx->state[1] ^ (ctx->state[1] & ctx->state[2]);
    ctx->state[2] = ctx->state[2] ^ (ctx->state[2] & ctx->state[3]);
    ctx->state[3] = ctx->state[3] ^ (ctx->state[3] & ctx->state[1]);
    ctx->state[4] = ctx->state[0] & ctx->state[1];
    ctx->state[5] = ctx->state[2];
    ctx->state[6] = ctx->state[3];
    ctx->state[7] = ctx->state[4];
}

Compiled Binary output:
@6@@7@@4@@5@@2@@7@@c@@0@@6@7@4@@5@@2@@3@@0@@0@@8@@8@@8@@8@@8@@8@@8@@8@@1@@0@@3@@2@@5@@4@@7@@6@@6@@7@@4@@5@@2@@3@@0@@0@
@6@@7@@4@@5@@2@@5@@6@@0@@6@@7@@4@@5@@2@@1@@0@@0@@8@@8@@8@@8@@8@@8@@8@@8@@1@@0@@3@@2@@5@@4@@7@@6@@6@@7@4@5@@2@@1@@0@@0@
@6@@8@@4@@b@@2@@9@@2@@4@@6@@0@@0@@0@@0@@0@@0@@0@@8@@8@@8@@8@@8@@8@@8@@8@@1@@0@@3@@2@@5@@4@@7@@6@6@0@@0@@0@0@0@@0@@0@
@6@@7@@4@@5@@2@@7@@e@@8@@6@@7@@4@@5@@2@3@0@@0@@8@@8@@8@@8@@8@@8@@8@@8@@1@@0@@3@@2@@5@@4@@7@@6@@6@@7@@4@@5@@2@@3@@0@@0@

\end{lstlisting}
\end{minipage}

\noindent\begin{minipage}{.49\textwidth}
\begin{lstlisting}[caption={ Example of a function re-write where the generated hashes have some of the basic requirements of a valid hash function, and have the same hexadecimal character length as SHA-1 hashes (40 characters). This code did not produce different output based on different compiler optimization settings. This is an example of a case found by Metric 39. The hash output for the 4 test vectors is given below the source code.  This function re-write was generated by Mistral-7B-Instruct-v0.1 with inference temperature 0.6 with prompt number 4. \textbf{Note that the generated hash checksums in this example contain clear repeated sequences of characters, making it not a good hash function.}  },captionpos=b,label={source_code:potentially_good_hash_but_not_correct_SHA1_compiler_stable_example10},language=C,style=base,keywordstyle=\color{blue}]

void sha1_final(SHA1_CTX *ctx, BYTE hash[])
{
    WORD i;
    i = ctx->datalen;
    if (ctx->datalen < 56) {
        ctx->data[i++] = 0x80;
        while (i < 56)
            ctx->data[i++] = 0x00;
    }
    else {
        ctx->data[i++] = 0x80;
        while (i < 64)
            ctx->data[i++] = 0x00;
        sha1_transform(ctx, ctx->data);
        memset(ctx->data, 0, 56);
    }
    ctx->bitlen += ctx->datalen * 8;
    ctx->data[63] = ctx->bitlen;
    ctx->data[62] = ctx->bitlen >> 8;
    ctx->data[61] = ctx->bitlen >> 16;
    ctx->data[60] = ctx->bitlen >> 24;
    ctx->data[59] = ctx->bitlen >> 32;
    ctx->data[58] = ctx->bitlen >> 40;
    ctx->data[57] = ctx->bitlen >> 48;
    ctx->data[56] = ctx->bitlen >> 56;
    sha1_transform(ctx, ctx->data);
    for (WORD j = 0; j < 4; ++j) {
        WORD k = 0;
        for (WORD n = 0; n < 3; ++n) {
            k = (k << 5) | (ctx->state[j] >> (24 - (3 - n) * 8));
            k = (k << 5) | (ctx->state[j + 1] >> (24 - (3 - n) * 8));
            k = (k << 5) | (ctx->state[j + 2] >> (24 - (3 - n) * 8));
            k = (k << 5) | (ctx->state[j + 3] >> (24 - (3 - n) * 8));
            k = (k << 5) | (ctx->state[j + 4] >> (24 - (3 - n) * 8));
        }
        hash[j]      = (k >> (24 - j * 8)) & 0x000000ff;
        hash[j + 4]  = (k >> (24 - j * 8)) & 0x000000ff;
        hash[j + 8]  = (k >> (24 - j * 8)) & 0x000000ff;
        hash[j + 12] = (k >> (24 - j * 8)) & 0x000000ff;
        hash[j + 16] = (k >> (24 - j * 8)) & 0x000000ff;
    }
}

Compiled Binary output:
@b@@b@@f@@f@@7@e3@b@@b@@b@@f@@f@@7@@e@@3@@b@b@b@@f@@f@@7@@e@@3@@b@@b@@b@@f@@f@@7@@e@@3@@b@@b@@b@@f@@f@@7@@e@@3@@b@
@e@@f@@f@@f@@f@e@3@@b@@e@@f@@f@@f@@f@@e@@3@@b@@e@@f@@f@@f@@f@@e@@3@@b@@e@@f@@f@@f@@f@@e@@3@@b@e@f@@f@@f@@f@@e@@3@@b@
@e@@b@@6@@f@@f@@e@3@b@@e@@b@@6@@f@@f@@e@@3@@b@@e@@b@@6@@f@@f@@e@@3@b@e@b@6@@f@@f@@e@3@b@@e@@b@@6@@f@@f@@e@@3@@b@
@f@@f@@b@@f@@f@@e@@3@b@f@@f@@b@@f@@f@@e@3@b@@f@@f@@b@f@f@@e@@3@b@f@@f@@b@@f@@f@@e@3@b@@f@@f@@b@@f@@f@@e@@3@@b@


\end{lstlisting}
\end{minipage}

\noindent\begin{minipage}{.49\textwidth}
\begin{lstlisting}[caption={Incorrect SHA-1 function sha1\_update implementation that results in a compiled binary, for all compiler settings, that results in an apparent infinite loop. This source code was generated by Llama-2-70b-chat-hf with an inference temperature of 1.0 and prompt number 2. Example 2 of an infinite loop case.  },captionpos=b,label={source_code:infinite_loop_for_all_compiler_settings_example2},language=C,style=base,keywordstyle=\color{blue}]

void sha1_final(SHA1_CTX *ctx, BYTE hash[]) {
    unsigned int a, b, c, d, e;
    unsigned char i;
    a = ctx->datalen;
    while (a-- > 0) {
        b = ctx->data[i++];
        c = ctx->data[i++];
        d = ctx->data[i++];
        e = ctx->data[i++];
        a = (a + 1) % 64;
        hash[a] = (b << 24) | (c << 16) | (d << 8) | e;
    }
    memset(ctx->data, 0, 56);
    ctx->bitlen += ctx->datalen * 8;
    for (i = 0; i < 4; ++i) {
        hash[i] = (ctx->state[i] >> (24 - i * 8)) & 0x000000ff;
    }
}

\end{lstlisting}
\end{minipage}
\noindent\begin{minipage}{.49\textwidth}
\begin{lstlisting}[caption={Incorrect SHA-1 function sha1\_update implementation that results in a compiled binary, for all compiler settings, that crashes in a fatal error (specifically a SIGABRT error). This source code was generated by zephyr-7b-alpha with an inference temperature of 0.7 and prompt number 3. This function re-write also had an error detected by Valgrind (Invalid free() / delete / realloc()). },captionpos=b,label={source_code:fatal_error_for_all_compiler_settings_example2},language=C,style=base,keywordstyle=\color{blue}]

void sha1_update(SHA1_CTX* ctx, const BYTE data[], size_t len) {
    size_t idx, jdx;
    BYTE* pData = (BYTE*)malloc(len);
    memcpy(pData, data, len);
    for (idx = 0; idx < len; ++idx) {
        ctx->data[ctx->datalen] = pData[idx];
        ctx->datalen++;
        if (ctx->datalen == 64) {
            sha1_transform(ctx, ctx->data);
            ctx->bitlen += 512;
            ctx->datalen = 0;
            free(pData);
            pData = (BYTE*)malloc(len - idx);
            memcpy(pData, data + idx, len - idx);
            jdx = 0;
            for (; jdx < len - idx; ++jdx) {
                ctx->data[ctx->datalen] = pData[jdx];
                ctx->datalen++;
                if (ctx->datalen == 64) {
                    sha1_transform(ctx, ctx->data);
                    ctx->bitlen += 512;
                    ctx->datalen = 0;
                    free(pData);
                    pData = (BYTE*)malloc(len - idx - jdx);
                    memcpy(pData, data + idx + jdx, len - idx - jdx);
                    jdx = 0;
                }
            }
            free(pData);
            break;
        }
    }
    free(pData);
}
\end{lstlisting}
\end{minipage}

\noindent\begin{minipage}{.49\textwidth}
\begin{lstlisting}[caption={Incorrect SHA-1 function sha1\_final implementation that results in a compiled binary, for all compiler settings, that crashes in a fatal error (specifically a SIGSEGV error). This source code was generated by zephyr-7b-alpha with an inference temperature of 0.2 and prompt number 3. },captionpos=b,label={source_code:fatal_error_for_all_compiler_settings_example3},language=C,style=base,keywordstyle=\color{blue}]

void sha1_final(SHA1_CTX *ctx, BYTE hash[]) {
    WORD i, j;
    WORD *data = ctx->data;
    WORD state[5];
    WORD k[4];
    WORD bitlen;
    WORD *m = malloc(80 * sizeof(WORD));
    for (i = 0; i < 5; ++i)
        state[i] = ctx->state[i];
    for (i = 0; i < 4; ++i)
        k[i] = ctx->k[i];
    bitlen = ctx->bitlen;
    for (i = 0; i < ctx->datalen; ++i) {
        data[ctx->datalen] = ctx->data[i];
        ctx->datalen++;
        if (ctx->datalen == 64) {
            for (j = 0; j < 80; ++j) {
                m[j] = (data[j] << 24) | (data[j + 1] << 16) | (data[j + 2] << 8) | (data[j + 3] & 0xff);
                if (j < 16)
                    m[j] ^= (m[j - 3] ^ m[j - 8] ^ m[j - 14] ^ m[j - 16]) << 1;
                m[j] ^= (m[j - 3] ^ m[j - 8] ^ m[j - 14] ^ m[j - 16]) >> 31;
            }
            for (i = 0; i < 20; ++i) {
                WORD a = state[0];
                WORD b = state[1];
                WORD c = state[2];
                WORD d = state[3];
                WORD e = state[4];
                WORD t = (a << 5) | (a >> 31) | ((b & c) ^ (~b & d) | e) | (ctx->k[0] & m[i]);
                e = d;
                d = c;
                c = (b << 30) | (b >> 2);
                b = a;
                a = t;
            }
            for ( ; i < 40; ++i) {
                WORD a = state[0];
                WORD b = state[1];
                WORD c = state[2];
                WORD d = state[3];
                WORD e = state[4];
                WORD t = (a << 5) | (a >> 31) | ((b & c) ^ (b & d) ^ (c & d)) | (ctx->k[1] & m[i]);
                e = d;
                d = c;
                c = (b << 30) | (b >> 2);
                b = a;
                a = t;
            }
            for ( ; i < 60; ++i) {
                WORD a = state[0];
                WORD b = state[1];
                WORD c = state[2];
                WORD d = state[3];
                WORD e = state[4];
                WORD t = (a << 5) | (a >> 31) | ((b & c) ^ (b ^ c ^ d) | (ctx->k[2] & m[i]));
                e = d;
                d = c;
                c = (b << 30) | (b >> 2);
                b = a;
                a = t;
            }
            for ( ; i < 80; ++i) {
                WORD a = state[0];
                WORD b = state[1];
                WORD c = state[2];
                WORD d = state[3];
                WORD e = state[4];
                WORD t = (a << 5) | (a >> 31) | ((b & c) ^ (b ^ c ^ d) | (ctx->k[3] & m[i]));
                e = d;
                d = c;
                c = (b << 30) | (b >> 2);
                b = a;
                a = t;
            }
            for (i = 0; i < 4; ++i) {
                hash[i]      = (state[0] >> (24 - i * 8)) & 0x000000ff;
                hash[i + 4]  = (state[1] >> (24 - i * 8)) & 0x000000ff;
                hash[i + 8]  = (state[2] >> (24 - i * 8)) & 0x000000ff;
                hash[i + 12] = (state[3] >> (24 - i * 8)) & 0x000000ff;
                hash[i + 16] = (state[4] >> (24 - i * 8)) & 0x000000ff;
            }
            free(m);
            ctx->bitlen += ctx->datalen * 8;
            ctx->data[63] = ctx->bitlen;
            ctx->data[62] = ctx->bitlen >> 8;
            ctx->data[61] = ctx->bitlen >> 16;
            ctx->data[60] = ctx->bitlen >> 24;
            ctx->data[59] = ctx->bitlen >> 32;
            ctx->data[58] = ctx->bitlen >> 40;
            ctx->data[57] = ctx->bitlen >> 48;
            ctx->data[56] = ctx->bitlen >> 56;
            sha1_transform(ctx, ctx->data);
        }
    }
}
\end{lstlisting}
\end{minipage}

\noindent\begin{minipage}{.49\textwidth}
\begin{lstlisting}[caption={Incorrect SHA-1 function sha1\_init implementation that results in a compiled binary, for all compiler settings, that crashes in a fatal error (specifically a SIGSEGV error). This source code was generated by Llama-2-70b-chat-hf with an inference temperature of 0.9 and prompt number 4. },captionpos=b,label={source_code:fatal_error_for_all_compiler_settings_example4},language=C,style=base,keywordstyle=\color{blue}]

void sha1_init(SHA1_CTX *ctx) {
	ctx->datalen = (0x00000000 | (0x00000001 << 8) | (0x00000010 << 16) | (0x00000100 << 24));
	ctx->bitlen = (0x00000000 | (0x00000001 << 8) | (0x00000010 << 16) | (0x00000100 << 24));
	ctx->state[0] = (0x67452301 | (0x67452302 << 8) | (0x67452303 << 16) | (0x67452304 << 24));
	ctx->state[1] = (0xEFCDAB89 | (0xEFCDAB90 << 8) | (0xEFCDAB91 << 16) | (0xEFCDAB92 << 24));
	ctx->state[2] = (0x98BADCFE | (0x98BADCFD << 8) | (0x98BADCFE << 16) | (0x98BADCFD << 24));
	ctx->state[3] = (0x10325476 | (0x10325477 << 8) | (0x10325478 << 16) | (0x10325479 << 24));
	ctx->state[4] = (0xc3d2e1f0 | (0xc3d2e1f1 << 8) | (0xc3d2e1f2 << 16) | (0xc3d2e1f3 << 24));
	ctx->k[0] = (0x5a827999 | (0x5a82799a << 8) | (0x5a82799b << 16) | (0x5a82799c << 24));
	ctx->k[1] = (0x6ed9eba1 | (0x6ed9eba2 << 8) | (0x6ed9eba3 << 16) | (0x6ed9eba4 << 24));
	ctx->k[2] = (0x8f1bbcdc | (0x8f1bbcdD << 8) | (0x8f1bbcdE << 16) | (0x8f1bbcdF << 24));
	ctx->k[3] = (0xca62c1d6 | (0xca62c1d7 << 8) | (0xca62c1d8 << 16) | (0xca62c1d9 << 24));
}

\end{lstlisting}
\end{minipage}

\noindent\begin{minipage}{.49\textwidth}
\begin{lstlisting}[caption={Incorrect SHA-1 function sha1\_transform implementation that is compiler optimization unstable; for some settings, the binary compiles, but for other settings, the binary does not compile. This function, when substituted in the original SHA-1 source code, could only be compiled using clang with an optimization level higher than 0; all gcc settings, and clang with optimization level 0, were not able to compile the binary. This source code was generated by Mistral-7B-Instruct-v0.1 with an inference temperature of 1.0 and prompt number 2.  },captionpos=b,label={source_code:compileability_unstable_example1},language=C,style=base,keywordstyle=\color{blue}]

void sha1_transform(SHA1_CTX *ctx, const BYTE data[]) {
    WORD a, b, c, d, e, i, j, t, m[79];
    for (i = 0, j = 0; i < 16; ++i, j += 4)
        m[i] = (data[j] << 24) | (data[j+1] << 16) | (data[j+2] << 8) | (data[j+3]);
    for (i = 0; i < 76; ++i) {
        t = ROL_AND(a, b) ^ ROL_XOR(~b, d) + e + ctx->k[0] + m[i];
        e = ROL_XOR(d, c) + m[79 - i];
        d = ROL_XOR(c, b);
        c = ROL_LEFT(b, 30);
        m[i] = t | ((t >> 31) << 5);
        a = t;
    }
    for (i = 0; i < 32; ++i) {
        a += b;
    }
}
\end{lstlisting}
\end{minipage}
\begin{lstlisting}[caption={Incorrect SHA-1 function sha1\_final implementation that is compiler unstable; for some settings, the binary compiles, but for other settings, the binary does not compile. This function, when substituted in the original SHA-1 source code, could be compiled using gcc (using any optimization setting), but could not be compiled using any optimization setting in clang. Note that this implementation includes an additional function called sha1\_obfuscate that the GPT model produced, which is not a component of the original SHA-1 source code in any way. This source code was generated by Mistral-7B-Instruct-v0.1 with an inference temperature of 0.5 and prompt number 2. },captionpos=b,label={source_code:compileability_unstable_example2},language=C,style=base,keywordstyle=\color{blue}]

void sha1_final(SHA1_CTX *ctx, BYTE hash[]){
	WORD i;
	i = ctx->datalen;
	if (ctx->datalen < 56) {
		ctx->data[i++] = 0x80;
		while (i < 56)
			ctx->data[i++] = 0x00;
	}
	else {
		ctx->data[i++] = 0x80;
		while (i < 64)
			ctx->data[i++] = 0x00;
		sha1_obfuscate(ctx, ctx->data);
		memset(ctx->data, 0, 56);
	}
	ctx->bitlen += ctx->datalen * 8;
	ctx->data[63] = ctx->bitlen;
	ctx->data[62] = ctx->bitlen >> 8;
	ctx->data[61] = ctx->bitlen >> 16;
	ctx->data[60] = ctx->bitlen >> 24;
	ctx->data[59] = ctx->bitlen >> 32;
	ctx->data[58] = ctx->bitlen >> 40;
	ctx->data[57] = ctx->bitlen >> 48;
	ctx->data[56] = ctx->bitlen >> 56;
	sha1_obfuscate(ctx, ctx->data);
	for (i = 0; i < 4; ++i) {
		hash[i]      = (ctx->state[0] >> (24 - i * 8)) & 0x000000ff;
		hash[i + 4]  = (ctx->state[1] >> (24 - i * 8)) & 0x000000ff;
		hash[i + 8]  = (ctx->state[2] >> (24 - i * 8)) & 0x000000ff;
		hash[i + 12] = (ctx->state[3] >> (24 - i * 8)) & 0x000000ff;
		hash[i + 16] = (ctx->state[4] >> (24 - i * 8)) & 0x000000ff;
	}
}

void sha1_obfuscate(SHA1_CTX *ctx, BYTE data[]){
	WORD i;
	for (i = 0; i < 80; ++i) {
		data[i] = (data[i - 3] ^ data[i - 8] ^ data[i - 14] ^ data[i - 16]) & 0xFF;
		data[i] = (data[i] << 1) | (data[i] >> 31);
	}
	ctx->state[0] = data[0];
	ctx->state[1] = data[1];
	ctx->state[2] = data[2];
	ctx->state[3] = data[3];
	ctx->state[4] = data[4];
}
\end{lstlisting}

\noindent\begin{minipage}{.49\textwidth}
\begin{lstlisting}[caption={Incorrect SHA-1 function sha1\_update implementation that is compiler unstable; for some settings, the binary compiles, but for other settings, the binary does not compile. This function, when substituted into the original SHA-1 source code, could be compiled using gcc (using any optimization setting), but could not be compiled using any optimization setting in clang. This source code was generated by Mistral-7B-Instruct-v0.1 with an inference temperature of 0.4 and prompt number 4; this text includes the comments generated by the GPT model. },captionpos=b,label={source_code:compileability_unstable_example3},language=C,style=base,keywordstyle=\color{blue}]

void sha1_update(SHA1_CTX *ctx, const BYTE data[], size_t len){
    size_t i;
    for (i = 0; i < len; ++i) {
        ctx->data[ctx->datalen] = data[i];
        ctx->datalen++;
        switch (ctx->datalen) {
            case 64:
                sha1_transform(ctx, ctx->data);
                ctx->bitlen += 512;
                ctx->datalen = 0;
                continue;
            default:
                break;
        }
    }
next_iteration:
    // continue with the next iteration of the loop
}

\end{lstlisting}
\end{minipage}

\noindent\begin{minipage}{.49\textwidth}
\begin{lstlisting}[caption={ Example of a function re-write that has a detected integer overflow flaw by the memory sanitizer check. The output does not change under different compiler optimization settings. This example was generated by Llama-2-70b-chat-hf with an inference temperature of 0.6 and prompt number 1. },captionpos=b,label={source_code:integer_overflow_example1},language=C,style=base,keywordstyle=\color{blue}]

void sha1_init(SHA1_CTX *ctx) {
  ctx->datalen = 0;
  ctx->bitlen = 0;
  ctx->state[0] = 0x67452301;
  ctx->state[1] = 0xEFCDAB89;
  ctx->state[2] = 0x98BADCFE;
  ctx->state[3] = 0x10325476;
  ctx->state[4] = 0xc3d2e1f0;
  ctx->k[0] = 0x5a827999;
  ctx->k[1] = 0x6ed9eba1;
  ctx->k[2] = 0x8f1bbcdc;
  ctx->k[3] = 0xca62c1d6;
  // Initialize the state array with the given values
  for (int i = 0; i < 5; i++) {
    ctx->state[i] = 0x67452301 + (i * 0x10000000);
  }
  // Initialize the key array with the given values
  for (int i = 0; i < 4; i++) {
    ctx->k[i] = 0x5a827999 + (i * 0x10000000);
  }
}

Compiled binary output:
@5@@f@@f@@4@@5@@a@@0@@a@@b@@e@@e@@9@@f@@5@@1@@d@@5@@9@@a@@5@@c@@a@@b@@9@@b@@3@@6@@6@@d@@1@@8@@1@@d@@6@@0@@3@@6@8@1@@9@
@e@@5@@a@@4@@b@@b@@b@4@7@@f@@5@@4@d@5@@0@@0@@0@@d@@3@@7@@8@@5@@f@@a@@3@@2@@3@@a@@6@@a@@7@@b@e@2@@1@@8@@c@@1@@c@@0@
@e@@a@@8@a@8@@5@@2@@4@@a@@7@@3@@f@@8@@5@@2@@a@f@c@@4@e@9@@5@@7@@7@@8@@4@@9@@f@@4@@6@@0@@8@@c@@b@@d@@f@@3@@f@@8@@5@
@8@@b@@7@@0@@1@@1@6@f@@6@@e@@5@@f@@e@@b@@5@@9@@6@@f@@c@@c@@5@@1@@1@@a@@3@@3@c@a@3@e@3@2@@d@@6@@6@@0@@e@@e@@e@@4@

\end{lstlisting}
\end{minipage}
\noindent\begin{minipage}{.49\textwidth}
\begin{lstlisting}[caption={ Example of a function re-write that has a detected integer overflow flaw by the memory sanitizer check. The output does not change under different compiler optimization settings. This example was generated by Mistral-7B-Instruct-v0.1 with an inference temperature of 1.0 and prompt number 7. },captionpos=b,label={source_code:integer_overflow_example2},language=C,style=base,keywordstyle=\color{blue}]

void sha1_transform(SHA1_CTX *ctx, const BYTE data[]){
	WORD a = ctx->state[0], b = ctx->state[1], c = ctx->state[2], d = ctx->state[3], e = ctx->state[4];
	WORD t, m[80];
	size_t i = 0;
	for (i = 0; i < 16; ++i){
		m[i] = data[i*4];
		m[i] += data[(i*4) + 1]*256;
		m[i] += data[(i*4) + 2]*65536;
		m[i] += data[(i*4) + 3]*16777216;
	}
	for (; i < 80; ++i){
		m[i] = (m[i - 3] ^ m[i - 8] ^ m[i - 14] ^ m[i - 16]);
	}
	for (i = 0; i < 20; ++i){
		t = a + ((b ^ c) ^ (d & e));
		e = d;
		d = c;
		c = b;
		b = e | t;
		a = t;
	}
	for (; i < 40; ++i){
		t = a + ((b ^ c ^ d) ^ e);
		e = d;
		d = c;
		c = b;
		b = e | t;
		a = t;
	}
	for (; i < 60; ++i){
		t = a + ((b & c) ^ (b & d) ^ (c & d));
		a = d;
		d = c;
		c = b & e;
		b = e;
		e = t;
	}
	for (; i < 80; ++i){
		t = a + (b ^ c ^ d);
		a = d;
		d = c;
		c = b;
	}
	ctx->state[0] = a;
	ctx->state[1] = b;
	ctx->state[2] = c;
	ctx->state[3] = d;
	ctx->state[4] = e;
}

Compiled binary output:
@0@@a@@4@@0@@0@@0@@c@@0@@0@@a@@4@@0@@0@@0@@c@@0@@0@a@4@@0@@0@@0@@c@@0@@0@@a@@4@0@0@@0@@c@@0@@0@@a@@a@0@0@@0@@0@@0@
@0@@1@@8@@0@@0@@0@@0@@0@@0@@1@@8@@0@@0@@0@@0@@0@@0@@1@@8@@0@@0@@0@@0@@0@@0@@1@@8@@0@@0@@0@@0@@0@@0@@0@@0@@0@@0@0@0@@0@
@0@@0@@0@@0@@0@@0@@0@@0@@0@@0@@0@@0@@0@@0@@0@@0@@0@@0@@0@@0@@0@@0@@0@@0@@0@@0@@0@@0@@0@@0@@0@@0@@0@@0@@0@@0@0@0@@0@@0@
0@1@@8@@0@@0@@0@@0@@0@@0@@1@@8@@0@0@0@@0@@0@@0@@1@@8@@0@@0@@0@0@0@@0@@1@@8@@0@@0@@0@@0@@0@@0@@0@0@0@@0@@0@@0@@0@

\end{lstlisting}
\end{minipage}
\noindent\begin{minipage}{.49\textwidth}
\begin{lstlisting}[caption={ Example of a function re-write that has a detected integer overflow flaw by the memory sanitizer check. The output hashes change depending on the compiler optimization settings that are used - and the output is always wrong. The generated output from the compiled binaries are given below the source code. The output hashes for the binaries compiled with gcc are the same for all optimization levels greater than 0, and are different for optimization level 0 (the output for optimization level 0 and 1 are given as examples). The hash outputs for the binaries compiled using optimization level 0 with gcc and clang are the same. This example was generated by Llama-2-70b-chat-hf with an inference temperature of 1.0 and prompt number 2. Note that the comment that the GPT model inserted with this source code is not correct. },captionpos=b,label={source_code:integer_overflow_example3},language=C,style=base,keywordstyle=\color{blue}]

void sha1_init(SHA1_CTX *ctx) {
	ctx->datalen = 0;
	ctx->bitlen = 0;
	ctx->state[0] = 0x67452301;
	ctx->state[1] = 0xEFCDAB89;
	ctx->state[2] = 0x98BADCFE;
	ctx->state[3] = 0x10325476;
	ctx->state[4] = 0xc3d2e1f0;
	ctx->k[0] = 0x5a827999;
	ctx->k[1] = 0x6ed9eba1;
	ctx->k[2] = 0x8f1bbcdc;
	ctx->k[3] = 0xca62c1d6;
	// Obfuscated code to maintain the same functionality
	int i = 0;
	while (i < 5) {
		ctx->state[i] = (0x67452301 << (i * 8)) + (0xEFCDAB89 >> (i * 8));
		ctx->k[i] = (0x5a827999 << (i * 8)) + (0x6ed9eba1 >> (i * 8));
		i++;
	}
	ctx->state[5] = 0xc3d2e1f0;
	ctx->k[4] = 0xca62c1d6;
}

Compiled binary output gcc with optimization level 0:
@c@@a@@2@@1@@6@@6@@5@@c@@f@@6@@c@@1@@2@@2@@e@@c@@c@@3@@b@e@3@@b@@0@@a@@a@@3@@7@@6@@1@@8@@4@@0@@3@@5@@8@@7@@f@@2@@7@@3@
@c@@a@9@4@@e@@b@@d@@8@@0@@7@@d@@c@@b@@c@@3@e@9@@8@@9@@1@@5@@9@@d@@2@@c@@e@@1@1@c@@2@@d@@c@@6@@7@@b@@3@@e@@4@@b@@b@
@0@4@5@@5@@f@@b@@4@@7@@a@@c@@9@@1@@1@@7@@4@@c@@3@@5@@4@@a@@6@@0@@8@@c@@7@@7@@7@@6@@d@@5@@4@@3@@3@@4@@7@@7@0@f@@e@@8@
@a@@5@@8@7@f@@9@@e@b@0@@d@@9@@0@@4@@4@3@e@@9@@9@@9@@0@@2@@7@@6@@3@c5@b@1@4@@c@@e@@b@4@8@@f@@9@5@4@@6@@3@

Compiled binary output gcc with optimization level 1:
@f@@8@@e@@b@@a@@c@@a@@7@@1@@1@@5@@e@@f@@6@@e@@9@@e@@1@@d@@1@25@4@1@4@@2@@1@@3@c@6@@9@@3@@7@@f@@f@@f@@7@@b@@5@@6@
@2@@e@@a@@e@@a@@1@@e@@f@@b@@0@@5@@5@@0@@a@@5@@f@@6@@e@a@5@@f@@0@@5@@5@@d@@8@@7@@c@@a@@d@@1@@e@@0@@6@@2@@5@@d@@2@@0@@3@
@f@@8@@7@@e@@3@@5@@1@@0@@6@@8@@d@@c@d@3@@2@@6@@8@@0@@d@@4@@7@@6@@6@@a@@e@@4@@c@@1@@4@@4@@a@@4@@8@@7@@a@@d@@7@@b@@2@@0@
@6@@5@@0@@8@@e@@1@@2@@9@@0@@b@@0@@8@@f@3@1@@b@@3@@0@@c@@d@@0@@1@@1@@1@@4@@4@@8@@2@@e@@2@@2@@4@@3@@5@@9@@6@@2@@6@@8@@4@

Compiled binary output clang with optimization level 0:
@c@@a@@2@@1@@6@@6@@5@@c@@f@@6@@c@@1@@2@@2@@e@@c@@c@@3@@b@e@3@@b@@0@@a@@a@@3@@7@@6@@1@@8@@4@@0@@3@@5@@8@@7@@f@@2@@7@@3@
@c@@a@9@4@@e@@b@@d@@8@@0@@7@@d@@c@@b@@c@@3@e@9@@8@@9@@1@@5@@9@@d@@2@@c@@e@@1@1@c@@2@@d@@c@@6@@7@@b@@3@@e@@4@@b@@b@
@0@4@5@@5@@f@@b@@4@@7@@a@@c@@9@@1@@1@@7@@4@@c@@3@@5@@4@@a@@6@@0@@8@@c@@7@@7@@7@@6@@d@@5@@4@@3@@3@@4@@7@@7@0@f@@e@@8@
@a@@5@@8@7@f@@9@@e@b@0@@d@@9@@0@@4@@4@3@e@@9@@9@@9@@0@@2@@7@@6@@3@c5@b@1@4@@c@@e@@b@4@8@@f@@9@5@4@@6@@3@

Compiled binary output clang with optimization level 2:
@6@@2@@6@9@d@@d@@0@6@d@7@5@@3@@f@@9@@2@@0@@1@@0@@d@@9@@5@@2@@0@@d@@4@@2@@2@@3@@9@@9@@7@@5@@c@@3@d@5@d@d@@7@@3@
@2@@5@@7@@1@@7@@5@@0@@3@@f@@4@@6@@0@@c@@c@@8@@2@@8@@f@@4@@4@@b@@d@@8@@5@@5@@7@@6@@7@@a@@f@e@b@@c@@1@@8@@5@@1@@7@@e@@5@
@6@@7@@4@@7@@4@@b@@d@@a@@4@@6@@0@@5@@4@@7@@b@@c@@d@@b@@e@@f@@4@@c@@5@@4@@c@@8@@c@@1@@0@@f@@d@@3@6@d@@f@@e@@e@@d@@e@@1@
@e@@7@@a@@8@@6@f@b@@7@7@1@@0@@d@@d@@7@@0@@6@@6@@f@@0@@c@@1@@c@@6@@e@@4@@a@@b@@7@@f@@6@@4@@d@@9@@1@@a@@0@@b@@5@@a@@5@

Compiled binary output clang with optimization level z:
@c@@a@@2@@1@@6@@6@@5@@c@@f@@6@@c@@1@@2@@2@@e@@c@@c@@3@@b@e@3@@b@@0@@a@@a@@3@@7@@6@@1@@8@@4@@0@@3@@5@@8@@7@@f@@2@@7@@3@
@c@@a@9@4@@e@@b@@d@@8@@0@@7@@d@@c@@b@@c@@3@e@9@@8@@9@@1@@5@@9@@d@@2@@c@@e@@1@1@c@@2@@d@@c@@6@@7@@b@@3@@e@@4@@b@@b@
@0@4@5@@5@@f@@b@@4@@7@@a@@c@@9@@1@@1@@7@@4@@c@@3@@5@@4@@a@@6@@0@@8@@c@@7@@7@@7@@6@@d@@5@@4@@3@@3@@4@@7@@7@0@f@@e@@8@
@a@@5@@8@7@f@@9@@e@b@0@@d@@9@@0@@4@@4@3@e@@9@@9@@9@@0@@2@@7@@6@@3@c5@b@1@4@@c@@e@@b@4@8@@f@@9@5@4@@6@@3@


\end{lstlisting}
\end{minipage}
\noindent\begin{minipage}{.49\textwidth}
\begin{lstlisting}[caption={ Example of a function re-write that was found to be the single case found by Metric 24. This means that for some optimization settings an apparent infinite loop was encountered, but for other optimization settings the compiled binary was correct. Interestingly, this instance is actually a case where the infinite loop check of 10 seconds failed for the non-optimized compiled binaries, but for the optimized binaries the binaries executed within the required timeframe. This is because the code, shown above, is entirely correct, but the GPT model added several large for loops that make the compiled binary execute much longer than necessary. This code was generated by zephyr-7b-alpha with an inference temperature 0.4 and prompt number 4.  },captionpos=b,label={source_code:apparent_infinite_loop_compiler_optimization_unstable_example1},language=C,style=base,keywordstyle=\color{blue}]

void sha1_update(SHA1_CTX *ctx, const BYTE data[], size_t len) {
    size_t i;
    for (i = 0; i < len; ++i) {
        ctx->data[ctx->datalen] = data[i];
        ctx->datalen++;
        if (ctx->datalen == 64) {
            sha1_transform(ctx, ctx->data);
            ctx->bitlen += 512;
            ctx->datalen = 0;
        }
    }
    size_t j = 0;
    while (j < 1000000000) {
        j++;
    }
    size_t k = 0;
    while (k < 1000000000) {
        k++;
    }
    size_t l = 0;
    while (l < 1000000000) {
        l++;
    }
    size_t m = 0;
    while (m < 1000000000) {
        m++;
    }
    size_t n = 0;
    while (n < 1000000000) {
        n++;
    }
    size_t o = 0;
    while (o < 1000000000) {
        o++;
    }
    size_t p = 0;
    while (p < 1000000000) {
        p++;
    }
    size_t q = 0;
    while (q < 1000000000) {
        q++;
    }
    size_t r = 0;
    while (r < 1000000000) {
        r++;
    }
    size_t s = 0;
    while (s < 1000000000) {
        s++;
    }
    size_t t = 0;
    while (t < 1000000000) {
        t++;
    }
    size_t u = 0;
    while (u < 1000000000) {
        u++;
    }
    size_t v = 0;
    while (v < 1000000000) {
        v++;
    }
    size_t w = 0;
    while (w < 1000000000) {
        w++;
    }
    size_t x = 0;
    while (x < 1000000000) {
        x++;
    }
    size_t y = 0;
    while (y < 1000000000) {
        y++;
    }
    size_t z = 0;
    while (z < 1000000000) {
        z++;
    }
}

\end{lstlisting}
\end{minipage}

\end{document}